\documentclass[preprint,aps]{revtex4-1}
\pdfoutput=1

\usepackage{amsmath,amssymb,amsfonts,dcolumn,color,graphicx,graphics,latexsym,placeins,epsfig}
\usepackage{epsfig}
\usepackage{bm}
\usepackage{slashed}
\usepackage{latexsym}
\usepackage{natbib}
\usepackage{url}
\usepackage{dcolumn}
\usepackage{color}
\usepackage{amsfonts,amssymb,amsmath}
\usepackage{graphicx,epsfig}
\usepackage{psfrag}
\usepackage{subfigure}
\usepackage{tabularx}
\usepackage{hyperref}
\hypersetup{colorlinks=true}

\newcommand{\be}{\begin{equation}}
\newcommand{\ee}{\end{equation}}
\newcommand{\ba}{\begin{eqnarray}}
\newcommand{\ea}{\end{eqnarray}}

\begin{document}

\title{Listening to dark sirens from gravitational waves: \\ \it{Combined effects of fifth force, ultralight particle radiation, and eccentricity}  }
\author{Tanmay Kumar Poddar$^{1}$\footnote{tanmay.poddar@tifr.res.in}}
\author{Anish Ghoshal$^{2}$\footnote{anish.ghoshal@fuw.edu.pl}}
\author{Gaetano Lambiase$^{3,4}$\footnote{lambiase@sa.infn.it}}

\affiliation { $^{1}$\it Department of Theoretical Physics, Tata Institute of Fundamental Research, Mumbai-400005, India}

\affiliation{ $^{2}$ Institute  of  Theoretical  Physics,  Faculty  of  Physics,University  of  Warsaw,ul.   Pasteura  5,  02-093  Warsaw,  Poland}

\affiliation{$^{3}$INFN Sezione di Napoli, Gruppo collegato di Salerno, I-84084 Fisciano (SA), Italy}

\affiliation{ $^{4}$ Dipartimento di Fisica ``E.R. Caianiello”, Universit'a di Salerno, I-84084 Fisciano (SA), Italy} 

\begin{abstract}  
We derive in detail the orbital period loss of a compact binary system in presence of a fifth force and radiation of ultralight particles for a general eccentric Keplerian orbit. We obtain constraints on fifth force strength $\alpha\lesssim 1.11\times 10^{-3}$ from the orbital period decay of compact binary systems. We derive constraints on the gauge coupling of ultralight scalar $(g_S\lesssim 3.06\times 10^{-20})$ and vector $(g_V\lesssim 2.29\times 10^{-20})$ particles from orbital period loss and the constraints get stronger in presence of a fifth force $(\alpha=0.9)$. In addition, we also obtain constraints on the axion decay constant $(7.94\times 10^{10}~\rm{GeV}\lesssim f_a\lesssim 3.16\times 10^{17}~\rm{GeV}, \alpha=0.9)$ if the orbital period decays due to the combined effects of axionic fifth force and axion radiation. We also achieve constraints on the strengths of the fifth force $(\alpha\lesssim 0.025)$ and radiation $(\beta\lesssim 10^{-3})$ from GW170817. The constraints on new force parameters depend on the choice of the initial eccentricity which we include in our analysis $(\epsilon_0=10^{-6}, 0.1)$. We do the model independent estimate of the capture of dark matter mass fraction by a binary system. Lastly, we obtain constraints on fifth force strength due to Brans-Dicke mediated scalar between two compact stars in a binary system $(\omega_{\rm{BD}}>266)$ and from the Nordtvedt effect $(\omega_{\rm{BD}}>75858)$. The bound on Brans-Dicke coupling gets stronger if one includes the effect of eccentricity. Our constraints can be generalized to any alternative theories of gravity and will be within the reach of second and third generation gravitational wave detectors. 
\end{abstract}

\maketitle
\section{Introduction}
The Standard Model (SM) of particle physics and Einstein's General Relativity (GR) theory are the two pillars to understand the four fundamental forces in nature. There are several observations and experiments which are done to test these theories with a great level of accuracy. The success of these two theories has been consolidated with the discoveries of Higgs Boson in 2012 at the Large Hadron Collider (LHC) \cite{CMS:2012qbp,ATLAS:2012yve} and the direct detection of Gravitational Waves (GW) in 2015 at LIGO/Virgo \cite{LIGOScientific:2016aoc}. However, there are several observations and indications such as neutrino mass \cite{ParticleDataGroup:2020ssz}, matter-antimatter asymmetry \cite{Sakharov:1967dj}, strong CP problem \cite{Peccei:1977hh}, Dark Matter (DM) \cite{Planck:2015fie,Planck:2018vyg}, fifth force \cite{Fischbach:1985tk}, cosmological fine-tuning problem \cite{Joyce:2014kja}, Black Hole (BH) singularity \cite{Wald:1984rg}, etc. which cannot be explained from these two theories. Hence, stepping beyond SM and Einstein's GR theory is necessary to encounter these signals. In this paper, we obtain constraints on some of these New Physics (NP) scenarios from indirect and direct detection of GW.  

The measurement of orbital period loss of Hulse-Taylor compact binary system \cite{Hulse:1974eb} is the first indirect evidence of GW radiation that validates Einstein's GR theory with less than $ 0.1\%$ uncertainty \cite{Weisberg:2016jye}. 
Ultralight scalar, pseudoscalar (axion), and vector particles can mediate between the two compact stars in the binary system and give rise to a long range Yukawa type fifth force \cite{Hook:2017psm,Huang:2018pbu,Kopp:2018jom,Poddar:2021ose,Poddar:2020qft,Poddar:2021sbc,KumarPoddar:2020kdz,KumarPoddar:2019ceq,KumarPoddar:2019jxe}. The range of the fifth force is equal to the inverse of the mediator mass. The mediator mass is constrained by the distance between the two stars in the binary system. The strength of the fifth force should be less than that of gravity otherwise we could detect it. The orbital period loss of the compact binary system can be affected due to the presence of a fifth force. Besides the fifth force, the radiation of ultralight particles from the binary system also contributes to the orbital period loss. However, the contribution of fifth force and ultralight particle radiation is limited to be no larger than the measurement uncertainty $(< 0.1\%)$ \cite{Weisberg:2016jye}. 

The first direct detection of GW by coalescence of two stellar mass BHs has strengthened Einstein's GR theory further. For GW170817, the coalescence time of two neutron stars (NSs) is also modified due to the effect of the fifth force and radiation of ultralight particles. We consider that the contribution of the fifth force and radiation should be within the uncertainty of $0.4\%$ in reconstructing the Chirp mass due to the unknown source distance \cite{LIGOScientific:2017vwq}.

In deriving the fifth force strengths and gauge couplings of scalar and vector gauge bosons from orbital period loss of compact binary systems, we consider eccentric orbits. The contribution of eccentricity is important as it can enhance the energy loss of gravitational wave radiation by one order of magnitude compared to $\epsilon\rightarrow 0$ limit for Hulse-Taylor compact binary system. The Hulse-Taylor binary system (PSR B1913+16) \cite{Hulse:1974eb,Taylor:1982zz,Weisberg:1984zz} is an NS-pulsar binary system and the eccentricity of this Keplerian orbit is $\epsilon=0.617127$. We have also considered the orbital period loss of a double pulsar binary PSR J0737-3039 \cite{Kramer:2006nb}, and two NS-White-Dwarf (WD) binary systems such as PSR J0348+0432 \cite{Antoniadis:2013pzd} and PSR J1738+0333 \cite{Freire:2012mg}. The amplitude of the GWs emitted from such quasi-stable compact binary systems is so small that LIGO/Virgo cannot detect such signals. The orbital period loss of the compact binary systems is a consequence of GW radiation which can be indirectly tested from the Pulsar Timing Array (PTA).

We also obtain constraints on axion parameters i.e; mass $(m_a)$ and axion decay constant $(f_a)$ from the orbital period loss of compact binary systems. The axion is a pseudoscalar boson which is a promising candidate for DM \cite{Preskill:1982cy,Abbott:1982af,Dine:1982ah}. The QCD (quantum chromodynamics) axion was first proposed by Peccei and Quinn to solve the strong CP problem \cite{Peccei:1977hh,Weinberg:1977ma,Wilczek:1977pj,Peccei:1977ur}. There are other ultralight axion like particles (ALPs) which arise from string compactifications \cite{Svrcek:2006yi}. The constraints on axion parameters from astrophysics and laboratory experiments are discussed in \cite{Inoue:2008zp,CAST:2008ixs,Hannestad:2005df,Melchiorri:2007cd,Hannestad:2008js,Hamann:2009yf,Semertzidis:1990qc,Cameron:1993mr,Robilliard:2007bq,GammeVT-969:2007pci,Sikivie:2007qm,Kim:1986ax,Cheng:1987gp,Rosenberg:2000wb,Hertzberg:2008wr,Visinelli:2009zm,Battye:1994au,Yamaguchi:1998gx,Hagmann:2000ja}. The axion can mediate a fifth force between two compact stars of the binary system if the mass of the axion is smaller than the inverse distance between the two stars. The axion can also radiate from the binary system if the mass of the axion is smaller than the orbital frequency of the binary system. From the orbital period loss of compact binary systems, we obtain constraints on the axion decay constant and axion mass from the fifth force and combined effects of the fifth force and radiation of ultralight axions.

The observation of GW from the NS-NS merger has opened up new possibilities to seek into the dark sectors and its effect in the postmerger frequency spectrum might be detectable in the future GW signals from NS mergers. The dark sector could have a significant effect on the NS mass radius relation \cite{Karkevandi:2021ygv}, and more recently the possible effect of DM on the tidal deformability of a NS has been considered in several kinds of literature \cite{Nelson:2018xtr,Chatziioannou:2020pqz}.

Observing binaries at a lower frequency in the millihertz range may give us lots of useful information about the formation channel, see Refs \cite{Nishizawa:2016jji,Nishizawa:2016eza}. Isolated binaries typically have very tiny eccentricity while dynamically formed binaries may possess observable large eccentricity. Therefore, measurements of eccentricity in LIGO could be an important way to differentiate among various formation channels. One way to differentiate the two channels is to measure the statistical distributions of binary parameters, including the spin alignment and the eccentricity (for such prospects in LIGO and LISA, see Refs \cite{Nishizawa:2016jji,Nishizawa:2016eza,Breivik:2016ddj,Randall:2017jop,Randall:2018qna,Randall:2018lnh}).

For the inspiral binary, the value of the eccentricity is not fixed and it varies with time and orbital frequency. Here, we consider different values of initial eccentricity. Hence, one can obtain different constraints on fifth force parameters and gauge couplings of ultralight particles for different initial eccentricity values. The shift of coalescence time with respect to the pure gravity scenario changes with different choices of initial eccentricity. We also calculate the GW amplitude for different initial eccentricity values and obtain the variation of amplitude with time for different values of initial eccentricity, fifth force and coupling parameters. These results can be generalized to other gravitational wave detectors such as advanced LIGO, Einstein Telescope, LISA, Cosmic Explorer, KAGRA etc \cite{KAGRA:2013rdx,Hild:2009ns,Will:2014kxa}. 

The constraints on NP parameters are derived from indirect and direct detection of GW, keeping in mind that Einstein's GR theory is the correct theory of gravity. In fact, Einstein's GR theory is well-tested in low energy and weak gravity regime. However, in high energy and a strong gravity limit, Einstein's GR theory fails. The motivation for studying alternative theories of gravity such as scalar-tensor theory is to explain gravity in a strong field and high energy regime. In the Brans-Dicke (BD) scalar-tensor theory of gravity, the scalar and the tensor field are coupled non minimally and gravity is mediated by both scalar and tensor fields \cite{Will:1993hxu,Clifton:2011jh,Esposito-Farese:2009ouh,Fujii:2003pa,Brans:1961sx}. We obtain constraints on coupling and scalar mass from orbital period loss of compact binary systems in BD theory. Here, we only consider that the BD scalar mediates the fifth force which causes the orbital period loss of compact binary systems. We also obtain constraints on BD coupling from the Nordtvedt effect. In both cases, we include the effect of eccentricity. The bounds on the BD parameters from other experiments are stated in \cite{Alsing:2011er,Perivolaropoulos:2009ak,Will:1989sk,Berti:2012bp,Seymour:2019tir}.

The constraints on gauge coupling due to the radiation of ultralight scalar, vector, and axion particles from orbital period loss of compact binary systems are derived in \cite{Hook:2017psm, KumarPoddar:2019jxe,KumarPoddar:2019ceq}. However, the effect of the fifth force is not been included in obtaining these gauge coupling constraints. In this paper, we obtain constraints on these gauge coupling due to the combined effect of the fifth force and radiation of ultralight particles. In \cite{Alexander:2018qzg}, the constraints on fifth force strength are obtained for the fifth force range $\lambda\lesssim 10^{6}~\rm{km}$. In this paper, we obtain the constraints on fifth force strength for $\lambda\gtrsim 10^{6}~\rm{km}$ from orbital period loss of compact binary systems. We properly include the effect of eccentricity in deriving these bounds. The constraints on fifth force strength and gauge coupling for the ultralight particles are obtained in \cite{Alexander:2018qzg, Kopp:2018jom} from the coalescence of NSs in a binary system. The effect of eccentricity is neglected in these studies. In this paper, we obtain constraints on fifth force strength and gauge coupling due to the effects of fifth force and radiation of scalar and vector particles from LIGO/Virgo data. We obtain these bounds for different choices of initial eccentricity values. In \cite{Will:1989sk,Alsing:2011er}, the authors obtained constraints on the BD parameters due to the radiation of massive BD scalar particles from orbital period loss of compact binary systems. In our paper, we obtain constraints on BD parameters due to the fifth force effect for an eccentric orbit, causing the orbital period loss of the binary system. The bound on BD coupling can also be derived from the Nordtvedt effect as it is discussed in \cite{Will:1989sk,Alsing:2011er}. However, the effect of eccentricity was not taken into account in these studies. Here, We obtain constraints on BD coupling from the Nordtvedt effect including the eccentricity contribution.  

The rest of the paper consists of the following sections. In Section \ref{sec2}, we derive the expression for orbital period loss due to the effect of the fifth force for an eccentric compact binary system. We also obtain constraints on the fifth force strength and range by considering that orbital period loss decreases due to only the fifth force effect together with the gravitational wave radiation. We also derive the bounds on axion decay constant and axion mass from orbital period decay if the fifth force is mediated by ultralight pseudoscalar axion. In Section \ref{radiation}, we derive the expression for orbital period loss due to the combined effect of fifth force and ultralight particle radiation. We obtain constraints on gauge coupling from the orbital period loss due to the combined effect of radiation of scalar/vector particles and fifth force effects. We derive bounds on axion decay constant and axion mass as a special case if the mediator particle is the pseudoscalar axion. In section \ref{direct}, we obtain constraints on fifth force strength, gauge coupling, and mass of ultralight particles from LIGO/Virgo data. In deriving these bounds, we include the effects of eccentricity. In Section \ref{amplitudeGW}, we obtain a variation of GW amplitude with time for different choices of initial eccentricity and couplings of fifth force and radiation. In Section \ref{capd}, we qualitatively obtain the capture of DM mass fraction by a compact binary system. In Section \ref{bd}, we also obtain constraints on fifth force parameters from the orbital period decay for an eccentric orbit and from the Nordtvedt effect in the Brans-Dicke theory of gravity. Finally, in section \ref{con} we conclude and discuss our results. 

We use natural units $(c=\hbar=1)$ throughout the paper.

\section{Orbital period loss due to the presence of a fifth force}\label{sec2}
The total energy of the compact binary system per unit mass is modified due to the presence of a long range Yukawa potential $V(r)=\frac{g^2 Qq}{4\pi r}e^{-M_{Z^\prime}r}$, which can give rise to a fifth force. Here, $g$ denotes the strength of the fifth force, Q and q are the dark matter charges in two compact stars of the binary system, and $\lambda\sim\frac{1}{M_{Z^\prime}}$ denotes the range of the fifth force. The effect of the fifth force vanishes if any of the Q or q or both are zero.  In presence of Yukawa type fifth force, the total energy per unit mass is \cite{KumarPoddar:2020kdz} 
\begin{equation}
E\simeq 1-\frac{GM}{2r}+\frac{g^2 Qq}{4\pi m}\left(\frac{u_{+}u_{-}^2e^{-M_{Z'}/u_{+}}-u_{+}^2u_{-}e^{-M_{Z'}/u_{-}}}{u_{+}^2-u_{-}^2}\right),
\label{tot_energy}
\end{equation}
where $M$ and $m$ denote the masses of the two compact stars in the binary system, $\epsilon$ is the eccentricity of the orbit, $u=u_{+}=1/{r(1+\epsilon)}$ (aphelion), $u=u_{-}=1/{r(1-\epsilon)}$ (perihelion), and $1$ in the right hand side is the rest energy per unit mass in the Minkowski background. We obtain the expression of $E$ from the condition $\frac{du}{d\phi}=0$ at $u_+$ and $u_-$.

Using Eq. \ref{tot_energy}, we can write the total potential as
\begin{equation}
\begin{split}
V(r)=-\frac{GMm}{r}-\frac{g^2Qq}{8\pi \epsilon r}\Big[(1+\epsilon)e^{-M_{Z^\prime}r(1+\epsilon)}-(1-\epsilon)e^{-M_{Z^\prime}r(1-\epsilon)}+\\
(1+\epsilon)^2M_{Z^\prime}r Ei(-M_{Z^\prime}r(1+\epsilon))-(1-\epsilon)^2M_{Z^\prime}r Ei(-M_{Z^\prime}r(1-\epsilon))\Big],
\end{split}
\label{Tot_pot}
\end{equation}
where $Ei(x)=-\int _{-x}^\infty \frac{e^{-t}}{t}dt$, denotes the exponential integral. Since we are only concerned with the long range Yukawa potential, we take the leading order term for the potential as
\begin{equation}
V(r)\approx-\frac{GMm}{r}-\frac{g^2Qq}{8\pi \epsilon r}\Big[(1+\epsilon)e^{-M_{Z^\prime}r(1+\epsilon)}-(1-\epsilon)e^{-M_{Z^\prime}r(1-\epsilon)}\Big]=-\frac{GMm}{r}-\frac{g^2Qq}{4\pi r}K(\epsilon),
\label{approxpot}
\end{equation}
where $K(\epsilon)=\frac{1}{2\epsilon}\Big[(1+\epsilon)e^{-M_{Z^\prime} r(1+\epsilon)}-(1-\epsilon)e^{-M_{Z^\prime} r(1+\epsilon)}\Big]$. In the $\epsilon\rightarrow 0$ limit, $K(\epsilon)$ becomes
\begin{equation}
\lim_{\epsilon\rightarrow 0}K(\epsilon)=-\Big[e^{-M_{Z^\prime}r}(-1+M_{Z^\prime}r)\Big]-\frac{1}{6}\Big[M_{Z^\prime}^2r^2(-3+M_{Z^\prime}r)\Big]\epsilon^2+\mathcal{O}(\epsilon^3).
\end{equation}
\begin{figure}[h]
\includegraphics[height=8cm]{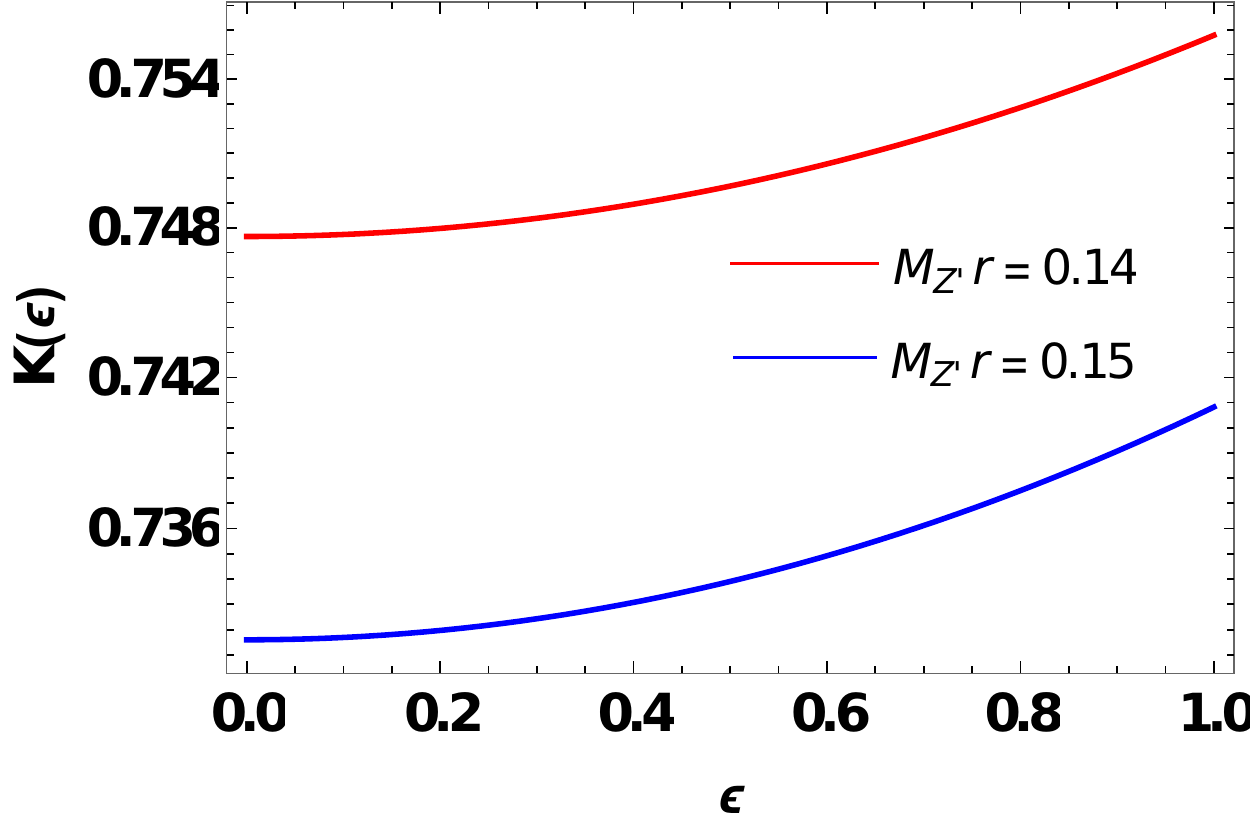}
\caption{\it Variation of $K(\epsilon)$ with $\epsilon$ for different values of $M_{Z^\prime}r$.}
\label{ec}
\end{figure} 
In FIG. \ref{ec} we obtain the variation of $K(\epsilon)$ with eccentricity for different values of $M_{Z^\prime}r$. The factor $K(\epsilon)$ is an increasing function of eccentricity. Hence, one should obtain stronger bounds on fifth force strength for high eccentric orbits. The function $K(\epsilon)$ also increases with decreasing $M_{Z^\prime}$. Therefore, the bound on fifth force strength becomes stronger for smaller values of $M_{Z^\prime}$.
In the limit $M_{Z^\prime}\rightarrow0$, we can write the total potential from Eq. \ref{Tot_pot} as
\begin{equation}
V(r)=-\frac{GMm}{r}(1+\alpha),
\end{equation}
where 
\begin{equation}
\alpha=\frac{g^2Qq}{4\pi GMm},
\label{strength}
\end{equation}
denotes the strength of the long range Yukawa type fifth force relative to Newton's gravitational force. For $M_{Z^\prime}r<<1$, we can write the total potential from Eq. \ref{Tot_pot} as
\begin{equation}
V(r)\simeq-\frac{GMm}{r}\Big[1+\alpha\{1-2M_{Z^\prime}r+(3+\epsilon^2)\frac{M^2_{Z^\prime}r^2}{2}\}\Big]+\mathcal{O}(M_{Z^\prime}^3).
\end{equation}
The orbital frequency of a binary system is related to the potential as $\Omega^2_f=\frac{M+m}{Mmr}\frac{dV}{dr}$ which changes due to the presence of the fifth force. Taking the derivative of Eq. \ref{Tot_pot} with respect to $r$, we can write the orbital frequency of a binary system as
\begin{equation}
\begin{split}
\Omega^2_f=\Omega^2\Big[1+\frac{\alpha}{2\epsilon}\Big\{(1+\epsilon)e^{-M_{Z^\prime}r(1+\epsilon)}-(1-\epsilon)e^{-M_{Z^\prime}r(1-\epsilon)}\Big\}+\\
\frac{\alpha M_{Z^\prime}r}{2\epsilon}\Big\{(1+\epsilon)^2e^{-M_{Z^\prime}r(1+\epsilon)}-(1-\epsilon)^2e^{-M_{Z^\prime}r(1-\epsilon)}\Big\}\Big].
\end{split}
\label{Tot_orb_freq}
\end{equation}
For $M_{Z^\prime}\rightarrow 0$, we obtain $\Omega^2_f=\frac{G(M+m)}{r^3}(1+\alpha)=\Omega^2(1+\alpha)$ as expected. $\Omega=\Big[\frac{G(M+m)}{r^3}\Big]^\frac{1}{2}$ is the fundamental frequency which does not include any fifth force effect.

The total energy loss due to the gravitational wave radiation is \cite{Poddar:2021yjd}
\begin{equation}
\frac{dE_{GW}}{dt}=\frac{32}{5}G\mu^2\Omega^6r^4(1-\epsilon^2)^{-7/2}\Big(1+\frac{73}{24}\epsilon^2+\frac{37}{96}\epsilon^4\Big),
\label{standard_GR}
\end{equation}
which modifies due to the presence of long range Yukawa potential as 
\begin{equation}
\begin{split}
\frac{dE}{dt}=\frac{dE_{GW}}{dt}\times\Big[1+\frac{\alpha}{2\epsilon}\Big\{(1+\epsilon)e^{-M_{Z^\prime}r(1+\epsilon)}-(1-\epsilon)e^{-M_{Z^\prime}r(1-\epsilon)}\Big\}+\\
\frac{\alpha M_{Z^\prime}r}{2\epsilon}\Big\{(1+\epsilon)^2e^{-M_{Z^\prime}r(1+\epsilon)}-(1-\epsilon)^2e^{-M_{Z^\prime}r(1-\epsilon)}\Big\}\Big]^3,
\end{split}
\label{nonzeroalpha}
\end{equation}
where $\mu=\frac{mM}{m+M}$ is the reduced mass of the compact binary system. If there is no long range Yukawa mediated fifth force then $\alpha=0$, and we get back the standard GR result (Eq. \ref{standard_GR}), although the $M_{Z^\prime}$ need not be zero in that case. However, if $M_{Z^\prime}\rightarrow 0$ (infinite range fifth force) and $\alpha\neq 0$, then the energy loss in GR will be modified as 
\begin{equation}
\Big(\frac{dE}{dt}\Big)^{M_{Z^\prime}\rightarrow 0}_{\alpha\neq 0}=\frac{32}{5}G\mu^2\Omega^6r^4(1-\epsilon^2)^{-7/2}\Big(1+\frac{73}{24}\epsilon^2+\frac{37}{96}\epsilon^4\Big)(1+\alpha)^3.
\end{equation}
The rate of orbital period decreases due to the presence of long range Yukawa potential as 
\begin{equation}
 \dot{P_b}=-6\pi G^{-3/2}(m_1m_2)^{-1}(m_1+m_2)^{-1/2}r^{5/2}\Big(\frac{dE}{dt}\Big),
\end{equation}
where $\frac{dE}{dt}$ is given by Eq. \ref{nonzeroalpha}. In the following, we consider four compact binary systems and obtain constraints on the strength $(\alpha)$, and the range $\Big(\lambda=\frac{1}{M_{Z^\prime}}\Big)$ of the fifth force from the orbital period loss of these compact binary systems.
\begin{itemize}
 \item \textbf{PSR B1913+16:} The Hulse-Taylor binary system or PSR B1913+16 \cite{Taylor:1993an} consists of one NS and one pulsar. The masses of the two stars are $M=1.42~M_{\odot}$ and $m=1.4 ~M_{\odot}$. The binary orbit is highly eccentric with eccentricity $\epsilon=0.617127$. The average orbital frequency of the binary system in natural units is $\Omega=1.48\times 10^{-19}~\rm{eV}$. The semi major axis of the orbit can be calculated from Kepler's law as $r=1.087\times 10^{16}~\rm{eV^{-1}}$. The GR predicted value of the orbital period loss is $\dot{P_b}_{\rm{GR}}=-2.4025\pm 0.0001\times 10^{-12}~\rm {s\hspace{0.1cm}s^{-1}}$ whereas the observed value of the orbital period loss is $\dot{P_b}=-2.4225\pm 0.0056\times 10^{-12}~\rm {s\hspace{0.1cm}s^{-1}}$.
 \item \textbf{PSR J0737-3039:} PSR J0737-3039  \cite{Kramer:2006nb} consists of two pulsars whose masses are $M=1.338 ~M_{\odot}$ and $m=1.250 ~M_{\odot}$. The eccentricity of the binary orbit is $\epsilon=0.087$. The orbital period of the binary system is $P_b=2.4~\rm{h}$ and the orbital frequency of the binary system is $\Omega=4.79\times 10^{-19}~\rm{eV}$. The semi major axis of the binary system can be calculated using Kepler's law and its value is $r=4.83\times 10^{15}~\rm{eV^{-1}}$. The GR predicted value of the orbital period loss is $\dot{P_b}_{\rm{GR}}=1.24787(13)\times 10^{-12}~\rm {s\hspace{0.1cm}s^{-1}}$ whereas the observed value of the orbital period loss is $\dot{P_b}=1.252(17)\times 10^{-12}~\rm {s\hspace{0.1cm}s^{-1}}$. The numbers in the brackets denote the uncertainty in the last significant digit.
 \item \textbf{PSR J0348+0432:} PSR J0348+0432 \cite{Antoniadis:2013pzd} compact binary system consists of one pulsar and one WD where the mass of the pulsar is $M=2.01 ~M_{\odot}$ and the mass of the WD is $m=0.172 ~M_{\odot}$. The eccentricity of the binary orbit is $\epsilon=10^{-6}$. The average orbital period of the binary system is $P_b=2.46~\rm{h}$, and the corresponding orbital frequency is $\Omega=4.67\times 10^{-19}~\rm{eV}$. The semi major axis of the binary orbit is $r=4.64\times 10^{15}~\rm{eV^{-1}}$. The GR predicted value of the orbital period loss is $\dot{P_b}_{\rm{GR}}=-0.258^{+0.008}_{-0.011}\times 10^{-12}~\rm {s\hspace{0.1cm}s^{-1}}$ and the observed value of the orbital period loss is $\dot{P_{b}}=-0.273(45)\times 10^{-12}~\rm {s\hspace{0.1cm}s^{-1}}$.
 \item \textbf{PSR J1738+0333:} PSR J1738+0333 \cite{Freire:2012mg} is also a pulsar WD binary system where the mass of the pulsar is $M=1.46 ~M_{\odot}$ and the mass of the WD is $m=0.181 ~M_{\odot}$. The binary orbit has a very low eccentricity as $\epsilon<3.4\times 10^{-7}$. The average orbital period of the binary system is $P_b=8.5~\rm{h}$ with the corresponding orbital frequency $\Omega=1.35\times 10^{-19}~\rm{eV}$. The semi major axis of the binary system is $r=9.647\times 10^{15}~\rm{eV^{-1}}$. The intrinsic orbital period decay is $\dot{P_b}=(-25.9\pm 3.2)\times 10^{-15}~\rm {s\hspace{0.1cm}s^{-1}}$ whereas the GR predicted value of the orbital period loss is $\dot{P_b}_{\rm{GR}}=-27.7^{+1.5}_{-1.9}\times 10^{-15}~\rm {s\hspace{0.1cm}s^{-1}}$.
\end{itemize}
\subsection{Constraints on strength and range of fifth force from orbital period loss of compact binary systems}
\begin{figure}[h]
\includegraphics[height=8cm]{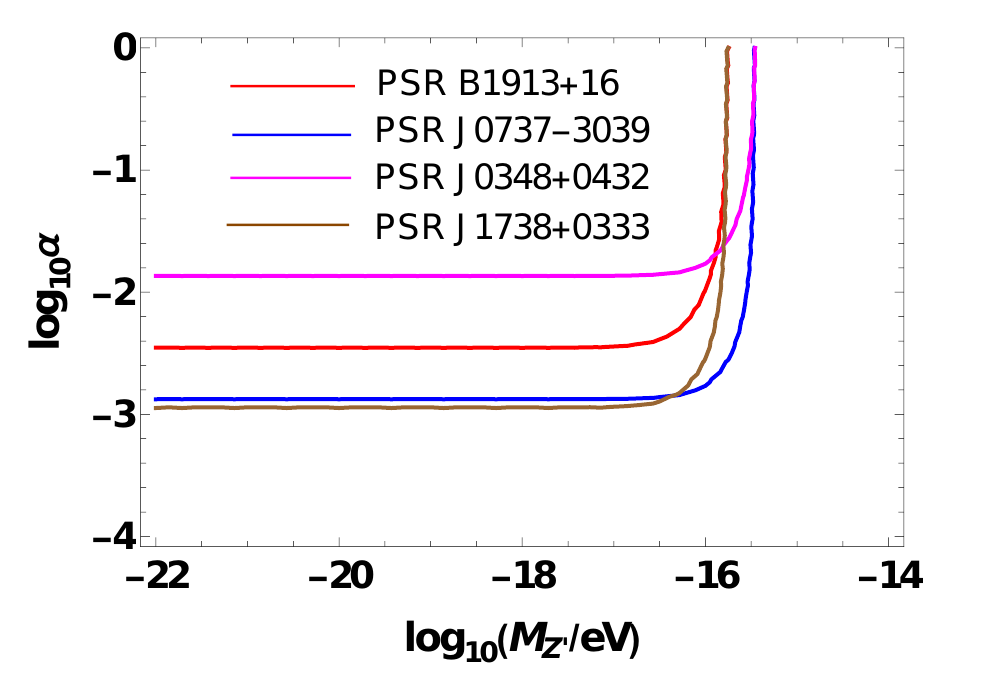}
\caption{\it Constraints on fifth force strength and range for four compact binary systems}
\label{plot1}
\end{figure} 
We consider the above four compact binary systems to constrain the fifth force parameters $(\alpha,M_{Z^\prime})$. The fifth force can only contribute to the orbital period loss if the mass of the fifth force mediator is less than the inverse of the binary separation and both compact stars contain dark charge particles. We consider that the contribution of the fifth force should be within the experimental uncertainty in the measurement of the orbital period loss. We use Eq. \ref{standard_GR} and Eq. \ref{nonzeroalpha} to obtain the constraints on the fifth force parameters. In this section, we consider that the orbital period only decreases due to the gravitational wave radiation and the fifth force effect. The orbital period can also decay due to the radiation of ultralight particles. We will discuss this possibility in the next section. Here, we consider the charge to mass asymmetry in the two stars of the binary system to be zero to eliminate the radiation effect. In FIG. \ref{plot1} we obtain constraints on $\alpha$ and $M_{Z^\prime}$ from the orbital period loss (indirect evidence of gravitational waves) of these four compact binary systems. The red line denotes the variation of the relative Yukawa fifth force strength $(\alpha)$ with the inverse of its range $(\lambda^{-1}= M_{Z^\prime})$ for PSR B1913+16. Similarly, the blue, magenta, and brown lines denote the same variations for PSR J0737-3039, PSR J0348+0432, and PSR J1738+0333 respectively. The regions above these lines are excluded. The upper bounds on the fifth force strength are 
\begin{eqnarray}
\alpha &\lesssim & 3.51\times 10^{-3} \quad \text{for PSR B1913+16}, \\
\alpha &\lesssim & 1.33\times 10^{-3} \quad \text{for PSR J0737-3039}, \\
\alpha &\lesssim & 1.36\times 10^{-2} \quad \text{for PSR J0348+0432}, \\
\alpha &\lesssim & 1.11\times 10^{-3} \quad \text{for PSR J1738+0333}. \label{e16}
\end{eqnarray}
We obtain the stronger bound on $\alpha$ as $\alpha\lesssim 1.11\times 10^{-3}$ from PSR J1738+0333. These bounds are only valid for the mass of the mediator $M_{Z^\prime}\lesssim 9.19\times 10^{-17}~\rm{eV}$ $(\lambda\gtrsim 2.15\times 10^6~\rm{km})$. The range of the fifth force is constrained by the distance between the two stars of the binary systems. In \cite{Alexander:2018qzg}, the authors obtain the constraints on fifth force strength from the projected sensitivity of the Einstein Telescope. Their bounds are only valid for the fifth force range $5~\rm{km}<\lambda<10^6~\rm{km}$ whereas our bounds on fifth force strength are valid for $\lambda\gtrsim 2.15\times 10^6~\rm{km}$ which we obtain from the orbital period loss of compact binary systems. 
\subsection{Constraints on axion decay constant from fifth force effect in orbital period loss of compact binary systems}
\begin{figure}[h]
\includegraphics[height=8cm]{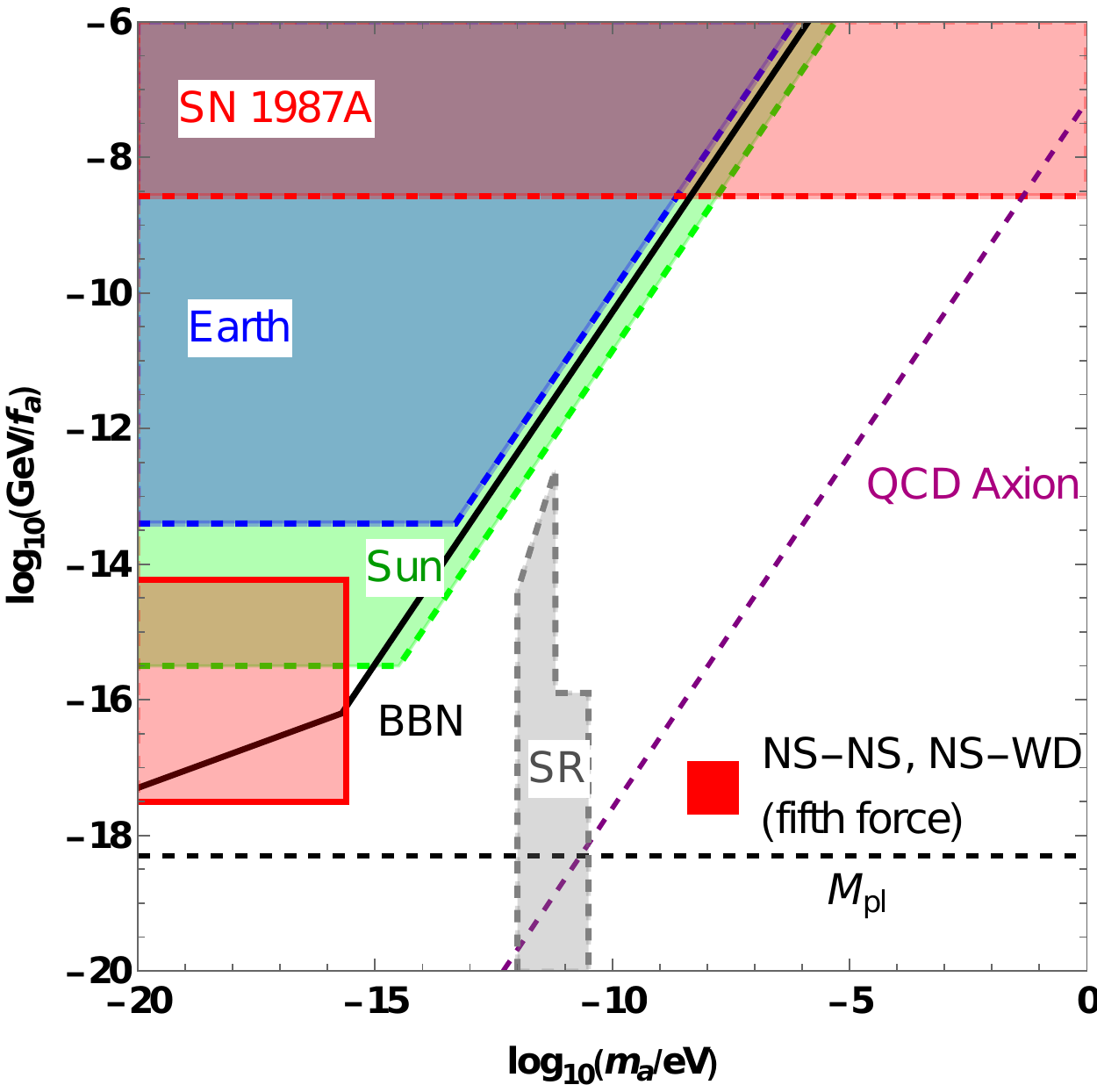}
\caption{\it Constraints on axion decay constant from SN1987A \cite{Hook:2017psm}, Sun \cite{Hook:2017psm}, Earth \cite{Hook:2017psm}, BBN \cite{Blum:2014vsa}, Superradiance \cite{Arvanitaki:2010sy,Arvanitaki:2014wva}, and orbital period loss of compact binary systems due to fifth force (this work).}
\label{pcap3}
\end{figure}
If compact stars (NS, WD) are immersed in a low mass axionic potential and if axions have coupling with the nucleons then axions are sourced by compact stars \cite{Hook:2017psm}. The finite density correction in the axion potential results in a long range axion field outside of the star. The axionic charge $(Q=-\frac{8\pi G Mf_a}{\ln\Big(1-\frac{2GM}{R}\Big)}$, where $R$ is the radius of the compact star, and $M$ is the mass of the compact star \cite{KumarPoddar:2019jxe}) in the compact star results in a long range Yukawa type fifth force between the two stars in the binary system. The orbital period of the compact binary system decreases primarily due to the gravitational wave radiation which matches quite well with Einstein's GR prediction. If the compact stars contain an axion charge then the axion mediated fifth force can also contribute to the orbital period loss. However, the contribution of the fifth force should be within the measurement uncertainty. The compact star can be the source of axions if the radius of the compact star is greater than a critical radius $(r_c)$, given as \cite{Hook:2017psm} 
\begin{equation}
r_c>\frac{1}{m_T},
\label{cond}
\end{equation}
where $m_T$ is the tachyonic mass of the axion inside the star that is given as 
\begin{equation}
m_T=\frac{m_\pi f_\pi}{2f_a}\sqrt{\frac{\sigma_N n_N}{m^2_\pi f^2_\pi}-\epsilon},
\end{equation}
where $m_\pi$ is the pion mass, $f_\pi$ is the pion decay constant, $m_a$ is the axion mass, and $f_a$ is the axion decay constant. We choose the nucleon $\sigma$ term as $\sigma_N\sim 59~\rm{MeV}$ from lattice simulation, and $n_N$ denotes the nucleon number density. Considering $\epsilon<0.1$, the mass of the axion ($m_a=\frac{m_\pi f_\pi}{2f_a}\sqrt{\epsilon}$) becomes lighter than the QCD axion. In FIG. \ref{pcap3} we obtain constraints on axion parameters from the orbital period loss of four compact binary systems. The red shaded region bounded by the red dashed line is excluded from SN1987A \cite{Hook:2017psm}. The blue and the green shaded regions bounded by the blue and green dashed lines are excluded from the direct observation of the sun and the earth respectively \cite{Hook:2017psm}. The grey shaded region bounded by the grey dashed line is excluded from BH superradiance \cite{Arvanitaki:2010sy,Arvanitaki:2014wva}. The black solid line denotes the constraint from BBN if the axion constitutes all of the dark matter in the universe \cite{Blum:2014vsa}. The purple dashed line denotes the QCD axion and the black dashed line denotes the reduced Planck constant. The red shaded region bounded by the red solid line is excluded from the orbital period loss of PSR B1913+16, PSR J0737-3039, PSR J0348+0432, and PSR J1738+0333. We obtain these bounds by only considering the fifth force effect (no radiation). Using Eq. \ref{standard_GR}, Eq. \ref{nonzeroalpha}, and Eq. \ref{cond} we obtain that the orbital period loss of NS-NS, and NS-WD binary systems due to the fifth force effect rule out axions with decay constant $1.69\times 10^{14}~\rm{GeV}\lesssim f_a\lesssim 3.16\times 10^{17}~\rm{GeV}$ for masses $m_a\lesssim 2.51\times 10^{-16}~\rm{eV}$. Axions with decay constant $f_a\gtrsim 3.16\times 10^{17}~\rm{GeV}$ cannot be sourced by a compact star because in that case, the critical radius is so large that the compact star cannot trigger the phase transition. In other words, axions with $f_a\gtrsim 3.16\times 10^{17}~\rm{GeV}$ cannot be sourced by the compact stars even if they are free from orbital period loss due to fifth force constraints. In \cite{KumarPoddar:2019jxe}, the authors obtain constraints on axion parameters from the orbital period loss by only considering the radiation of axions. 
\section{Orbital period loss due to the combined effect of fifth force and radiation of ultralight particles}
\label{radiation}
Besides long range Yukawa type fifth force, the radiation of ultralight particles can also contribute to the orbital period loss of compact binary systems. The contribution of the fifth force and/or radiation is limited to be no larger than the measurement uncertainty of orbital period loss. If any one of $Q$ and $q$ is zero then the orbital period loss due to the Yukawa type fifth force is zero. However, in that case, there is always a possibility for the radiation of ultralight particles. The radiation of ultralight particles can contribute to the orbital period loss if the charge to the mass asymmetry of the binary stars is nonzero. Also, the mass of the ultralight particle should be less than the orbital frequency of the binary system for radiation to happen. We can write the energy loss due to the radiation of ultralight scalar particle as \cite{KumarPoddar:2019ceq} 
\begin{equation}
\Big(\frac{dE}{dt}\Big)_S=\frac{g^2_S}{12\pi}\Big(\frac{Mm}{M+m}\Big)^2 r^2\Omega^4\Big(\frac{Q}{M}-\frac{q}{m}\Big)^2\sum_{n>n_0}2n^2\Big[{J^\prime_n}^2(n\epsilon)+\Big(\frac{1-\epsilon^2}{\epsilon^2}\Big)J_n^2(n\epsilon)\Big]\Big(1-\frac{{n_0^S}^2}{n^2}\Big)^\frac{3}{2},
\label{rad_scalar}
\end{equation} 
and for the radiation of ultralight vector particle, the energy loss is
\begin{equation}
\Big(\frac{dE}{dt}\Big)_V=\frac{g^2_V}{6\pi}\Big(\frac{Mm}{M+m}\Big)^2 r^2\Omega^4\Big(\frac{Q}{M}-\frac{q}{m}\Big)^2\sum_{n>n_0}2n^2\Big[{J^\prime_n}^2(n\epsilon)+\Big(\frac{1-\epsilon^2}{\epsilon^2}\Big)J_n^2(n\epsilon)\Big]\sqrt{1-\frac{{n_0^V}^2}{n^2}}\Big(1+\frac{{n_0^V}^2}{2n^2}\Big),
\label{rad_vector}
\end{equation}
where $g_V(g_S)$ denotes the coupling, and $n^V_0(n^S_0)=\frac{M_{Z^\prime}}{\Omega}(\frac{M_S}{\Omega})$ is the ratio of the mass of the ultralight particle to the orbital frequency for vector (scalar). It is evident from Eq. \ref{rad_scalar} and Eq. \ref{rad_vector} that the radiation of scalar and vector particles can be possible if the charge to mass asymmetry $\Big(\frac{Q}{M}-\frac{q}{m}\Big)$ of the two stars in the binary system is nonzero. If $M_{S}\rightarrow 0$ (infinite range scalar radiation) then Eq. \ref{rad_scalar} becomes
\begin{equation}
\Big(\frac{dE}{dt}\Big)^{{M_S}\rightarrow 0}_S=\frac{g^2_S}{12\pi}\Big(\frac{Mm}{M+m}\Big)^2 r^2\Omega^4\Big(\frac{Q}{M}-\frac{q}{m}\Big)^2\frac{\Big(1+\frac{\epsilon^2}{2}\Big)}{(1-\epsilon^2)^\frac{5}{2}},
\label{rads}
\end{equation}
and if $M_{Z^\prime}\rightarrow 0$ (infinite range vector radiation), Eq. \ref{rad_vector} becomes
\begin{equation}
\Big(\frac{dE}{dt}\Big)^{M_{Z^\prime}\rightarrow 0}_V=\frac{g^2_V}{6\pi}\Big(\frac{Mm}{M+m}\Big)^2 r^2\Omega^4\Big(\frac{Q}{M}-\frac{q}{m}\Big)^2\frac{\Big(1+\frac{\epsilon^2}{2}\Big)}{(1-\epsilon^2)^\frac{5}{2}}.
\label{radv}
\end{equation}
If there is a contribution of both fifth force $(\alpha \neq 0)$, and radiation $\Big(\frac{Q}{M}-\frac{q}{m}\Big)\neq 0$ then Eq. \ref{rad_scalar} is modified as
\begin{equation}
\begin{split}
\Big(\frac{dE}{dt}\Big)_{Y+S}=\frac{g^2_S}{12\pi}\Big(\frac{Mm}{M+m}\Big)^2 r^2\Omega^4\Big(\frac{Q}{M}-\frac{q}{m}\Big)^2\sum_{n>n_{0\alpha}^S}2n^2\Big[{J^\prime_n}^2(n\epsilon)+\Big(\frac{1-\epsilon^2}{\epsilon^2}\Big)J_n^2(n\epsilon)\Big]\times\\
\Big(1-\frac{{n_{0\alpha}^S}^2}{n^2}\Big)^\frac{3}{2}
\Big[1+\frac{\alpha}{2\epsilon}\Big\{(1+\epsilon)e^{-M_{S}r(1+\epsilon)}-(1-\epsilon)e^{-M_{S}r(1-\epsilon)}\Big\}+\\
\frac{\alpha M_{S}r}{2\epsilon}\Big\{(1+\epsilon)^2e^{-M_{S}r(1+\epsilon)}-(1-\epsilon)^2e^{-M_{S}r(1-\epsilon)}\Big\}\Big]^2,
\end{split}
\label{modified_scalar}
\end{equation}
and the energy loss due to the combined effect of radiation of vector particles and fifth force becomes
\begin{equation}
\begin{split}
\Big(\frac{dE}{dt}\Big)_{Y+V}=\frac{g^2_V}{6\pi}\Big(\frac{Mm}{M+m}\Big)^2r^2\Omega^4\Big(\frac{Q}{M}-\frac{q}{m}\Big)^2\sum_{n>n_{0\alpha}^V}2n^2\Big[{J^\prime_n}^2(n\epsilon)+\Big(\frac{1-\epsilon^2}{e^2}\Big)J_n^2(n\epsilon)\Big]\times\\
\sqrt{1-\frac{{n_{0\alpha}^V}^2}{n^2}}\Big(1+\frac{{n_{0\alpha}^V}^2}{2n^2}\Big)
\Big[1+\frac{\alpha}{2\epsilon}\Big\{(1+\epsilon)e^{-M_{Z^\prime}r(1+\epsilon)}-(1-\epsilon)e^{-M_{Z^\prime}r(1-\epsilon)}\Big\}+\\
\frac{\alpha M_{Z^\prime}r}{2\epsilon}\Big\{(1+\epsilon)^2e^{-M_{Z^\prime}r(1+\epsilon)}-(1-\epsilon)^2e^{-M_{Z^\prime}r(1-\epsilon)}\Big\}\Big]^2,
\end{split}
\label{modified_vector}
\end{equation}
where the $(Y+S/V)$ in the subscripts of Eq. \ref{modified_scalar} and Eq. \ref{modified_vector} denote the presence of Yukawa potential together with the radiation of scalar/vector particles. The ratio of the mass of the ultralight particle and the orbital frequency is now modified as  $n_{0\alpha}^V$ which is given as
\begin{equation}
\begin{split}
n_{0\alpha}^V=\frac{M_{Z^\prime}}{\Omega}\Big[1+\frac{\alpha}{2\epsilon}\Big\{(1+\epsilon)e^{-M_{Z^\prime}r(1+\epsilon)}-(1-\epsilon)e^{-M_{Z^\prime}r(1-\epsilon)}\Big\}+\\
\frac{\alpha M_{Z^\prime}r}{2\epsilon}\Big\{(1+\epsilon)^2e^{-M_{Z^\prime}r(1+\epsilon)}-(1-\epsilon)^2e^{-M_{Z^\prime}r(1-\epsilon)}\Big\}\Big]^{-\frac{1}{2}}.
\end{split}
\end{equation}
The expression for $n_{0\alpha}^S$ is same as $n_{0\alpha}^V$ with replacing $M_{Z^\prime}$ as $M_S$.
If the mass of the scalar and vector particle is zero then Eq. \ref{modified_scalar} becomes
\begin{equation}
\Big(\frac{dE}{dt}\Big)^{M_{S}\rightarrow 0}_{Y+S}=\frac{g^2_S}{12\pi}\Big(\frac{Mm}{M+m}\Big)^2 r^2\Omega^4\Big(\frac{Q}{M}-\frac{q}{m}\Big)^2\frac{\Big(1+\frac{\epsilon^2}{2}\Big)}{(1-\epsilon^2)^\frac{5}{2}}(1+\alpha)^2.
\end{equation}
Similarly, Eq. \ref{modified_vector} becomes
\begin{equation}
\Big(\frac{dE}{dt}\Big)^{M_{Z^\prime}\rightarrow 0}_{Y+V}=\frac{g^2_V}{6\pi}\Big(\frac{Mm}{M+m}\Big)^2 r^2\Omega^4\Big(\frac{Q}{M}-\frac{q}{m}\Big)^2\frac{\Big(1+\frac{\epsilon^2}{2}\Big)}{(1-\epsilon^2)^\frac{5}{2}}(1+\alpha)^2.
\end{equation}
If one star of the compact binary system does not contain any charge (either $Q$ or $q$ is zero) then the Yukawa fifth force is zero but there is always an existence of charge to mass asymmetry and hence radiation. If both $Q$ and $q$ are zero or the charge to mass asymmetry is zero then there cannot be any radiation of particles for any value of $M_{Z^\prime}$. 

The inclusion of eccentricity in the calculation of orbital period loss is important as for a high eccentric orbit, the eccentricity enhancement factor takes a large value which increases the orbital period loss by several orders of magnitude. From Eq. \ref{standard_GR}, we obtain the eccentricity enhancement factor in the orbital period loss due to the gravitational wave radiation as
\begin{equation}
F_{\rm{GW}}(\epsilon)=\Big(1+\frac{73}{24}\epsilon^2+\frac{37}{96}\epsilon^4\Big)(1-\epsilon^2)^{-7/2}.
\end{equation}
Similarly, from Eq. \ref{rads}, Eq. \ref{radv} we obtain the eccentricity enhancement factor due to radiation of scalar/vector particles as
\begin{equation}
F_{S/V}(\epsilon)=\frac{\Big(1+\frac{\epsilon^2}{2}\Big)}{(1-\epsilon^2)^{5/2}}.
\end{equation}
\begin{figure}[h]
\includegraphics[height=8cm]{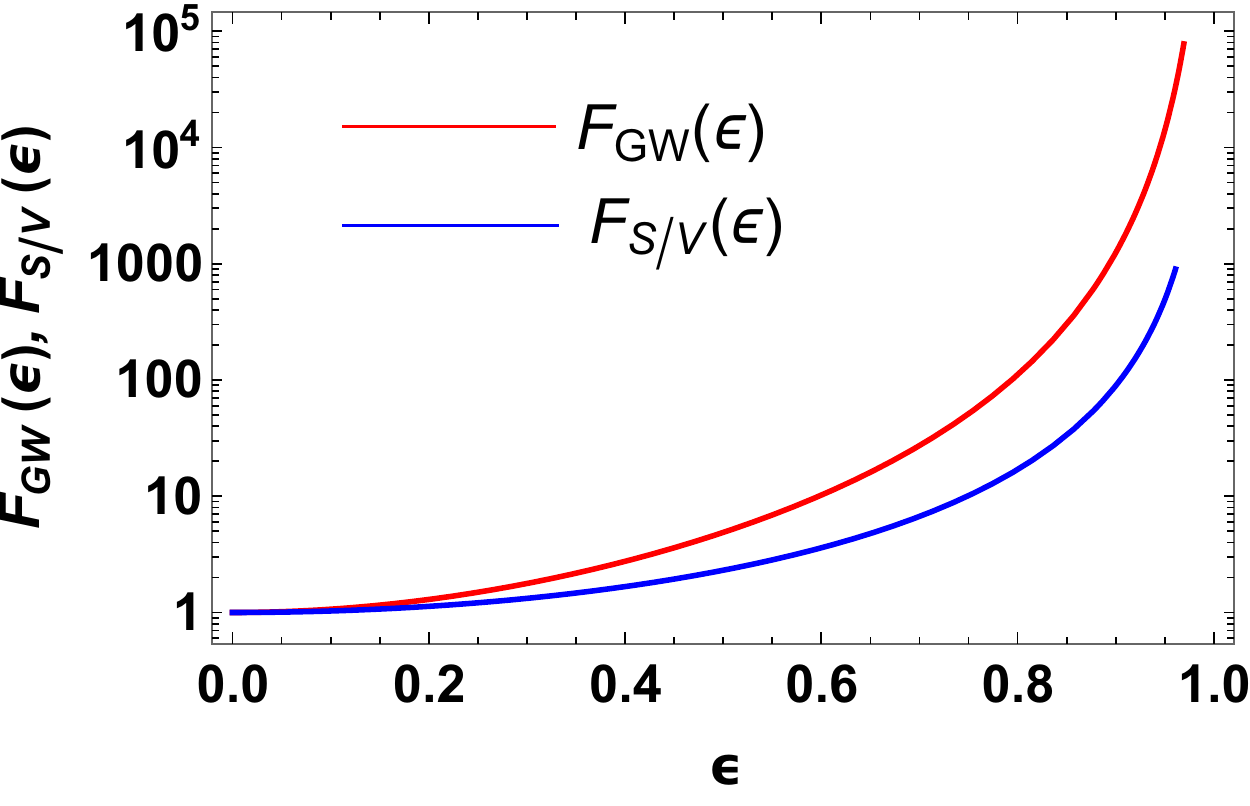}
\caption{\it Contribution of eccentricity enhancement factor in the measurement of orbital period loss}
\label{ploteccentricity}
\end{figure} 
In FIG. \ref{ploteccentricity} we have shown the variation of eccentricity enhancement factors $F_{\rm{GW}}(\epsilon)$ and $F_{\rm{S/V}}(\epsilon)$ with eccentricity $(\epsilon)$. The enhancement due to $F_{\rm{GW}}(\epsilon)$ is more than $F_{\rm{S/V}}(\epsilon)$. For HT binary system with $\epsilon=0.617127$, the enhancement factors are $F_{\rm{GW}}\sim 12$, and $F_{\rm{S/V}}\sim 4$. 
\subsection{Constraints on scalar and vector couplings from combined effect of fifth force and radiation in orbital period loss of compact binary systems}
\begin{figure}[!htbp]
\centering
\subfigure[$g_V$ vs. $M_{Z^\prime}$ for $\alpha=10^{-4}, 0.9$, and $Q= 10^{55}$]{\includegraphics[width=8cm]{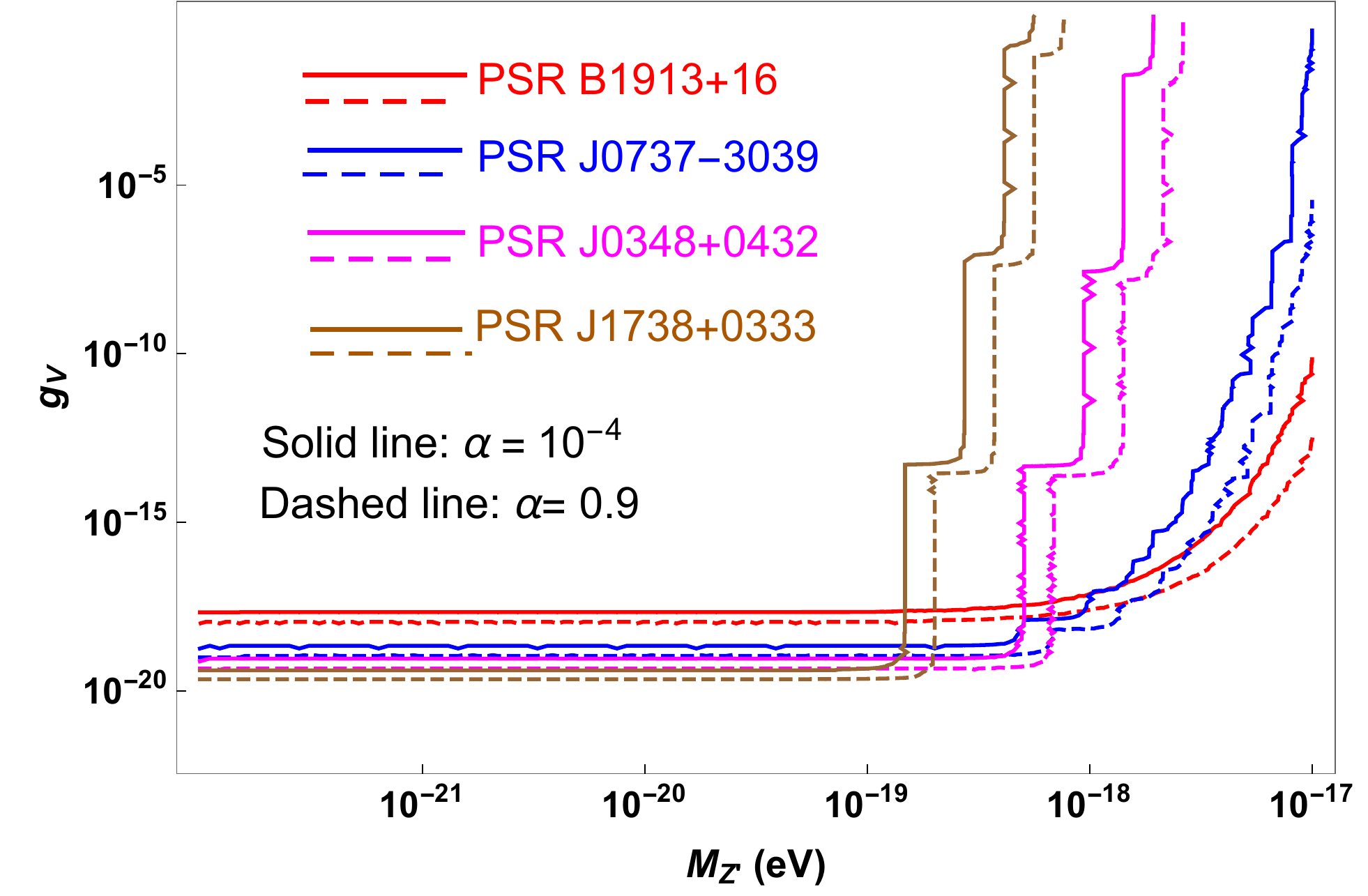}\label{plot2a}}
\subfigure[$g_S$ vs. $M_S$ for $\alpha=10^{-4}, 0.9$ and $Q=10^{55}$]{\includegraphics[width=8cm]{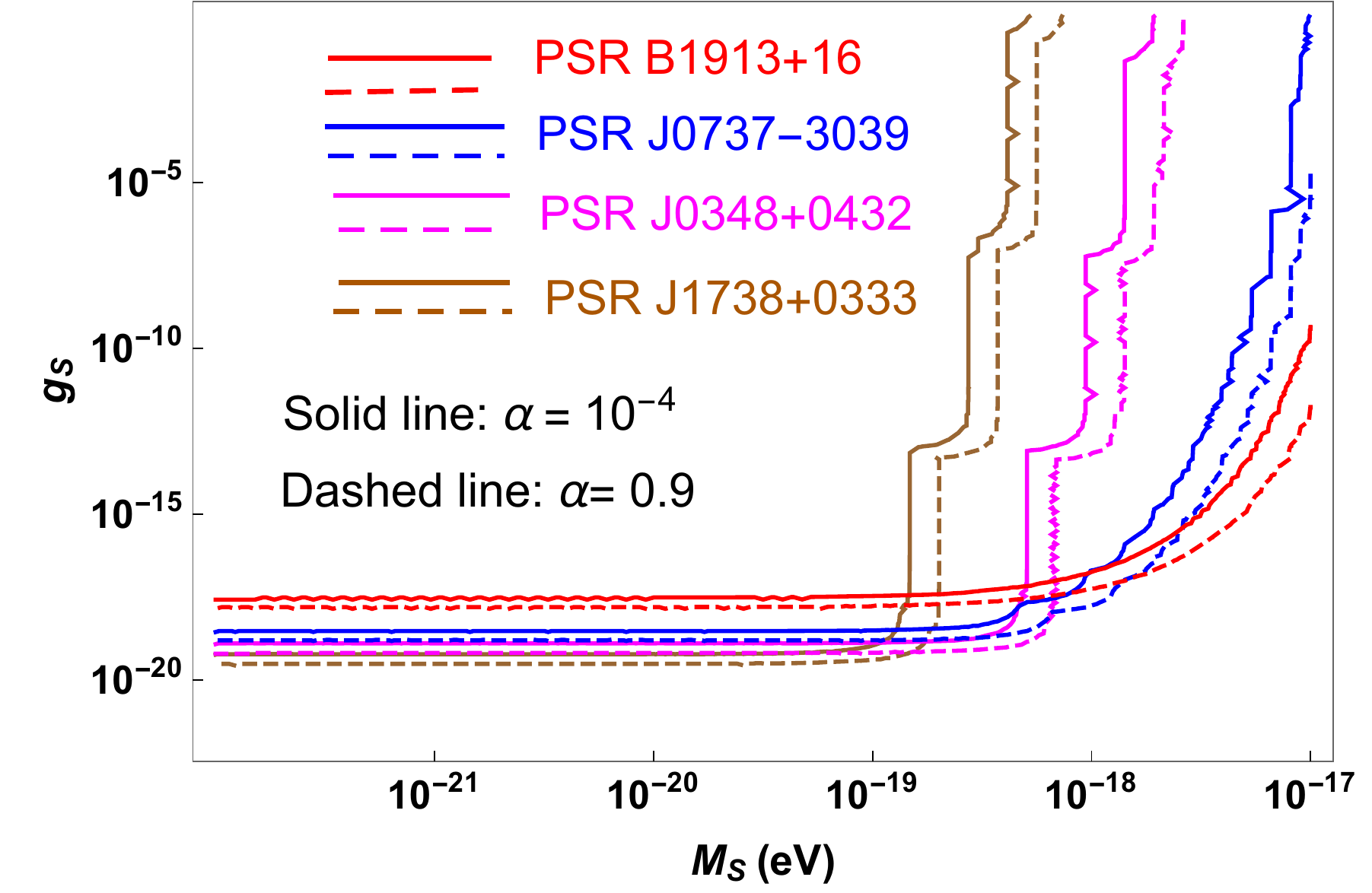}\label{plot2b}}
\subfigure[$g_V$ vs. $M_{Z^\prime}$ for $\alpha=10^{-4}, 0.9$ and $Q= 10^{44}$]{\includegraphics[width=8cm]{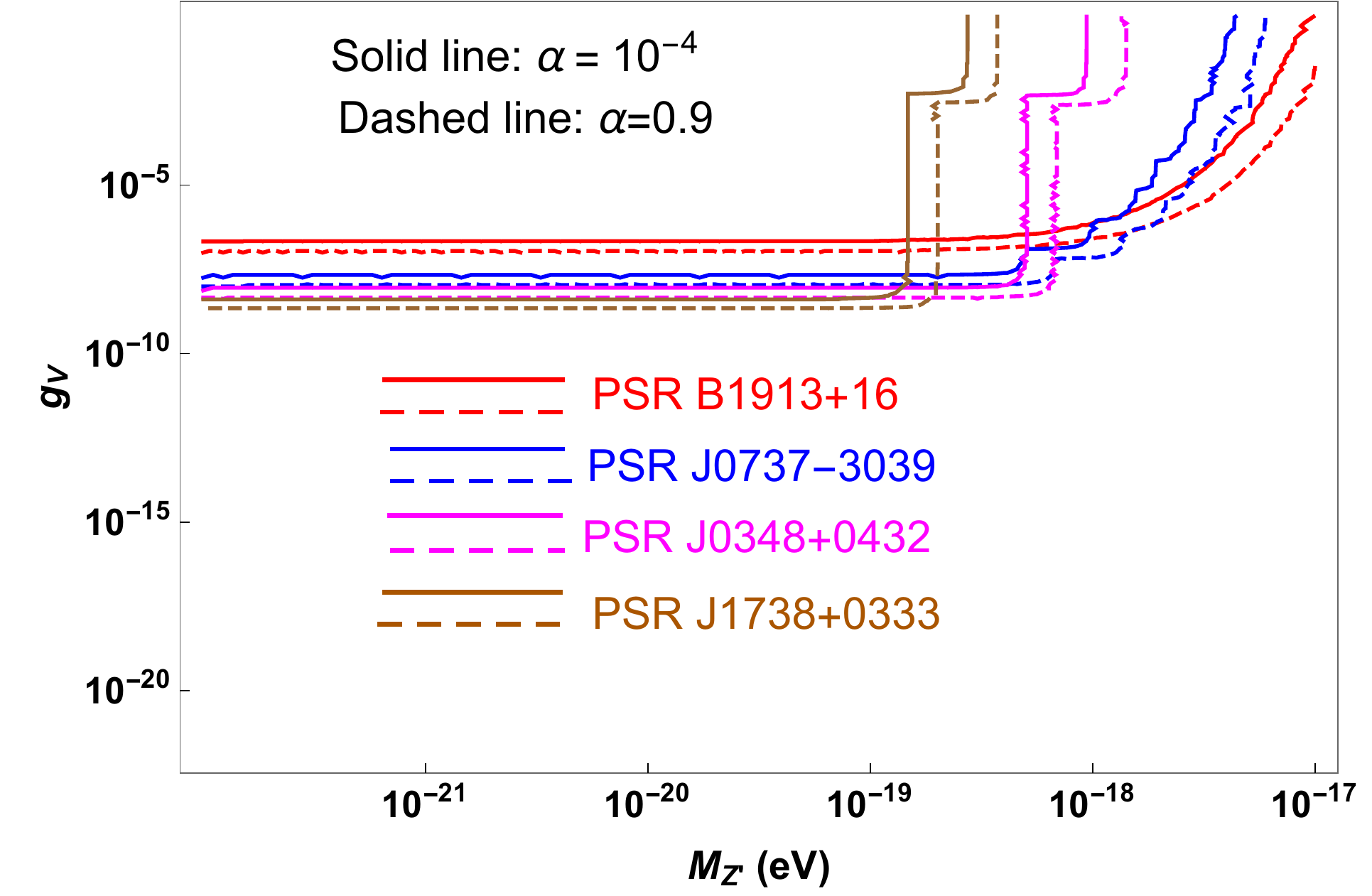}\label{plot3a}}
\subfigure[$g_S$ vs. $M_S$ for $\alpha=10^{-4}, 0.9$ and $Q= 10^{44}$]{\includegraphics[width=8cm]{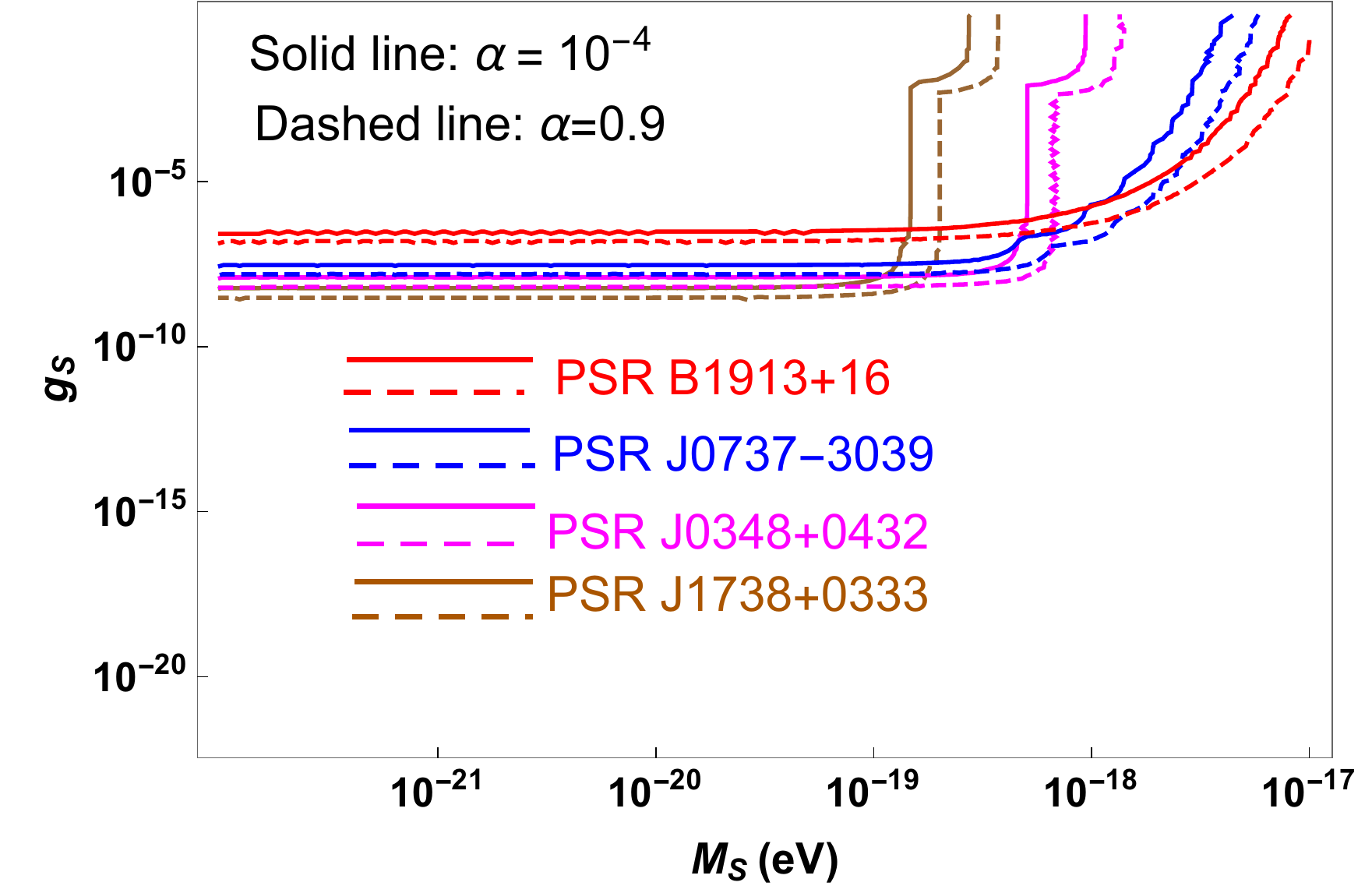}\label{plot3b}}
\caption{\it (a) Exclusion plot to constrain the coupling due to the radiation of vector particles for the fifth force strength $\alpha=10^{-4}, 0.9$ and the number of dark charge particles $Q=10^{55}$. (b) Exclusion plot to constrain the coupling due to the radiation of scalar particles for the fifth force strength $\alpha=10^{-4}, 0.9$ and the number of dark charge particles $Q=10^{55}$. (c) Exclusion plot to constrain the coupling due to the radiation of vector particles for the fifth force strength $\alpha=10^{-4}, 0.9$ and the number of dark charge particles $Q=10^{44}$. (d) Exclusion plot to constrain the coupling due to the radiation of scalar particles for the fifth force strength $\alpha=10^{-4}, 0.9$ and the number of dark charge particles $Q= 10^{44}$.}
\label{plot2}
\end{figure}
In FIG. \ref{plot2} we obtain constraints on the scalar $(g_S)$ and vector $(g_V)$ gauge couplings from the orbital period loss of compact binary systems. While deriving the constraints, we assume the combined effects of the fifth force between the two stars and the radiation of ultralight particles from the binary systems. We obtain the bounds on couplings for two NS-NS (PSR B1913+16, PSR J0737-3039) and two NS-WD (PSR J0348+0432, PSR J1738+0333) binary systems. The constraints on gauge couplings strongly depend on the charge to mass asymmetry $\Big(\frac{Q}{M}-\frac{q}{m}\Big)$ between the two stars in the binary systems (see Eq. \ref{modified_scalar}, Eq. \ref{modified_vector}). We obtain the bounds on gauge couplings for the number of dark charge particles $Q=10^{55},$ and $10^{44}$. We consider for the NS-NS binary systems $Q\sim q$ whereas for the NS-WD binary systems, $Q>>q$ (NS captures more number of DM particles than the WD), and $\Big(\frac{Q}{M}-\frac{q}{m}\Big)\sim \frac{Q}{M}$. The bounds on $g_V$ and $g_S$ get stronger for larger numbers of dark charge particles. Though the gauge coupling and the strength of the fifth force parameter depend on each other, here we take them as independent to check the dependence of the fifth force in deriving the bounds on gauge coupling from the radiation of ultralight particles. For larger values of $\alpha$, we obtain moderately better bounds on the couplings (see Eq. \ref{modified_scalar}, Eq.\ref{modified_vector}). In getting these bounds, we do not consider any capture mechanism of dark matter particles. The bounds on gauge couplings are only valid for the mass of ultralight particles less than the orbital frequency of the binary systems. The red, blue, magenta and brown lines denote the variation of gauge couplings with the mass of ultralight particles for PSR B1913+16, PSR J0737-3039, PSR J0348+0432, and PSR J1738+0333 respectively. The solid lines correspond to the variation for $\alpha=10^{-4}$, and the dashed lines correspond to the variation for $\alpha=0.9$. The NS-WD binary system, PSR J1738+0333 gives stronger bounds on the scalar and vector gauge couplings. The radiation of ultralight vector particles gives stronger bounds on the coupling than the radiation of scalar particles. The regions above the solid and dashed lines are excluded. 
\begin{table}[h]
\caption{\label{tableI} Summary of the upper bounds on vector gauge coupling for $\alpha=0.9$ and $\alpha=10^{-4}$. Here we assume the number of dark charges $Q=10^{55}$. For all the binaries we assume $M_{Z^\prime,S}\lesssim 1.35\times 10^{-19}$ eV.}
\centering
\begin{tabular}{ lcccc  }
 
 \hline
Compact binary system & $g_V (\alpha=0.9)$  & $g_V (\alpha=10^{-4})$& $g_S (\alpha=0.9)$  & $g_S (\alpha=10^{-4})$\\
 \hline
PSR B1913+16  & $\lesssim 1.04\times 10^{-18}$  & $\lesssim 2.08\times 10^{-18}$ & $\lesssim 1.55\times 10^{-18}$ & $\lesssim 2.53\times 10^{-18}$ \\
 PSR J0737-3039 & $\lesssim   9.86\times 10^{-20}$  &$\lesssim 1.84\times 10^{-19}$ & $\lesssim 1.53\times 10^{-19}$ & $\lesssim 2.69\times 10^{-19}$ \\
 PSR J0348+0432 & $\lesssim 4.93\times 10^{-20}$  & $\lesssim 8.58\times 10^{-20}$ & $\lesssim 6.18\times 10^{-20}$ & $\lesssim 1.24\times 10^{-19}$\\
PSR J1738+0333  & $\lesssim 2.29\times 10^{-20}$  & $\lesssim 4.00\times 10^{-20}$ &$\lesssim 3.06\times 10^{-20}$ & $\lesssim 5.76\times 10^{-20}$\\
 \hline
\end{tabular}
\end{table}

\begin{table}[h]
\caption{\label{tableII} Summary of the upper bounds on vector gauge coupling for $\alpha=0.9$ and $\alpha=10^{-4}$. Here we assume the number of dark charges $Q=10^{44}$. For all the binaries we assume $M_{Z^\prime,S}\lesssim 1.35\times 10^{-19}$ eV.}
\centering
\begin{tabular}{ lcccc  }
 
 \hline
Compact binary system & $g_V (\alpha=0.9)$  & $g_V (\alpha=10^{-4})$& $g_S (\alpha=0.9)$  & $g_S (\alpha=10^{-4})$\\
 \hline
PSR B1913+16  & $\lesssim 1.10\times 10^{-7}$  & $\lesssim 2.20\times 10^{-7}$ & $\lesssim 1.47\times 10^{-7}$ & $\lesssim 2.96\times 10^{-7}$ \\
 PSR J0737-3039 & $\lesssim   1.12\times 10^{-8}$  &$\lesssim 2.23\times 10^{-8}$ & $\lesssim 1.44\times 10^{-8}$ & $\lesssim 2.91\times 10^{-8}$ \\
 PSR J0348+0432 & $\lesssim 4.86\times 10^{-9}$  & $\lesssim 9.07\times 10^{-9}$ & $\lesssim 6.19\times 10^{-9}$ & $\lesssim 1.16\times 10^{-8}$\\
PSR J1738+0333  & $\lesssim 2.26\times 10^{-9}$  & $\lesssim 4.23\times 10^{-9}$ &$\lesssim 3.06\times 10^{-9}$ & $\lesssim 5.77\times 10^{-9}$\\
 \hline
\end{tabular}
\end{table}
In TABLE \ref{tableI} and \ref{tableII} we show the constraints on $g_V$ and $g_S$ for $\alpha=0.9, 10^{-4}$ and $Q=10^{55}, 10^{44}$ from the orbital period loss of compact binary systems. The constraints on vector gauge coupling due to only radiation effect are derived in \cite{KumarPoddar:2019ceq}. The constraints on $g_V$ and $g_S$ become stronger in presence of a fifth force $(\alpha\neq 0)$.
\subsection{Constraints on axion decay constant from combined effect of fifth force and axionic radiation in orbital period loss of compact binary systems}
\begin{figure}[!htbp]
\includegraphics[height=8cm]{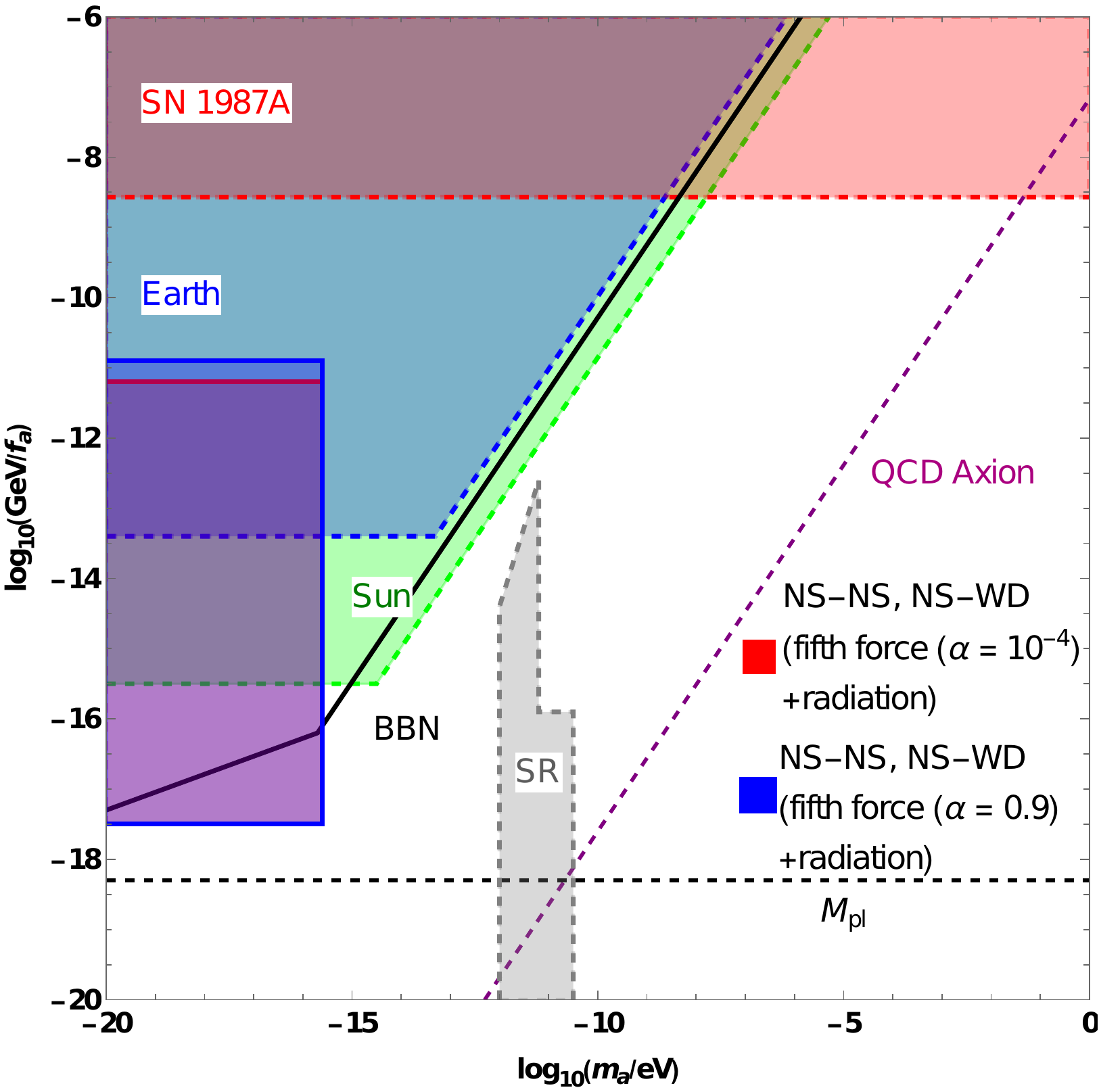}
\caption{\it Constraints on axion parameters in $f_a$ vs. $m_a$ plane from orbital period loss of compact binary systems due to the combined effect of fifth force and axionic radiation. We choose the fifth force strength as $\alpha=10^{-4}$ and $\alpha=0.9$. }
\label{radaxion}
\end{figure}
In FIG. \ref{radaxion} we obtain constraints on the axion decay constant from combined effects of fifth force and axionic radiation in orbital period loss of NS-NS (PSR B1913+16, PSR J0737-3039) and NS-WD (PSR J0348+0432, PSR J1738+0333) binary systems. We obtain these constraints for the fifth force strengths $\alpha=10^{-4}$, and $\alpha=0.9$. The combined effects of fifth force with $\alpha=10^{-4}(0.9)$ and radiation excludes axions with decay constant $1.58\times 10^{11}~\rm{GeV}\lesssim f_a\lesssim 3.16\times 10^{17}~\rm{GeV}(7.94\times 10^{10}~\rm{GeV}\lesssim f_a\lesssim 3.16\times 10^{17}~\rm{GeV})$ for axion mass $m_a\lesssim 2.51\times 10^{-16}~\rm{eV}$. The red (blue) shaded region bounded by the red (blue) solid line corresponds to the exclusion region for $\alpha=10^{-4}(0.9)$. Axions with decay constant $f_a\gtrsim 3.16\times 10^{17}~\rm{GeV}$ cannot be sourced by a compact star because in that case, the critical radius is so large that the compact star cannot trigger the phase transition even if they are free from orbital period loss due to fifth force and radiation constraints. Earlier, the authors of \cite{KumarPoddar:2019jxe} obtain the bound on axion decay constraints by only considering the radiation of ultralight particles. However, we consider the combined effect of axionic fifth force and axionic radiation. With increasing the fifth force strength one can exclude more parameter space in $m_a-\frac{1}{f_a}$ plane. Comparing with FIG. \ref{pcap3},  we also obtain that the axionic radiation excludes more parameter space than the axionic fifth force. Hence, the contribution of radiation is larger than that of the fifth force in the orbital period loss of compact binary systems.
\section{Constraints on fifth force strength and couplings of ultralight particles from LIGO/Virgo data }
\label{direct}
In this section, we obtain constraints on the coupling and mass of ultralight particles from the direct detection of GW using LIGO/Virgo data. From Eq. \ref{approxpot}, we can write the total energy of a compact binary system in presence of a long range Yukawa type fifth force as
\begin{equation}
\begin{split}
E_{\rm{tot}}=\frac{GMm}{2r}+\frac{GMm}{2r}\frac{\alpha}{2\epsilon}\{(1+\epsilon)e^{-M_{Z^\prime}r(1+\epsilon)}-(1-\epsilon)e^{-M_{Z^\prime}r(1-\epsilon)}\}+\frac{GMm\alpha M_{Z^\prime}}{4\epsilon}\times\\
\{(1+\epsilon)^2 e^{-M_{Z^\prime}r(1+\epsilon)}-(1-\epsilon)^2 e^{-M_{Z^\prime}r(1-\epsilon)}\},
\end{split}
\label{eq:l1}
\end{equation}
where the first term denotes the total Newtonian energy of the binary system and the rest of the terms denote the energy due to the contribution of Yukawa type fifth force. Hence, we can write the rate of change of total energy loss as
\begin{equation}
    \frac{dE_{\rm{tot}}}{dt}=\mathcal{E}_1+\mathcal{E}_2+\mathcal{E}_3,
\end{equation}
where the expressions of $\mathcal{E}_1$, $\mathcal{E}_2$, and $\mathcal{E}_3$ are given in Appendix \ref{appendix1}. We have kept terms up to $\mathcal{O}(\alpha)$. The terms $\mathcal{E}_2$ and $\mathcal{E}_3$ contribute to the secular decay of the eccentricity due to the back reaction of the GW that circularizes the binary orbit. The eccentricity of the orbit changes with time as the two stars of the binary come close to each other. Hence, we can write the rate of change of eccentricity in presence of a fifth force as 
\begin{equation}
\begin{split}
    \frac{d\epsilon}{dt}=\frac{304}{15}G\mu\Omega^4 r^2 \epsilon(1-\epsilon^2)^{-5/2}\Big(1+\frac{121}{304}\epsilon^2\Big)\Big[1+\frac{\alpha}{\epsilon}\Big\{(1+\epsilon)e^{-M_{Z^\prime}r(1+\epsilon)}-(1-\epsilon)e^{-M_{Z^\prime}r(1-\epsilon)}\Big\}+\\
    \frac{\alpha M_{Z^\prime}r}{\epsilon}\Big\{(1+\epsilon)^2e^{-M_{Z^\prime}r(1+\epsilon)}-(1-\epsilon)^2e^{-M_{Z^\prime}r(1-\epsilon)}\Big\}\Big]+\mathcal{O}(\alpha^2).
    \end{split}
\end{equation}
The energy of the binary system also decreases due to the radiation of gravitational waves and ultralight particles. Hence, we can write the rate of change of energy loss as 
\begin{equation}
\frac{dE_{\rm{tot}}}{dt}=\frac{dE}{dt}+\frac{dE_{Y+S/V}}{dt},
\end{equation}
where the first and second terms are governed by Eq.\ref{nonzeroalpha}, Eq. \ref{modified_scalar} and Eq. \ref{modified_vector}. We can write the rate of orbital separation $\frac{dr}{dt}$ as
\begin{equation}
    \frac{dr}{dt}=\frac{\frac{dE}{dt}+\frac{dE_{Y+S/V}}{dt}-\mathcal{E}_2-\mathcal{E}_3}{\mathcal{E}_4},
    \label{f1}
\end{equation}
where $\mathcal{E}_4$ is given in Appendix \ref{appendix1}. In absence of the fifth force and radiation, we get back the standard GR formula for the rate of decrease of orbital separation as 
\begin{equation}
\frac{dr}{dt}=\frac{\frac{dE}{dt}}{\mathcal{E}_4}=\frac{\frac{dE_{\rm{GW}}}{dt}}{\mathcal{E}_4}=-\frac{64}{5}\frac{G^3Mm(M+m)}{r^3}(1-\epsilon^2)^{-7/2}\Big(1+\frac{73}{24}\epsilon^2+\frac{37}{96}\epsilon^4\Big).
\end{equation}

The rate of change of orbital frequency of the binary system is also modified due to the presence of long range Yukawa potential as the binary stars come close to each other. Hence, using Eq.\ref{Tot_orb_freq} we can write the rate of change of orbital frequency as
\begin{equation}
2\Omega_f\dot{\Omega_f}=\mathcal{\xi}_1\frac{dr}{dt}+\mathcal{\xi}_2,
\label{der_omega}
\end{equation}
where $\mathcal{\xi}_1$ and $ \mathcal{\xi}_2$ are given in the Appendix \ref{appendix1}. We have substituted $m_1=G(M+m)$ in the expressions of $\mathcal{\xi}_1$ and $ \mathcal{\xi}_2$. The term $\mathcal{\xi}_2$ appears due to the variation in eccentricity.

Inserting Eq. \ref{f1} in Eq. \ref{der_omega} we obtain
\begin{equation}
    \Omega_f \dot{\Omega_f}=K_1+K^{S/V}_2+K_3+K_4+\frac{\mathcal{\xi}_2}{2},
\end{equation}
where the expressions of $K_1, K^{S/V}_2, K_3, K_4$ are given in Appendix \ref{appendix1}.

The strength of the fifth force is weaker than the gravitational force for the stars to merge in the binary system. From Eq. \ref{Tot_orb_freq} we can write $r$ as a function of $\Omega_f$ as
\begin{equation}
    \begin{split}
       r(\Omega_f)=\Big(\frac{m_1}{\Omega_f^2}\Big)^\frac{1}{3}\Big[1+\frac{\alpha}{6\epsilon}\{(1+\epsilon)e^{-M_{Z^\prime}(1+\epsilon)m_1(m_1\Omega_f)^{-\frac{2}{3}}}-(1-\epsilon)e^{-M_{Z^\prime}(1-\epsilon)m_1(m_1\Omega_f)^{-\frac{2}{3}}}\}+\\
\frac{\alpha M_{Z^\prime}}{6\epsilon}\Big(\frac{m_1}{\Omega_f^2}\Big)^\frac{1}{3}\{(1+\epsilon)^2e^{-M_{Z^\prime}(1+\epsilon)m_1(m_1\Omega_f)^{-\frac{2}{3}}}-(1-\epsilon)^2e^{-M_{Z^\prime}(1-\epsilon)m_1(m_1\Omega_f)^{-\frac{2}{3}}}\}\Big]+\mathcal{O}(\alpha^2).
    \end{split}
    \label{ri1}
\end{equation}
Putting the expressions of $r(\Omega_f)$ in Eq. \ref{ki1}, Eq. \ref{ki2}, Eq. \ref{ki3}, and Eq. \ref{ki4}, we obtain the time variation in the orbital frequency as 
\begin{equation}
    \dot{\Omega_f}=L_1+L^{S/V}_2+L_3+L_4+L_5,
    \label{final_freq}
\end{equation}
where the expressions of $L_1, L^{S/V}_2, L_3, L_4, L_5$ are given in the Appendix \ref{appendix1}. We have neglected the terms $\mathcal{O}(\alpha^2)$ and $\mathcal{O}(\delta\alpha)$ as they are very small compared to the leading order terms. We substitute $\delta=\Big(\frac{Q}{M}-\frac{q}{m}\Big)^2$ and $\mathcal{M}_{\rm{ch}}=\frac{(Mm)^{3/5}}{(M+m)^{1/5}}$ in the expressions of $L_1, L^{S/V}_2, L_3, L_4, L_5$. Here, $\mathcal{M}_{\rm{ch}}$ is called the Chirp mass. It measures the frequency evolution of the GW emitted during a binary's inspiral. The first term in  Eq. \ref{final_freq} arises due to the effect of only the fifth force. The second term $L^{S/V}_2$ contributes to the rate of change angular frequency due to the radiation of ultralight particles. The last three terms arise due to the eccentricity variation with time. In absence of the fifth force and radiation of ultralight particles, we get back the standard GR formula for the rate of change of angular frequency
\begin{equation}
\dot{\Omega}=\frac{96}{5}G^{\frac{5}{3}}\mathcal{M}_{\rm{ch}}^\frac{5}{3}\Omega^{\frac{11}{3}}(1-\epsilon^2)^{-\frac{7}{2}}\Big(1+\frac{73}{24}\epsilon^2+\frac{37}{96}\epsilon^4\Big).
\end{equation}
If there is no variation of eccentricity with time then $ L_3, L_4$ and $L_5$ will disappear from the expression of $\dot{\Omega_f}$. In, $M_{Z^\prime}\rightarrow 0$ limit, the expression of $L_1$ becomes
\begin{equation}
L_1=\frac{96}{5}(G\mathcal{M}_{\rm{ch}})^\frac{5}{3}\Omega^{\frac{11}{3}}_f(1-\epsilon^2)^{-\frac{7}{2}}\Big(1+\frac{73}{24}\epsilon^2+\frac{37}{96}\epsilon^4\Big)(1+\alpha)^\frac{2}{3},
\end{equation}
which matches quite well with Eq.2.8 of \cite{Kopp:2018jom}. In the Fourier space, the eccentricity of the compact binary inspiral also varies with the orbital frequency as $\epsilon\sim \epsilon_0\Big(\frac{\Omega_f}{\Omega}\Big)^{-\frac{19}{18}}$. Hence, if a binary inspiral enters the sensitivity band of a ground based interferometer at $30~\rm{Hz}$ orbital frequency with an initial eccentricity $\epsilon_0=0.1$, the eccentricity reduces to $\epsilon\sim 10^{-2}$ before the system's orbital frequency reaches to $300~\rm{Hz}$.
We solve Eq. \ref{ri1} and Eq. \ref{final_freq} numerically for non-zero $M_{Z^\prime}$ and express the eccentricity in terms of orbital frequency, we obtain the variation of gravitational wave frequency as a function of time. The contribution of the fifth force and the radiation of ultralight particles should be within the uncertainty in reconstructing the Chirp mass measurement. Earlier, the authors of \cite{Kopp:2018jom,Alexander:2018qzg} obtained constraints on fifth force parameters from the LIGO/Virgo data for a circular orbit. However, our calculations are true for a general eccentric orbit. We obtain constraints on the new force parameters from LIGO/Virgo data if there is an (A) fifth force and (B) radiation of ultralight particles.
\subsection{Fifth force effect in gravitational wave detection}
\begin{figure}[!htbp]
\centering
\subfigure[$\Omega_f$ vs. $t$ for $\epsilon_0=10^{-6}$ and $M_{Z^\prime}=1.98\times 10^{-12}~\rm{eV}$]{\includegraphics[width=8cm]{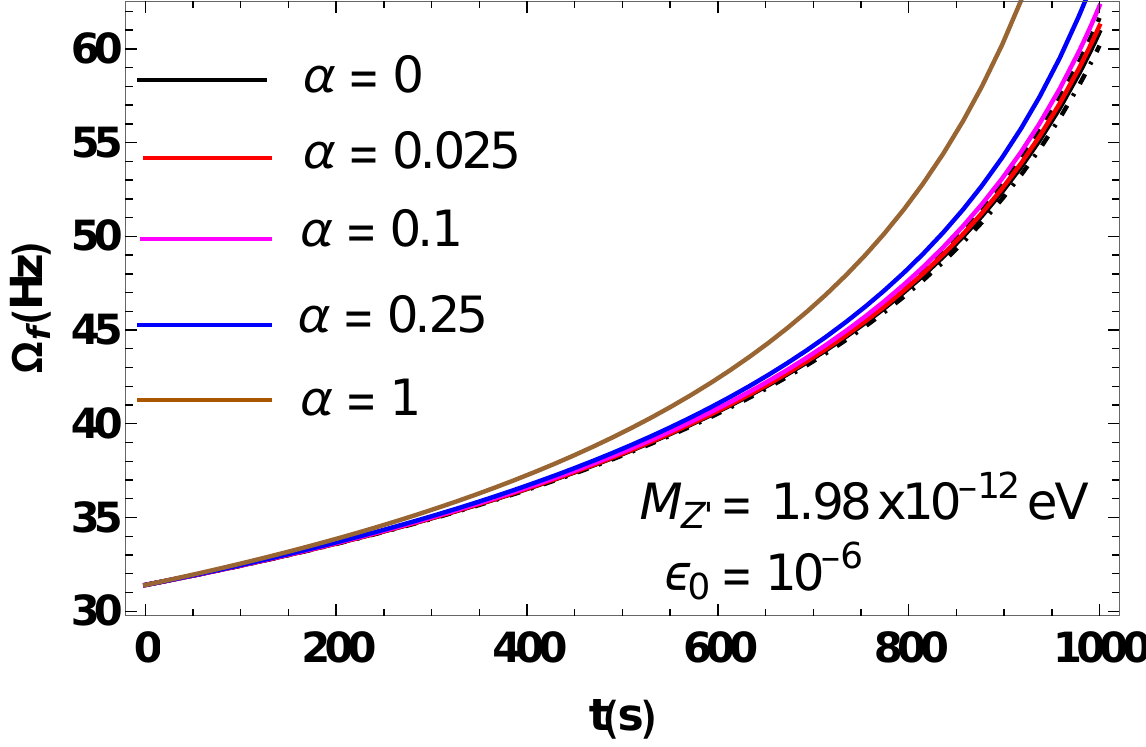}\label{plot6}}
\subfigure[$\Omega_f$ vs. $t$ for $\epsilon_0=0.1$ and $M_{Z^\prime}=1.98\times 10^{-12}~\rm{eV}$]{\includegraphics[width=8cm]{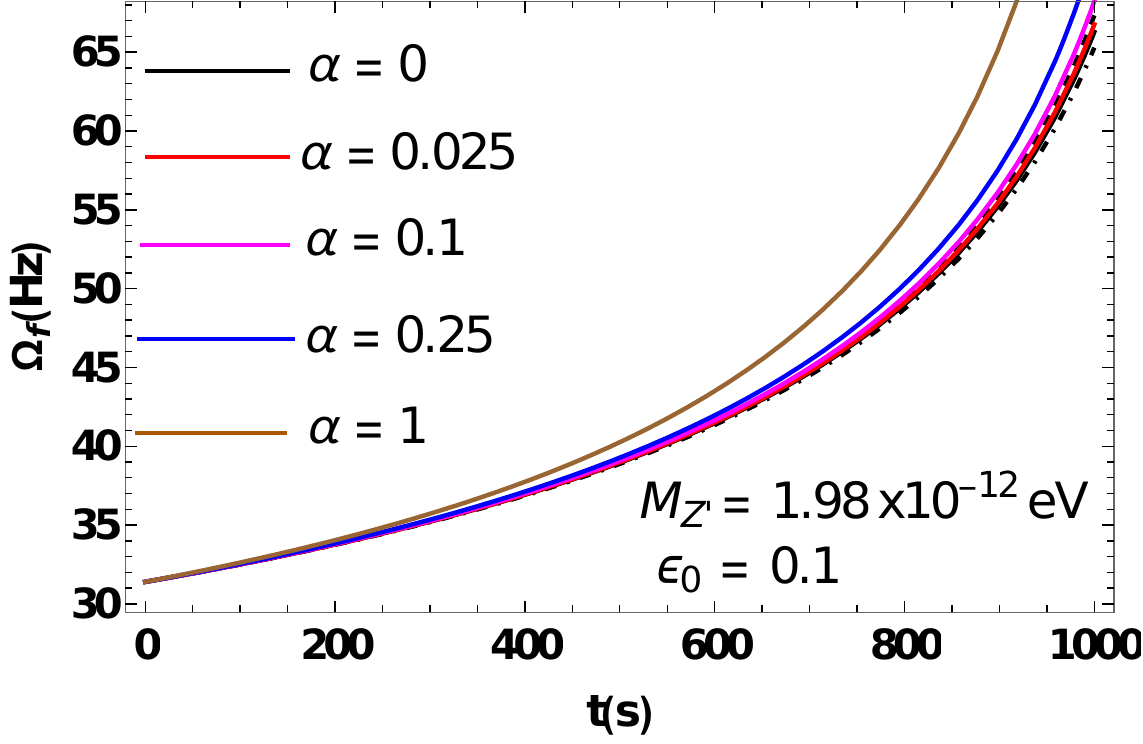}\label{plot7}}
\subfigure[$\Omega_f$ vs. $t$ for $\epsilon_0=10^{-6}$ and $M_{Z^\prime}=6.6\times 10^{-13}~\rm{eV}$]{\includegraphics[width=8cm]{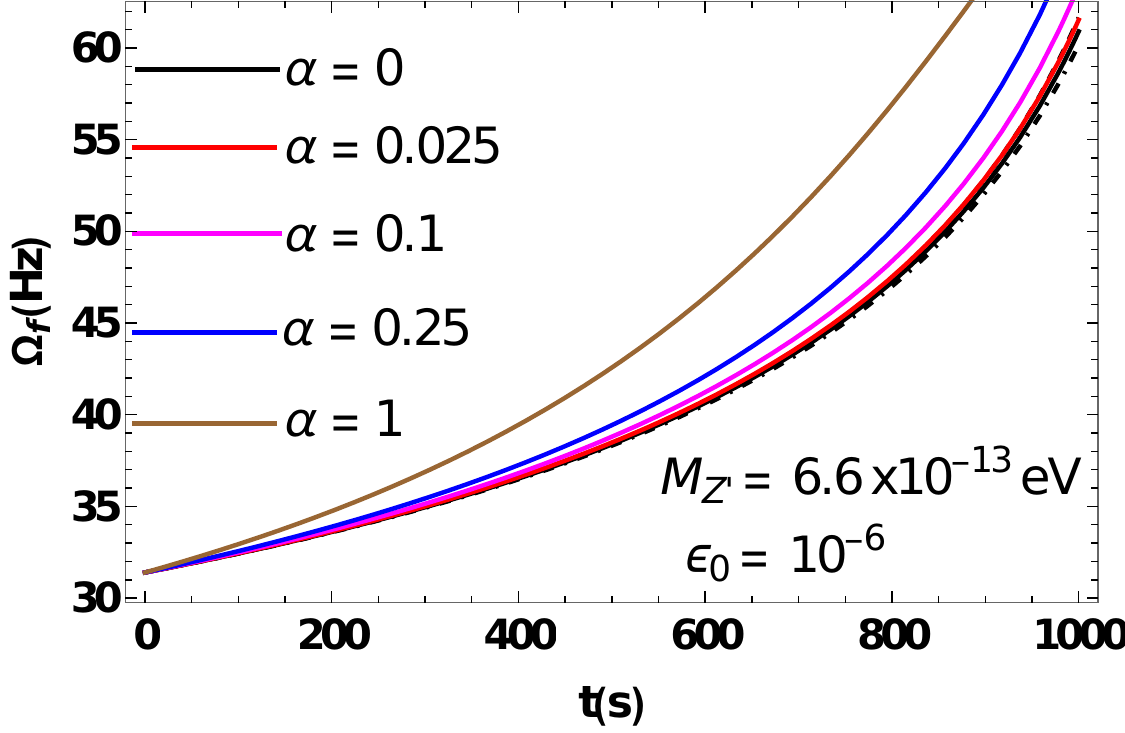}\label{plot8}}
\caption{\it Variation of gravitational wave frequency ($\Omega_f$) as a function of time (t) for a binary NS system with different fifth force strengths $(\alpha)$ varying from $0$ to $1$. We obtain the variations for two different choices of initial eccentricity values $\epsilon_0=10^{-6}, 0.1$ and the inverse ranges of fifth force values $M_{Z^\prime}=1.98\times 10^{-12}~\rm{eV}, 6.6\times 10^{-13}~\rm{eV}$. We omit the effect of radiation of ultralight particles in the orbital period loss by taking an equal charge to mass ratio of binary stars. Here, the energy loss of the binary system is only due to the gravitational wave radiation and the fifth force effect. The masses of the two NS are taken as $M=m=1.25~M_{\odot}$.}
\label{dplotfirst}
\end{figure}
The fifth force due to the presence of dark charge in the binary stars can contribute to the orbital period loss of the binary system if the mass of the fifth force mediator is less than the inverse of binary stars separation. However, the fifth force contribution should be within the uncertainty in the measurement of reconstructed Chirp mass. For the binary system with two NSs masses $M=m=1.25~\rm{M_{\odot}}$, the Chirp mas is $\mathcal{M}_{\rm{ch}}=1.088~\rm{M_{\odot}}$. From GW 170817, the largest uncertainty in reconstructing the Chirp mass is estimated as $0.4\%$ due to the unknown source distance \cite{LIGOScientific:2017vwq}. In FIG. \ref{dplotfirst} we obtain the variation of gravitational wave frequency $(\Omega_f)$ as a function of time $(t)$ for a binary NS system with NS masses $M=m=1.25~\rm{M_{\odot}}$. We obtain the variations for fifth force strengths $\alpha=\{0,0.025,0.1,0.25,1\}$, initial eccentricity values $\epsilon_0 =\{10^{-6}, 0.1\}$, and the inverse of the ranges of fifth force $M_{Z^\prime}=\{1.98\times 10^{-12}~\rm{eV}, 6.6\times 10^{-13}~\rm{eV}\}$. The black solid line corresponds to $\alpha=0$, denoting the pure gravity scenario. The region bounded by the black dot-dashed lines corresponds to the uncertainty band in reconstructing the Chirp mass. The orbital frequency of the gravitational wave is $\Omega_f=\pi\times f$. If the LIGO sensitivity begins at $f(t=0)= 10~\rm{Hz}$ corresponds to $\Omega_f(t=0)=31.4~\rm{Hz}$ then the two NSs with masses $1.25~\rm{M_{\odot}}$ enters into the LIGO sensitivity band when their spatial separation is $\sim 700~\rm{km} ~(M_{Z^\prime}=2.83\times 10^{-13}~\rm{eV})$. The red, magenta, blue, and brown lines denote the variation of the angular frequency of GW with time for $\alpha=0.025, 0.1, 0.25,$ and $1$ respectively. The gravitational wave frequency increases with time.

For $M_{Z^\prime}=1.98\times 10^{-12}$ (large) and $\alpha=0.025$ (small) with initial eccentricity values $\epsilon_0=10^{-6}$ (FIG. \ref{plot6}) and $\epsilon_0=0.1$ (FIG. \ref{plot7}), the orbital frequency falls into the LIGO sensitivity band in the entire time domain. For $\alpha=0.1$, the orbital frequency falls into the LIGO sensitivity band for lower values $(t\lesssim 800~\rm{s})$ of coalescence time. However, the shifts of coalescing time relative to the gravity alone scenario are different for different eccentric orbits. The coalescence time is shifted by $\sim 1.8~\rm{s}$ and $\sim 2.7~\rm{s}$ relative to the gravity only scenario for $\alpha=0.025$ for binary orbit with eccentricity $\epsilon_0=10^{-6}$ (FIG.\ref{plot6}) and $\epsilon_0=0.1$ (FIG.\ref{plot7}) respectively.

For $M_{Z^\prime}=6.6\times 10^{-13}~\rm{eV}$ (small) and $\epsilon_0=10^{-6}$ (FIG. \ref{plot8}), the orbital frequency falls into the LIGO sensitivity band for $\alpha=0.025$. Here, the coalescence time is shifted by $\sim 3.6~\rm{s}$ relative to the gravity only scenario for $\alpha=0.025$. For $\alpha=0.1$, the orbital frequency falls into the LIGO sensitivity band for lower values $(t\lesssim 450~\rm{s})$ of coalescence time.

Hence, larger values of $M_{Z^\prime}$ and smaller values of $\alpha$ can be probed from LIGO/Virgo data for a coalescing binary for large $t$. The degeneracy between the gravity only solution and the small $\alpha$ and large $M_{Z^\prime}$ solution can be further broken with higher order corrections to the gravitational wave emission. For smaller values of $M_{Z^\prime}$ and larger values of $\alpha$, the orbital frequency is out of the LIGO sensitivity band for large $t$. In that case, the exponential suppression goes away and the Yukawa type fifth force behaves like a Coulomb force. For these fifth force parameters, one can reconstruct the events with incorrect values of the Chirp mass. Hence, these values of fifth force parameters are not sensitive to the LIGO frequency band and alternative signals (electromagnetic) are required to probe these parameters. 
\begin{figure}[!htbp]
\centering
\subfigure[$\Omega_f$ vs. $t$ for $M_{Z^\prime}=1.98\times 10^{-12}~\rm{eV}$, and $(\epsilon_0, \alpha)= (10^{-6}, 0), (10^{-6}, 1), (0.1, 0), (0.1, 1)$]{\includegraphics[width=8cm]{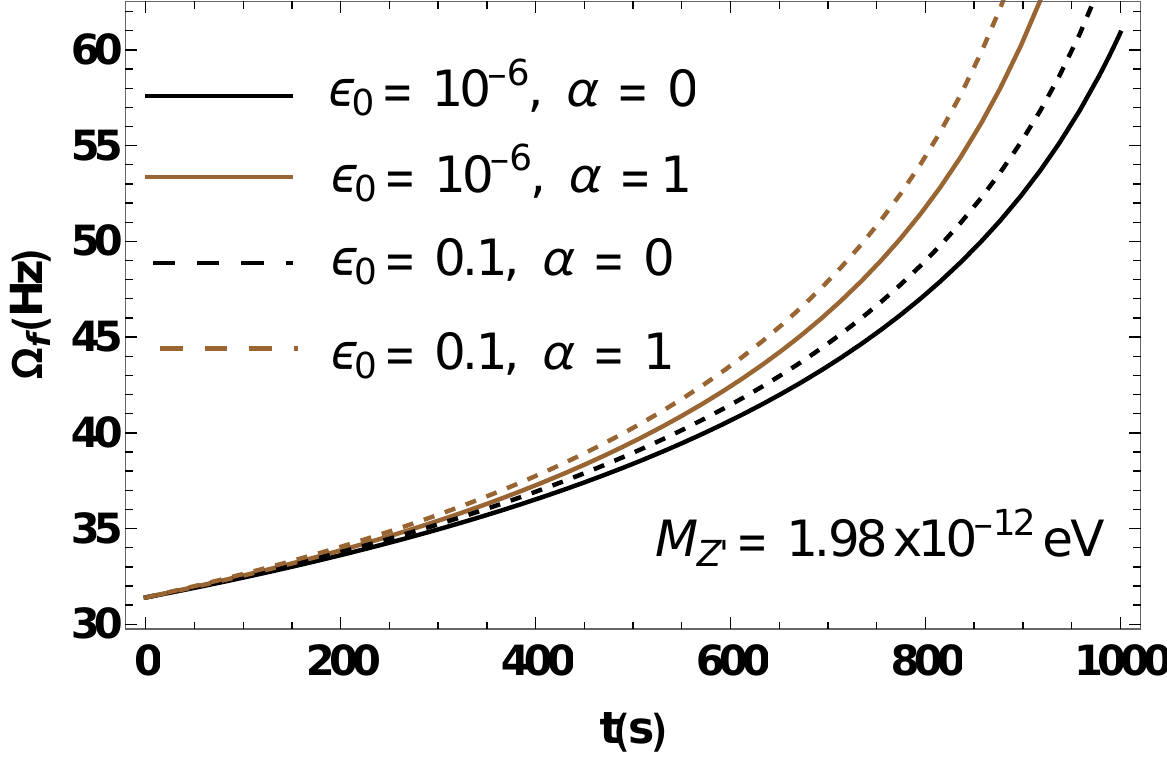}\label{plot10}}
\subfigure[$\Omega_f$ vs. $t$ for $M_{Z^\prime}=6.6\times 10^{-13}~\rm{eV}$, and $(\epsilon_0, \alpha)= (10^{-6}, 0), (10^{-6}, 1), (0.1, 0), (0.1, 1)$]{\includegraphics[width=8cm]{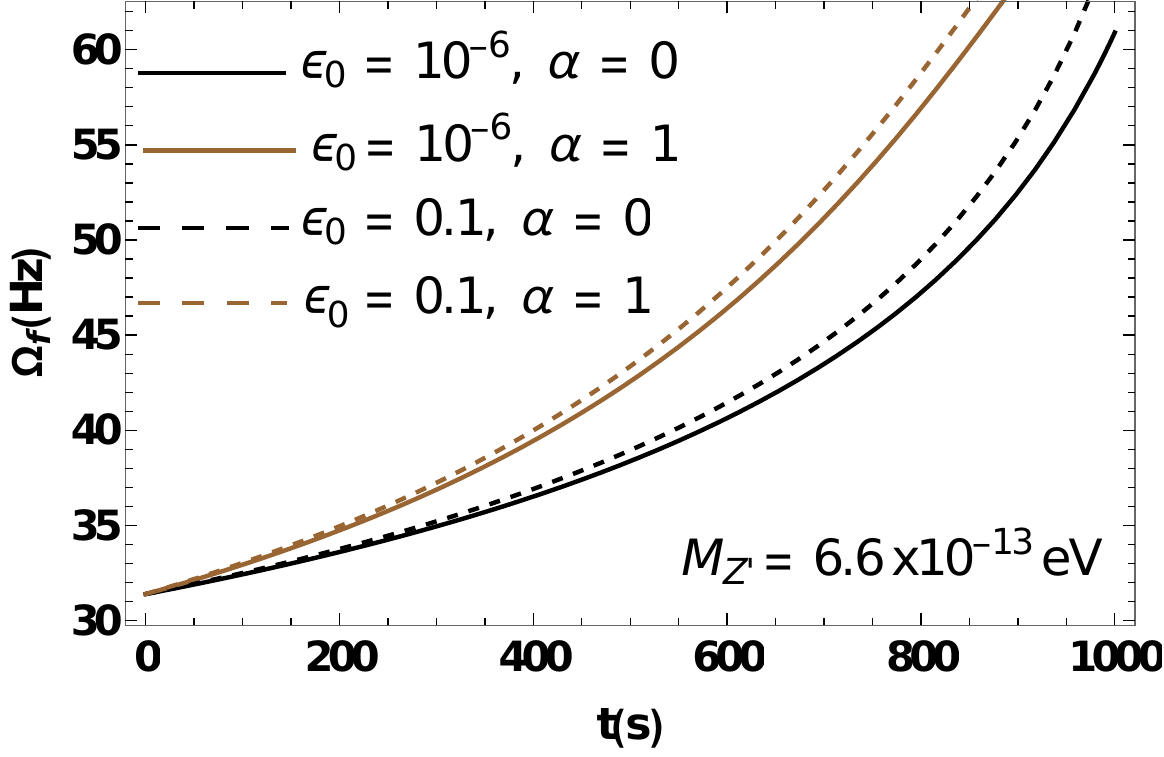}\label{plot11}}
\subfigure[$\Omega_f$ vs. $t$ for $\alpha=0.1$, and $(\epsilon_0, M_{Z^\prime})= (10^{-6}, 1.98\times 10^{-12}~\rm{eV}), (10^{-6}, 6.6\times 10^{-13}~\rm{eV}), (0.1, 1.98\times 10^{-12}~\rm{eV}), (0.1, 6.6\times 10^{-13}~\rm{eV})$]{\includegraphics[width=8cm]{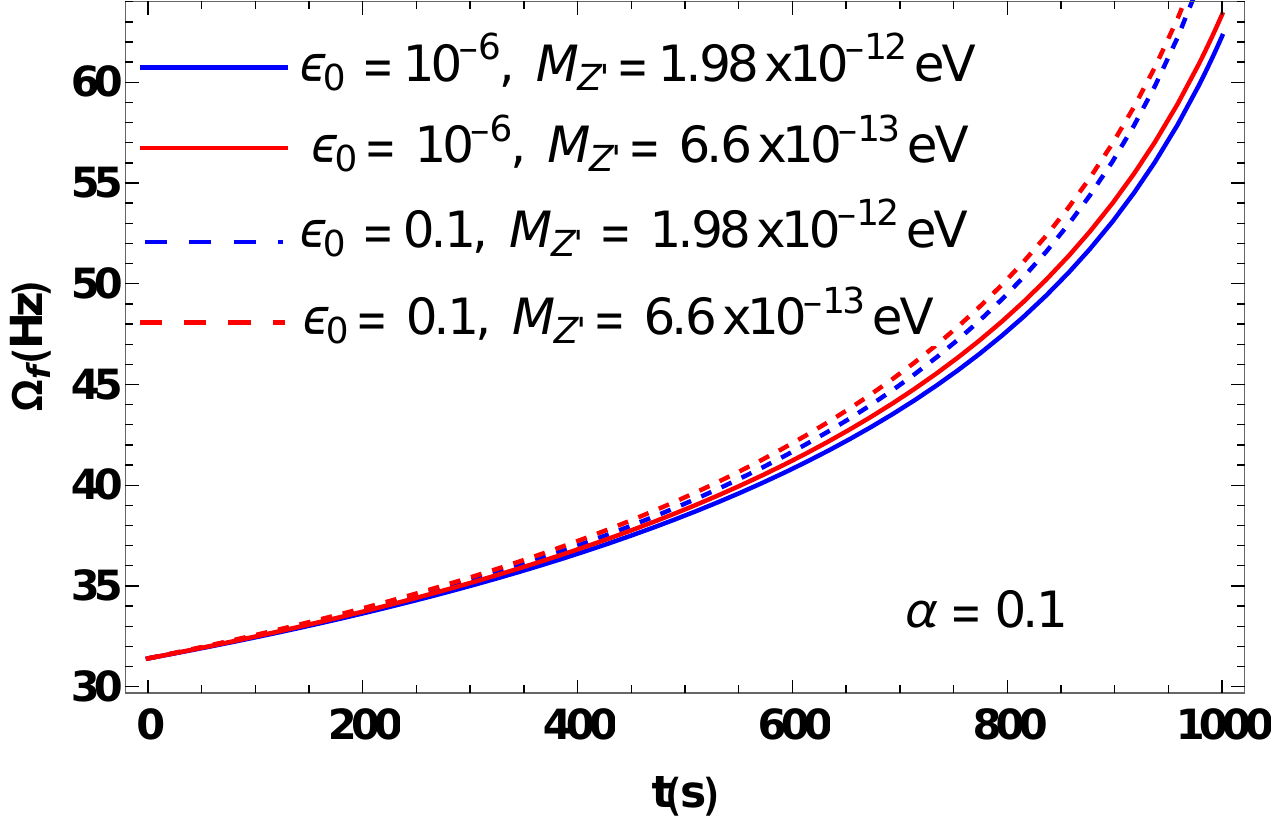}\label{plot9}}
\caption{\it Variation of gravitational wave angular frequency ($\Omega_f$) as a function of time (t) with different choice of parameter sets $(\epsilon_0, \alpha)$ with fixed $M_{Z^\prime}$ (a,b) and $(\epsilon_0, M_{Z^\prime})$ with fixed $\alpha$ (c). }
\label{dplotsecond}
\end{figure}

The variation of gravitational wave orbital frequency with time for different values of $\epsilon_0$, $\alpha$, and $M_{Z^\prime}$ is clearly depicted in FIG. \ref{dplotsecond}. In FIG. \ref{plot10} we obtain the variation of gravitational wave orbital frequency with time for $M_{Z^\prime}=1.98\times 10^{-12}~\rm{eV}$, and $(\epsilon_0, \alpha)= (10^{-6}, 0), (10^{-6}, 1), (0.1, 0), (0.1, 1)$. The frequency shifts towards the left of the LIGO frequency with increasing the initial eccentricity value and fifth force strength. We obtain the same variation in FIG. \ref{plot11} for $M_{Z^\prime}=6.6\times 10^{-13}~\rm{eV}$. 

In FIG. \ref{plot9}, we obtain the variation of gravitational wave frequency with time for $\alpha=0.1$, and $(\epsilon_0, M_{Z^\prime})= (10^{-6}, 1.98\times 10^{-12}~\rm{eV}), (10^{-6}, 6.6\times 10^{-13}~\rm{eV}), (0.1, 1.98\times 10^{-12}~\rm{eV}), (0.1, 6.6\times 10^{-13}~\rm{eV})$. The frequency shifts towards the left of the LIGO frequency for smaller values of $M_{Z^\prime}$. The gravitational wave frequency increases with increasing eccentricity and fifth force coupling. The frequency also increases with decreasing the mediator mass. Hence, the frequency with smaller values of $\alpha$, $\epsilon_0$ and larger values of $M_{Z^\prime}$ may stay inside of the LIGO frequency band and these parameters can be probed from direct detection of gravitational waves.  
\subsection{Ultralight particle radiation in gravitational wave detection}     
\begin{figure}[!htbp]
\centering
\subfigure[$\Omega_f$ vs. $t$ for $\epsilon_0=10^{-6}$, and $M_{Z^\prime}=1.98\times 10^{-14}~\rm{eV}$]{\includegraphics[width=8cm]{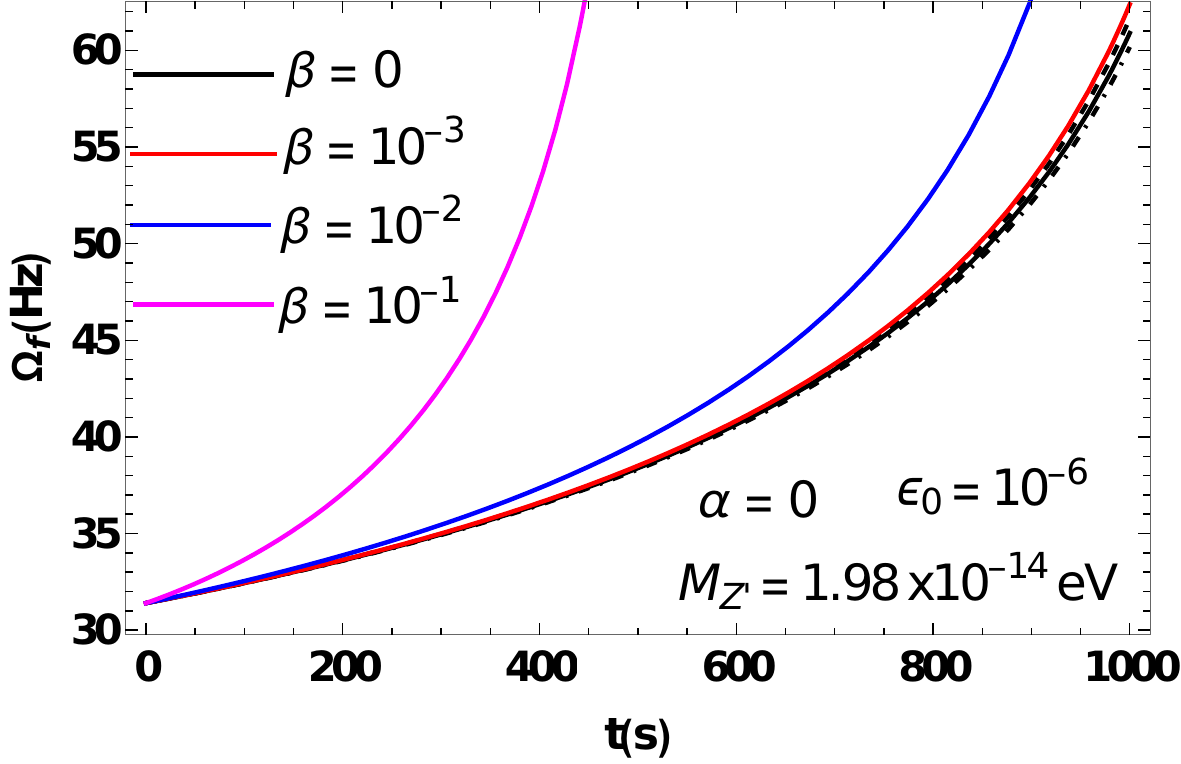}\label{dplot5}}
\subfigure[$\Omega_f$ vs. $t$ for $\epsilon_0=10^{-6}$, and $M_{Z^\prime}=2.2\times 10^{-14}~\rm{eV}$]{\includegraphics[width=8cm]{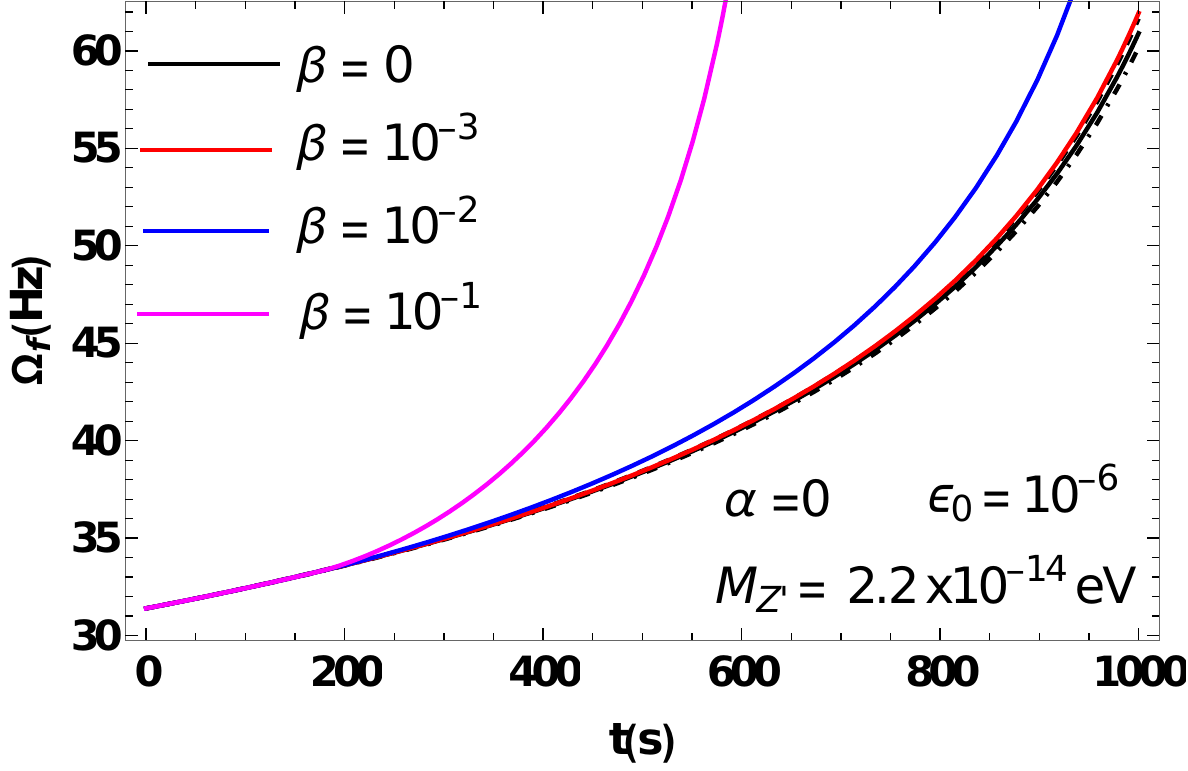}\label{dplot4}}
\subfigure[$\Omega_f$ vs. $t$ for $\epsilon_0=0.1$, and $M_{Z^\prime}=1.98\times 10^{-14}~\rm{eV}$]{\includegraphics[width=8cm]{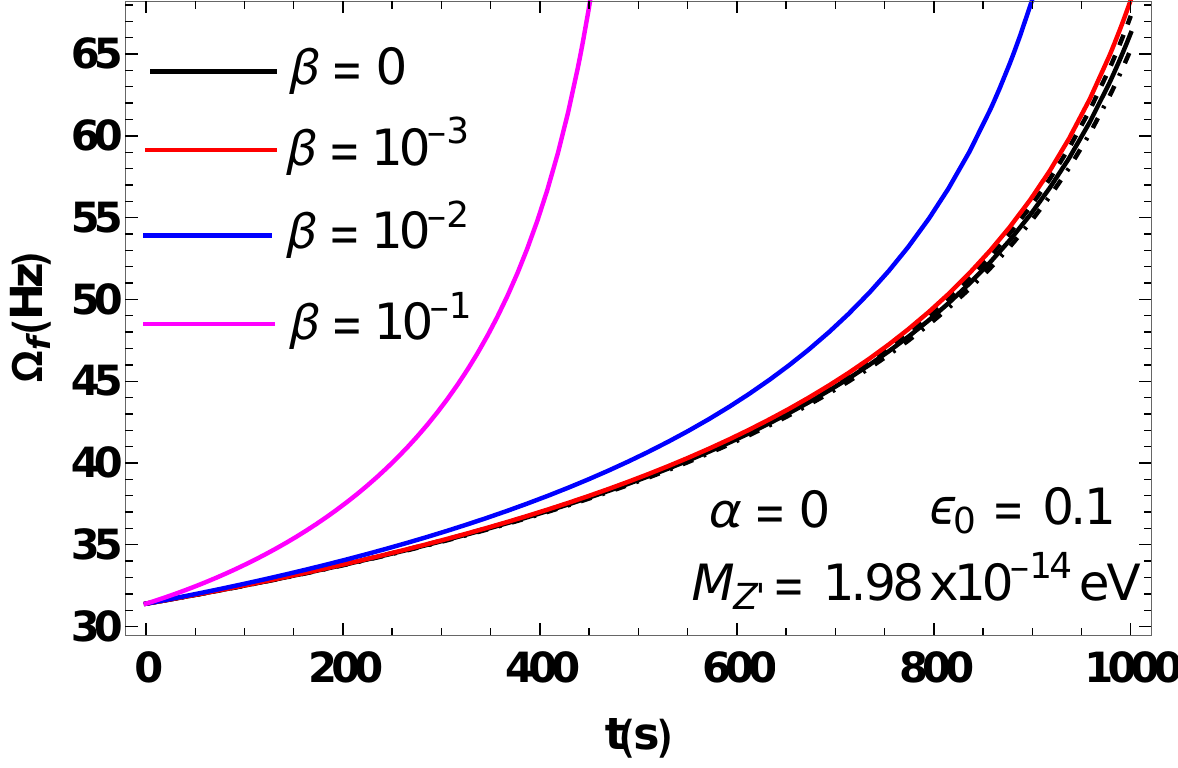}\label{dplot3}}
\subfigure[$\Omega_f$ vs. $t$ for $\epsilon_0=0.1$, and $M_{Z^\prime}=1.98\times 10^{-14}~\rm{eV}$]{\includegraphics[width=8cm]{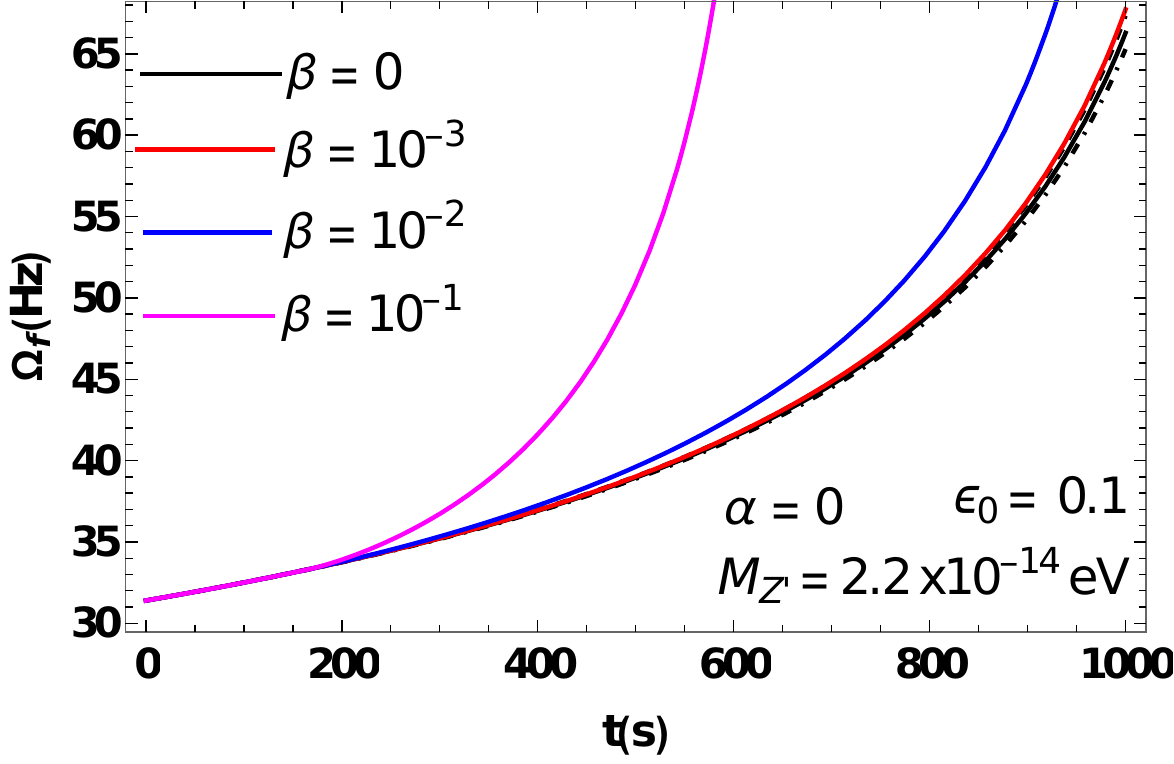}\label{dplot2}}
\caption{\it Variation of GW frequency $(\Omega_f)$ as a function of time $(t)$ for a binary NS system with different ultralight particle radiation strength $\beta$ varying from $0$ to $0.1$. We obtain the variation for two different values of ultralight particle mass $M_{Z^\prime}=1.98\times 10^{-14}~\rm{eV}$ (FIG. \ref{dplot5}, FIG. \ref{dplot3}) and $M_{Z^\prime}=2.2\times 10^{-14}~\rm{eV}$ (FIG. \ref{dplot4}, FIG.  \ref{dplot2}) and the initial eccentricity values $\epsilon_0=10^{-6}$ (FIG. \ref{dplot5}, FIG. \ref{dplot4}) and $\epsilon_0=0.1$ (FIG. \ref{dplot3}, FIG. \ref{dplot2}). We omit the effect of the fifth force by considering that only one neutron star of the binary system contains the dark charge. The masses of the two neutron stars are taken as $M=m=1.25~M_{\odot}$. }
\label{dplotthird}
\end{figure}
In FIG. \ref{dplotthird} we obtain a variation of GW orbital frequency with time for different values of mediator mass, initial eccentricity, and ultralight particle radiation strength $(\beta)$. The quantity $\beta$ is defined as
\begin{equation}
\beta=\frac{g^2_V Mm\Big(\frac{Q}{M}-\frac{q}{m}\Big)^2}{\pi G (M+m)^2}.
\end{equation}
We also consider that one of the two stars of the binary system does not carry any dark charge to eliminate the fifth force effect. Hence, the orbital period of the binary system decreases primarily due to gravitational wave radiation and the contribution of ultralight particle radiation being limited to be no larger than the measurement uncertainty. Specifically, we consider ultralight vector particle radiation in plotting FIG. \ref{dplotthird}. The black, red, blue, and magenta lines denote the variation of $\Omega_f$ with $t$ for $\beta=0, 10^{-3}, 10^{-2}, $ and $10^{-1}$ respectively. The black solid line $(\alpha=0, \beta=0)$ corresponds to the gravity only scenario. The region bounded by the dot-dashed lines denotes the uncertainty band in reconstructing the Chirp mass which we choose as $0.4\%$.

In FIG. \ref{dplotthird} we obtain that the GW frequency increases with time. The frequency with $\beta<10^{-3}$ falls into the LIGO sensitivity band and can be probed from direct detection of GW. However, the shifts in coalescing time compared to the gravity only scenario for FIG. \ref{dplot5}, FIG. \ref{dplot4}, FIG. \ref{dplot3}, and FIG. \ref{dplot2} are different. For $\beta=10^{-3}$ with $\epsilon_0=10^{-6}$ and $M_{Z^\prime}=1.98\times 10^{-14}~\rm{eV}$ (FIG. \ref{dplot5}), the coalescing time is shifted by $9.7~\rm{s}$ compared to the gravity only scenario. For an increased value of $M_{Z^\prime}=2.2\times 10^{-14}~\rm{eV}$ (FIG. \ref{dplot4}), the shift of coalescing time decreases to the value $6.8~\rm{s}$. For a larger initial eccentricity value $(\epsilon_0=0.1)$ and dark radiation strength $\beta=10^{-3}$ with $M_{Z^\prime}=1.98\times 10^{-14}~\rm{eV}$ (FIG. \ref{dplot3}) and $M_{Z^\prime}=2.2\times 10^{-14}~\rm{eV}$ (FIG. \ref{dplot2}) the shifts in the coalescing time compared to the gravity only scenario are $10.6~\rm{s}$ and $8~\rm{s}$ respectively. The frequency also increases with increasing $\beta$ and shifts towards the left of the LIGO sensitivity band. Hence, large values of $\beta$ cannot be probed from the LIGO/Virgo data. An alternative signal (electromagnetic) associated with the GW can be used to probe large values of $\beta$.

For the radiation case, smaller values of $M_{Z^\prime}$ can be probed from LIGO/Virgo data compared to the fifth force case (FIG. \ref{dplotfirst}). It means that the ultralight vector particle radiation is switched on before the waveform comes into the LIGO frequency band. It can be justified by looking at Eq. \ref{final_freq}. If there is no radiation then $L_2^{S/V}$ is zero and the dominant contribution of $\dot{\Omega}_f$ comes from $L_1$ which is proportional to $\Omega^{11/3}_f$. On the contrary, the radiation term is proportional to $\Omega^3_f$. In absence of fifth force $(\alpha=0)$, the GW orbital frequency becomes 
\begin{equation}
\begin{split}
\dot{\Omega}=\frac{96}{5}G^{5/3}\mathcal{M}_{\rm{ch}}^{5/3}\Omega^{11/3}(1-\epsilon^2)^{-7/2}\Big(1+\frac{73}{24}\epsilon^2+\frac{37}{96}\epsilon^4\Big)+\frac{g^2}{2\pi}\frac{\Omega_f^3\mathcal{M}_{\rm{ch}}^{5/3}}{(M+m)^{2/3}}\delta\times \\
\sum_{n>n_{0}^V}2n^2\Big[{J^\prime_n}^2(n\epsilon)+\Big(\frac{1-\epsilon^2}{\epsilon^2}\Big)J_n^2(n\epsilon)\Big]\sqrt{1-\frac{{n_{0}^V}^2}{n^2}}\Big(1+\frac{{n_{0}^V}^2}{2n^2}\Big),
\end{split}
\end{equation} 
where $n^V_0=\frac{M_{Z^\prime}}{\Omega}$.
\begin{figure}[!htbp]
\centering
\subfigure[$\Omega_f$ vs. $t$ for $\epsilon_0=10^{-6}$, $\beta=10^{-2}, 10^{-1}$ and $M_{Z^\prime}=1.98\times 10^{-14}~\rm{eV}, 2.2\times 10^{-14}~\rm{eV}$ ]{\includegraphics[width=8cm]{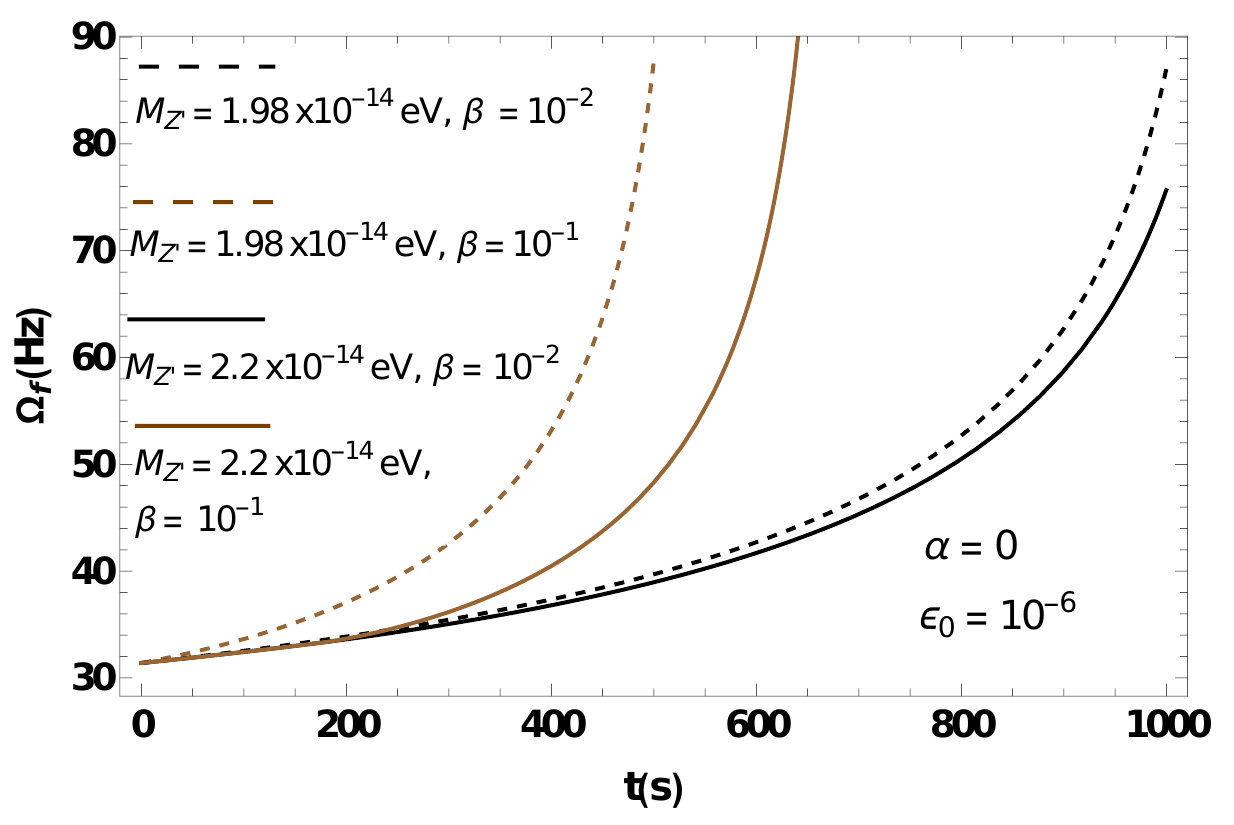}\label{dplot1}}
\subfigure[$\Omega_f$ vs. $t$ for $\epsilon_0=10^{-6}$, $\beta=0, 10^{-2}$, and $M_{Z^\prime}/M_{S}=1.98\times 10^{-14}~\rm{eV}$]{\includegraphics[width=8cm]{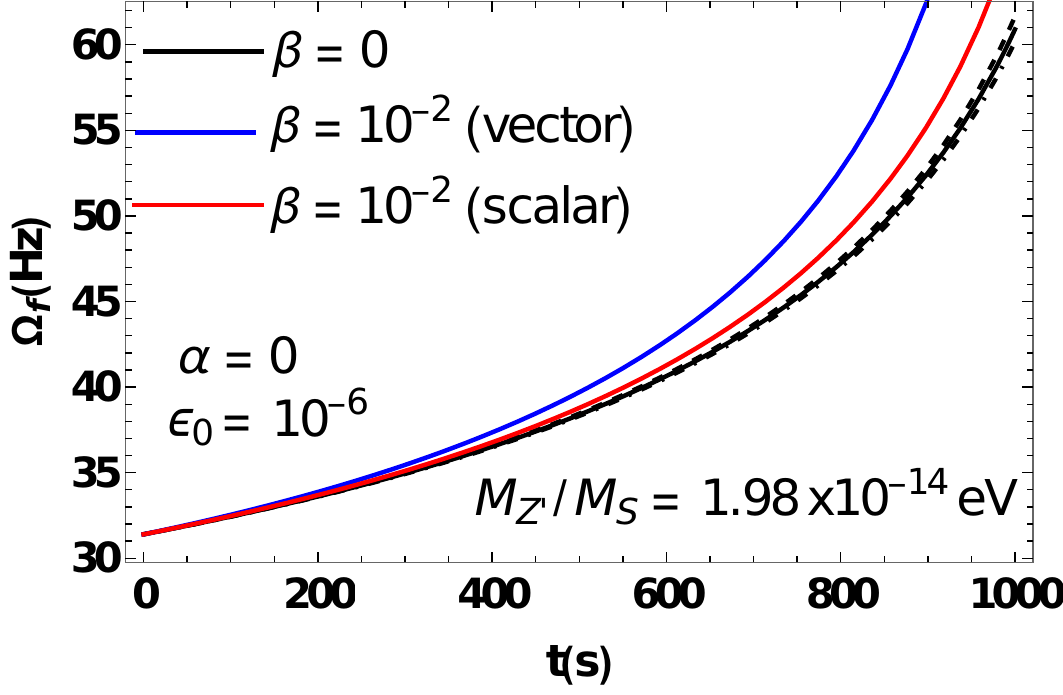}\label{dplot12}}
\caption{\it (a) Variation of GW orbital frequency as a function of time for $\epsilon_0=10^{-6}$, $\beta=10^{-2}, 10^{-1}$ and $M_{Z^\prime}=1.98\times 10^{-14}~\rm{eV}, 2.2\times 10^{-14}~\rm{eV}$. (b) Variation of GW orbital frequency as a function of time for $\epsilon_0=10^{-6}$, $\beta=0, 10^{-2}$ and $M_{Z^\prime}/M_{S}=1.98\times 10^{-14}~\rm{eV}$ for ultralight scalar and vector particle radiation.}
\label{dplotfourth}
\end{figure}

The variation of GW orbital frequency with time for different values of $M_{Z^\prime}$ and $\beta$ with an initial eccentricity value $\epsilon_0=10^{-6}$ is clearly depicted in FIG. \ref{dplot1}. The GW frequency increases with increasing $\beta$. However, the frequency decreases with increasing $M_{Z^\prime}$. The frequency also increases with increasing the initial value of eccentricity (FIG. \ref{dplotthird}). Hence, gravitational wave frequency with larger mediator mass, smaller eccentricity, and smaller dark radiation strength can fall in the LIGO sensitivity band that can be probed from the direct detection of GW for larger values of coalescence time. We can obtain plots of similar nature for $\Omega_f$ vs $t$ for the radiation of scalar particles. In FIG. \ref{dplot12} we obtain the GW frequency variation as a function of time for $M_{Z^\prime}/M_S=1.98\times 10^{-14}~\rm{eV}$, $\epsilon_0=10^{-6}$, and $\beta=10^{-2}$ for the scalar and vector particles. The frequency of the gravitational wave for vector radiation is larger than the scalar radiation. The frequency shifts towards the left OF LIGO frequency more for vector particle radiation than scalar particle radiation. The reason is $L^V_2\approx 2L^S_2$. Hence, a particular value of $\beta$ which cannot be probed for vector radiation can be probed for scalar radiation from the LIGO/Virgo data.  

Our method of calculating the GW frequency is general and can be used for eccentric compact binary systems (NS-NS, NS-WD, NS-BH, etc). As mentioned earlier, the eccentricity factor is important as it can enhance the value of the GW frequency. Hence, particular values of fifth force parameters $(\alpha, \beta, M_{Z^\prime}/M_S)$ which cannot be probed from LIGO/Virgo data with $\epsilon_0=0$ scenario, can be probed from $\epsilon_0\neq 0$ scenario and vice versa. Also, the shift in the coalescence time from the gravity only scenario is larger for orbits with higher values of eccentricity. 
\section{GW amplitude in presence of fifth force, radiaiton, and initial eccentricity}\label{amplitudeGW}
The long range Yukawa type fifth force and the radiation of ultralight particles also affect the amplitude of the GW signal. The GW amplitude varies with time as \cite{Croon:2017zcu} 
\begin{equation}
A_{\rm{GW}}(t)=\frac{4GMm}{d_L(M+m)}\Omega^2_f(t)r^2(t), 
\end{equation} 
where $d_L$ denotes the luminosity distance of the GW source. Using Eq. \ref{Tot_orb_freq} and Eq. \ref{f1} we obtain the variation of GW amplitude with time in presence of the fifth force and ultralight particle radiation.
\begin{figure}[h]
\includegraphics[height=8cm]{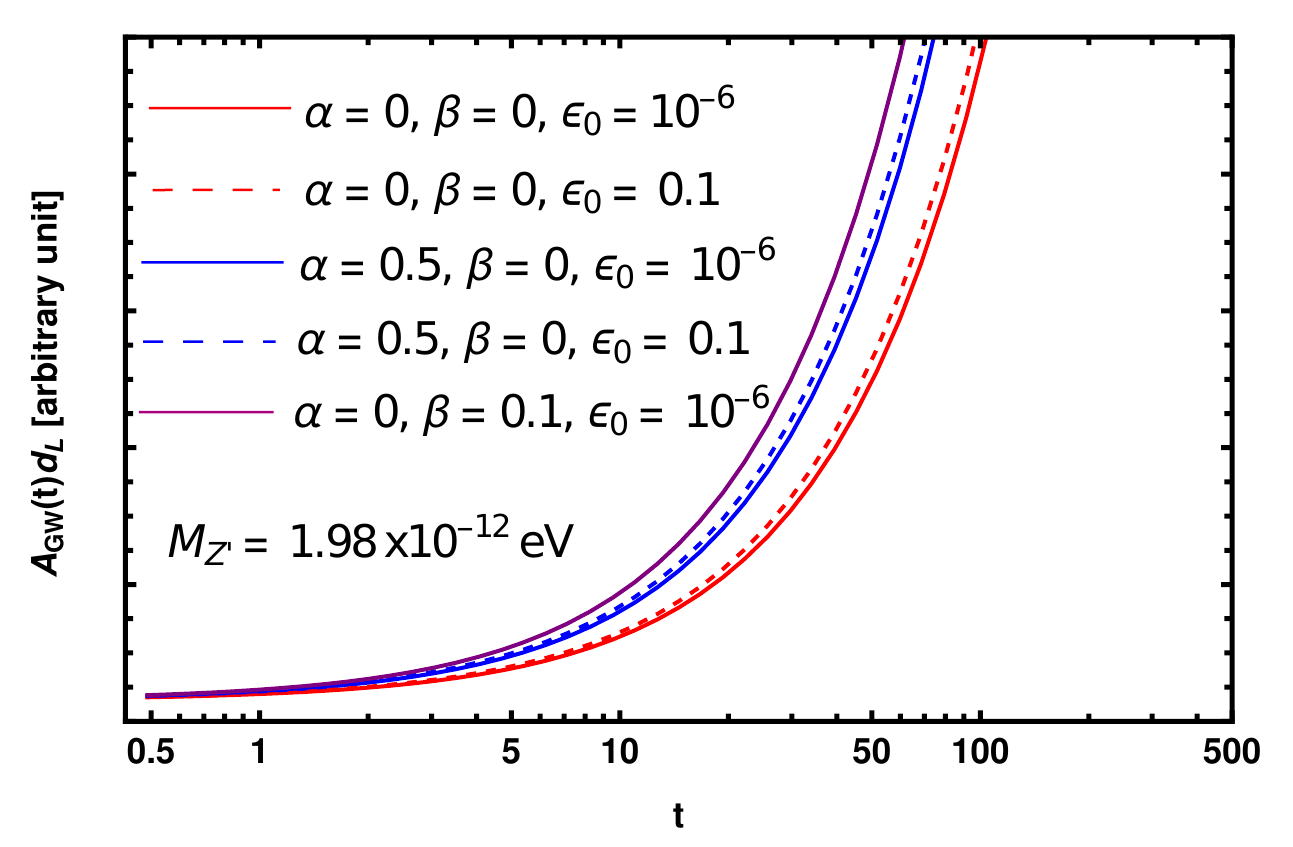}
\caption{\it Variation of GW amplitude times luminosity distance in megaparsec ($Mpc$) vs time in seconds $(s)$ for different values of fifth force strength $(\alpha)$, initial eccentricity $(\epsilon_0)$, and ultralight particle radiation strength $(\beta)$. }
\label{amplitude}
\end{figure}  
In FIG. \ref{amplitude} we obtain the variation of GW amplitude times luminosity distance with time for different values of fifth force strength $(\alpha)$, initial eccentricity $(\epsilon_0)$, and ultralight particle radiation strength $(\beta)$. The value of $A_{\rm{GW}}(t)d_L$ at $t=0$ is $7.42\times 10^{-22}$ in arbitrary units and each unit in $y$ axis is $2.94\times 10^{-24}$. The choice $\alpha=\beta=0$ denotes the GR only solution for the GW amplitude measurement. The solid red, blue, and purple lines denote the amplitude variation with time for $\epsilon_0=10^{-6}$ whereas the dashed red, and blue lines denote the corresponding variation for relatively larger values of initial eccentricity $(\epsilon_0=0.1)$. The amplitude increases with increasing $\alpha, \beta,$ and $\epsilon_0$. The contribution of $\beta$ in amplitude measurement is larger than $\alpha$.
\section{Capture of dark matter for a compact binary system}\label{capd}
\begin{figure}[h]
\includegraphics[height=8cm]{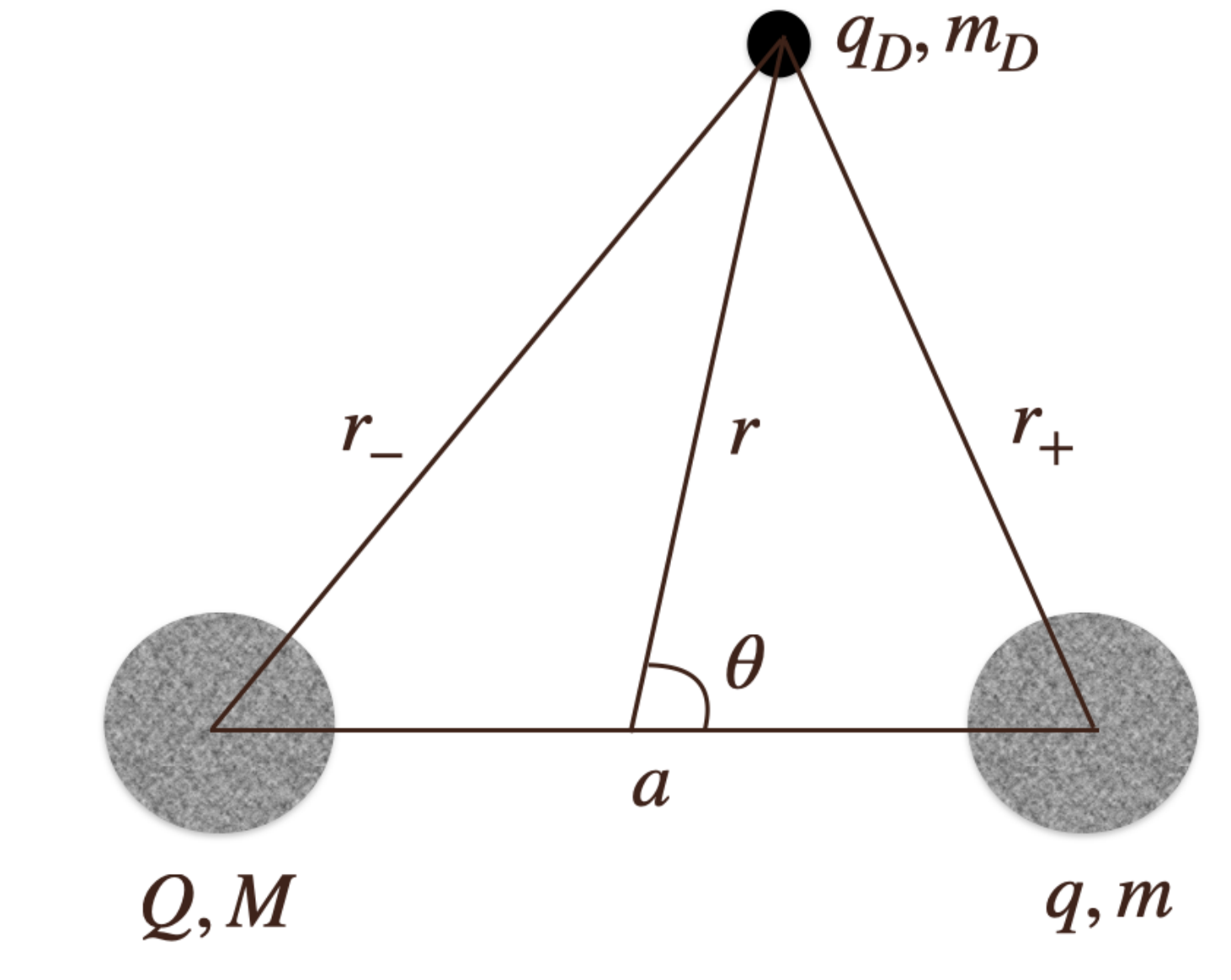}
\caption{\it Scematic diagram for the capture of a dark matter particle by a compact binary system}
\label{cap}
\end{figure} 
Besides baryonic charge, we assume that the compact stars of masses $M$ and $m$ in a binary system contain dark charges $Q$ and $q$ respectively and they are at a distance $a$ (FIG. \ref{cap}). These two charges can have the same or different signs. Now the potential energy required to bring a dark matter particle of charge $q_D$ and mass $m_D$ at a distance $r$ from the binary system is 
\begin{equation}
V(r)=V_{13}(r)+V_{23}(r),
\label{cap1}
\end{equation}
where
\begin{equation}
V_{13}(r)=-\frac{GMm_D}{r}+\frac{g^2Qq_D}{4\pi r_{-}}e^{-M_{Z^\prime}r},
\end{equation}
and 
\begin{equation}
V_{23}(r)=-\frac{Gm m_{D}}{r}+\frac{g^2qq_{D}}{4\pi r_{+}}e^{-M_{Z^\prime}r},
\end{equation}
where, $M_{Z^\prime}$ denotes the mediator mass, $\theta$ is the angle between the line joining two charges $Q$ and $q$ and the radius vector $\textbf{r}$ and
\begin{equation}
r_\pm=r\Big(1+\frac{r^2_{2(1)}}{r^2}\mp2\frac{r_{2(1)}}{r}\cos\theta\Big)^\frac{1}{2},
\end{equation}
where in the centre of mass frame $m_1 r_1=m_2 r_2$, $\mathbf{a}=\mathbf{r_1}+\mathbf{r_2}$, $r_1=\frac{\mu a}{m_1}$ and $r_2=-\frac{\mu a}{m_2}$.

If a DM particle is far away from the binary system $(r\gg a)$ and $Q$ and $q$ are equal in magnitude and sign then Eq. \ref{cap1} becomes
\begin{equation}
V(r)=-\frac{G(M+m)m_{D}}{r}+\frac{g^2Qq_D}{2\pi r}e^{-M_{Z^\prime}r}.
\label{cap2}
\end{equation}
On the contrary, if $Q$ and $q$ are equal in magnitude but have different sign then Eq. \ref{cap1} becomes
\begin{equation}
V(r)=-\frac{G(M+m)m_{D}}{r}+\frac{g^2Qq_D}{4\pi r^2}\mu a\Big(\frac{1}{M}-\frac{1}{m}\Big)e^{-M_{Z^\prime}r},
\label{cap3}
\end{equation}
where $\mu=\frac{Mm}{M+m}$ is the reduced mass of the binary system and we assume $\cos\theta\sim 1$ to get the stronger bound on DM capture.
The expressions of the potentials (Eq. \ref{cap2} and Eq. \ref{cap3}) are different as compared to the analysis done in \cite{Kopp:2018jom} where the authors consider DM capture by an isolated NS. In the case of a binary system, both the stars in the binary will contribute to the potential energy of bringing the dark matter particle. In the following, we calculate the maximum number of dark matter particles that can be captured by the binary system. The calculation is independent of dark matter production and capture mechanism.
\begin{itemize}
\item \textbf{Case I ($Q=q$):} Since compact stars contain both baryonic matter and dark matter, therefore we can write
\begin{equation}
M=M^1_{N}+M^1_{D}, \hspace{0.2cm} m= M^2_{N}+M^2_{D}, \hspace{0.2cm} M+m=M_{N}+M_{D},
\end{equation} 
where $M^{1,2}_{N(D)}$ denote the masses of the baryonic (dark) matter in the first and second compact stars of the binary system. $M_{N(D)}$ denotes the masses of total baryonic (dark) matter of the compact binary system. Hence, we can write Eq. \ref{cap2} as
\begin{equation}
V(r)=-\frac{G(M_{N}+M_{D})m_D}{r}+\frac{g^2Qq_D}{2\pi r}e^{-M_{Z^\prime}r}.
\label{cap5}
\end{equation}
The mass of the total dark matter in a star can be written as $M_D=N_Dm_D$, where $N_D$ is the total number of dark matter particles. Similarly, the total dark charge in a star is $Q=N_D q_D$. Hence, Eq. \ref{cap5} becomes
\begin{equation}
V(r)=-\frac{GM_N m_D}{r}-\frac{GN_Dm^2_D}{r}+\frac{g^2N_D q^2_D}{2\pi r}e^{-M_{Z^\prime}r}.
\label{cap6}
\end{equation} 
The binary system can capture the dark matter particles until the net potential of the binary system ceases to be attractive. Considering the net potential is attractive $(V(r)\leq 0)$ at length scale $M_{Z^\prime}r\ll 1$, we obtain the maximum number of captured dark matter particles by the binary system as 
\begin{equation}
N_D\leq \frac{M_N m_D}{\frac{g^2q^2_D}{2\pi G}-m^2_D}.
\label{cap7}
\end{equation}
We can also write the captured dark matter mass fraction for the binary system as 
\begin{equation}
\frac{M_D}{M_N}\leq \frac{1}{\frac{g^2 q^2_D}{2\pi G m^2_D}-1}.
\label{cap8}
\end{equation}
Putting Eq. \ref{cap7} in Eq. \ref{strength}, we can write the strength of the fifth force compared to gravity as 
\begin{equation}
\alpha\leq\frac{Gm^2_D\pi}{g^2q^2_D},
\end{equation}
where we assume $2\alpha q^2_D\gg Gm^2_D$ and $M_N\gg M_D$.
\item \textbf{Case II ($Q=-q$):} For the dipole potential of the binary system (Eq. \ref{cap3}) we similarly obtain the maximum number of captured dark matter particles as 
\begin{equation}
N_D\leq \frac{M_Nm_D}{\frac{g^2q^2_D\mu a}{4\pi G r}(\frac{1}{M}-\frac{1}{m})-m^2_D}.
\end{equation} 
We also obtain the captured dark matter mass fraction as 
\begin{equation}
\frac{M_D}{M_N}\leq \frac{1}{\frac{g^2q^2_D \mu a}{4\pi Gr m^2_D}(\frac{1}{M}-\frac{1}{m})-1}.
\end{equation}
In this scenario, the strength of the fifth force compared to gravity becomes 
\begin{equation}
\alpha \leq\frac{4\pi Gm^2_Dr^2}{g^2 q^2_D\mu^2 a^2\Big(\frac{1}{M}-\frac{1}{m}\Big)^2}.
\end{equation}
\end{itemize}
\begin{figure}[h]
\includegraphics[height=8cm]{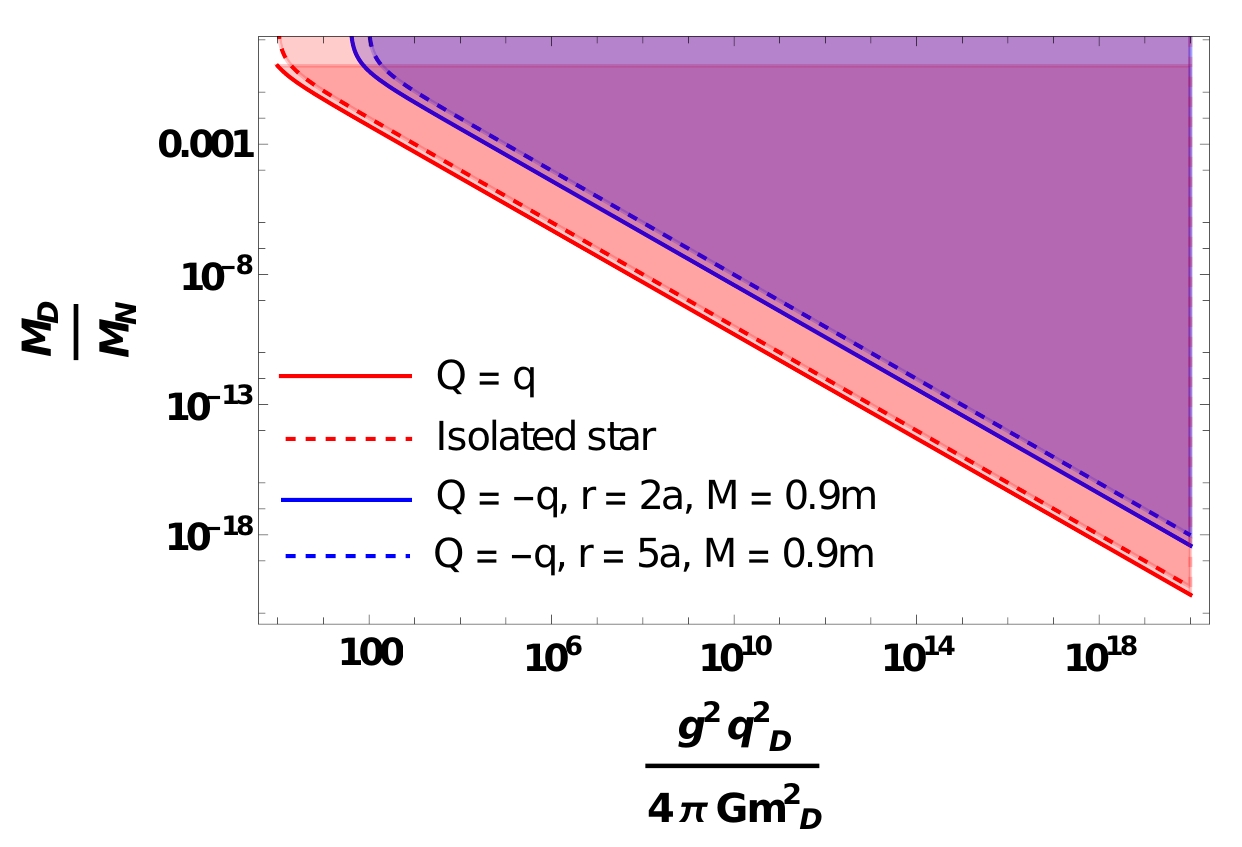}
\caption{\it Variation of dark matter mass fraction with the relative strength of dark force. }
\label{cap1}
\end{figure} 
In FIG. \ref{cap1} we obtain the variation of DM mass fraction with the relative strength of the dark force. The shaded regions are excluded. We obtain a stronger bound on the DM mass fraction if the two stars of the binary system have equal and the same sign of dark charge (red solid line). The bound is stronger than that for the isolated stars which is expected (red dashed line) \cite{Kopp:2018jom}. Also, we obtain a stronger bound on dark matter mass fraction if the DM particles are captured from a shorter distance (blue solid and dashed lines). The DM particles cannot be captured if the two stars of the binary system contain equal and opposite dark charges.  
\section{An application to scalar-tensor theories of gravity }\label{bd}
The massive scalar-tensor theories of gravity such as Brans-Dicke theory contains scalar and tensor degrees of freedom which are coupled non minimally in the action of BD theory as \cite{Will:1993hxu,Alsing:2011er}
\begin{equation}
\mathcal{S}=\frac{1}{16\pi  G}\int \Big[\phi R-\frac{\omega_{\rm{BD}}(\phi)}{\phi}g^{\mu\nu}\phi_{,\mu}\phi_{,\nu}+M_{\rm{BD}}(\phi)\Big](-g)^{\frac{1}{2}}d^4x-\sum_a\int m_a(\phi)d\tau_a,
\end{equation} 
where $\omega_{\rm{BD}}(\phi)$ is called the coupling function and $M_s(\phi)$ is called the cosmological function. The last integral denotes the matter fields for particles of inertial masses $m_a(\phi)$ and $\tau_a$ is the proper time of particle $a$ traversing along the worldline $x^\lambda_a$. $M_{\rm{BD}}(\phi)$ is the potential term in the action and the scalar mass from the potential can be derived as
\begin{equation}
m^2_{\rm{BD}}=-\frac{\phi_0}{3+2\omega_{\rm{BD}}}M^{''}_{\rm{BD}}(\phi_0),
\end{equation} 
where $\phi_0=\frac{4+2\omega_{\rm{BD}}}{3+2\omega_{\rm{BD}}}$ is the background value of the scalar field. The local value of the gravitational constant as felt by the compact object is 
\begin{equation}
\mathcal{G}_{\rm{local}}=\frac{\phi_0}{\phi}=\frac{G}{\phi}\Big(\frac{4+2\omega_{\rm{BD}}}{3+2\omega_{\rm{BD}}}\Big).
\end{equation}
Hence, the mass of the compact object depends on the scalar field as
\begin{equation}
m_a(\phi)=m_a(\ln \mathcal{G}_{\rm{local}})=m_a(\phi_0)\Big[1+s_a\Big(\frac{\eta}{\phi_0}\Big)-\frac{1}{2}(s^\prime_a-s^2_a+s_a)\Big(\frac{\eta}{\phi_0}\Big)^2+\mathcal{O}\Big(\frac{\eta}{\phi_0}\Big)^3\Big],
\end{equation}
where $\eta$ denotes the perturbation of the scalar field around $\phi_0$. The first $(s_a)$ and second $(s^\prime_a)$ sensitivities are defined as 
\begin{equation}
s_a=-\Big(\frac{\partial(\ln m_a)}{\partial(\ln \mathcal{G}_{\rm{local}})}\Big)_{\phi_0},\hspace{0.2cm}s^\prime_a=-\Big(\frac{\partial^2(\ln m_a)}{\partial(\ln \mathcal{G}_{\rm{local}})^2}\Big)_{\phi_0}.
\end{equation}
Therefore, the equation of motion of the tensor and scalar fields become \cite{Alsing:2011er}
\begin{equation}
\begin{split}
R_{\mu\nu}-\frac{1}{2}g_{\mu\nu}R=-\frac{3+2\omega_{\rm{BD}}}{4\phi_0\phi}m^2_{\rm{BD}}(\phi-\phi_0)^2g_{\mu\nu}+\frac{8\pi G}{\phi}T_{\mu\nu}+\frac{\omega_{\rm{BD}}}{\phi^2}\Big(\phi_{{},\mu}\phi_{{},\nu}-\frac{1}{2}g_{\mu\nu}\phi_{{},\lambda}\phi^{{},\lambda}\Big)\\
+\frac{1}{\phi}\Big(\phi_{{},\mu\nu}-g_{\mu\nu}\Box_g\phi\Big),
\end{split}
\end{equation}
and
\begin{equation}
\Box_g \phi-m^2_{\rm{BD}}(\phi-\phi_0)=\frac{8\pi G\tilde{T}}{3+2\omega_{\rm{BD}}},
\end{equation}
where 
\begin{equation}
\tilde{T}=T-2\phi\frac{\partial T}{\partial \phi}, \hspace{0.2cm} \Box_g=(-g)^{-\frac{1}{2}}\partial_\nu((-g)^{\frac{1}{2}}g^{\mu\nu}\partial_\mu).
\end{equation}
Here, we consider that $\omega_{\rm{BD}}$ does not vary with the scalar field. The stress energy tensor $T_{\mu\nu}$ is given as 
\begin{equation}
T^{\mu\nu}=(-g)^{-\frac{1}{2}}\sum _a m_a(\phi)\frac{u^\mu_au^\nu_a}{u^0_a}\delta^3(\bf{x}-\bf{x_a}),
\end{equation}
and $T=T^{\mu\nu}g_{\mu\nu}$. The typical values of the sensitivities for normal star is $10^{-6}$, for planet it is $10^{-9}$, for WD it is $10^{-4}$, for NS it is $0.2$, and for BH it is $\frac{1}{2}$ \cite{Zaglauer:1992bp}. In the following, we obtain constraints on fifth force strength by assuming that the BD scalar mediated fifth force contributes to the orbital period loss of the binary system, and its contribution is within the measurement uncertainty. We also obtain constraints on $\omega_{\rm{BD}}$ from the Nordtvedt effect.
\subsection{Constraints on fifth force strength from orbital period loss of compact binary systems}
The total energy of the compact binary system is also modified in alternative theories of gravity such as in Brans-Dicke (BD) theory due to the presence of a BD scalar mediated Yukawa potential $V(r)=\frac{G_{\rm{eff}}MmQq}{(3+2\omega_{\rm{BD}})r}e^{-m_{BD}r}$, which can give rise to a fifth force. Here, $\omega_{\rm{BD}}$ denotes the strength of the BD scalar mediated fifth force with respect to gravity, $m_{\rm{BD}}$ denotes the mass of the Brans-Dicke scalar, $Q$ and $q$ denote the scalar charges of the two compact objects, and $G_{\rm{eff}}=G\Big(\frac{3+2\omega_{\rm{BD}}}{4+2\omega_{\rm{BD}}}\Big)$, where $G$ denotes the standard Newton's gravitational constant. The scalar charge of an i'th object is related to the sensitivity as $q_i=(1-2s_i)$ where the sensitivity $(s_i)=\frac{U_i}{m_i}$ in the weak field limit is defined as the gravitational self energy $(U_i)$ per unit mass. The range of the Brans-Dicke scalar mediated fifth force is $\lambda\sim \frac{1}{m_{\rm{BD}}}$.

The equation of motion for a system of gravitating pointlike masses can be obtained from Einstein, Infeld, and Hoffmann Lagrangian and in the BD theory of gravity the corresponding gravitational coupling becomes \cite{Alsing:2011er}
\begin{equation}
\mathcal{G}=G\Big[1-\frac{1}{2}\xi+\frac{1}{2}\xi (1-2s_1)(1-2s_2)e^{-m_{\rm{BD}}r}\Big].
\label{bd1}
\end{equation} 
Putting $\xi=\frac{1}{2+\omega_{\rm{BD}}}$, $Q=(1-2s_1)$, and $q=(1-2s_2)$, we obtain the expression of $\mathcal{G}$ from Eq. \ref{bd1} as
\begin{equation}
\mathcal{G}=G_{\rm{eff}}+\frac{G_{\rm{eff}}}{3+2\omega_{\rm{BD}}}Qqe^{-m_{\rm{BD}}r},
\label{bd2}
\end{equation}
where $G_{\rm{eff}}$ has been previously defined.

Hence, the total potential of the eccentric compact binary system in BD theory of gravity is 
\begin{equation}
\begin{split}
V(r)=-\frac{GMm}{r}\Big(\frac{3+2\omega_{\rm{BD}}}{4+2\omega_{\rm{BD}}}\Big)-\frac{GMm}{4\epsilon r}\frac{1}{2+\omega_{\rm{BD}}}(1-2s_1)(1-2s_2)\Big[(1+\epsilon)e^{-m_{\rm{BD}}r(1+\epsilon)}-\\
(1-\epsilon)e^{-m_{\rm{BD}}r(1-\epsilon)}\Big].
\end{split}
\label{bd3}
\end{equation}
In the limit $\omega_{\rm{BD}}\rightarrow \infty$, we obtain the standard Newtonian potential $V(r)=-\frac{GMm}{r}$. In $m_{\rm{BD}}\rightarrow 0$ (infinite range fifth force) limit, the total potential becomes
\begin{equation}
V(r)=-\frac{G_{\rm{eff}}Mm}{r}\Big(1+\frac{1}{3+2\omega_{\rm{BD}}}Qq\Big).
\label{bd4}
\end{equation}
The orbital frequency of the eccentric compact binary system in BD theory is obtained as
\begin{equation}
\begin{split}
\Omega^2_f=\Omega^2\Big[\frac{3+2\omega_{\rm{BD}}}{4+2\omega_{\rm{BD}}}+\frac{1}{2\epsilon}\frac{(1-2s_1)(1-2s_2)}{4+2\omega_{\rm{BD}}}\Big\{(1+\epsilon)e^{-m_{\rm{BD}}r(1+\epsilon)}-(1-\epsilon)e^{-m_{\rm{BD}}r(1-\epsilon)}\Big\}+\\
\frac{1}{2\epsilon}\frac{(1-2s_1)(1-2s_2)m_{\rm{BD}}r}{4+2\omega_{\rm{BD}}}\Big\{(1+\epsilon)^2e^{-m_{\rm{BD}}r(1+\epsilon)}-(1-\epsilon)^2e^{-m_{\rm{BD}}r(1-\epsilon)}\Big\}\Big].
\end{split}
\label{bd5}
\end{equation} 
Hence, the total gravitational wave energy loss due to the BD scalar mediated Yukawa potential is 
\begin{equation}
\begin{split}
\frac{dE}{dt}=\frac{dE_{GW}}{dt}\times \Big[\frac{3+2\omega_{\rm{BD}}}{4+2\omega_{\rm{BD}}}+\frac{1}{2\epsilon}\frac{(1-2s_1)(1-2s_2)}{4+2\omega_{\rm{BD}}}\Big\{(1+\epsilon)e^{-m_{\rm{BD}}r(1+\epsilon)}-(1-\epsilon)e^{-m_{\rm{BD}}r(1-\epsilon)}\Big\}+\\
\frac{1}{2\epsilon}\frac{(1-2s_1)(1-2s_2)m_{\rm{BD}}r}{4+2\omega_{\rm{BD}}}\Big\{(1+\epsilon)^2e^{-m_{\rm{BD}}r(1+\epsilon)}-(1-\epsilon)^2e^{-m_{\rm{BD}}r(1-\epsilon)}\Big\}\Big]^3,
\end{split}
\label{bd6}
\end{equation}
where $\frac{dE_{GW}}{dt}$ is defined in Eq. \ref{standard_GR}. For infinite range $(m_{\rm{BD}}\rightarrow 0)$ BD scalar mediated fifth force, Eq. \ref{bd6} becomes
\begin{equation}
\Big(\frac{dE}{dt}\Big)^{m_{\rm{BD}}\rightarrow 0}_{\omega_{\rm{BD}}\neq \infty}=\frac{32}{5}G\mu^2\Omega^6r^4(1-\epsilon^2)^{-7/2}\Big(1+\frac{73}{24}\epsilon^2+\frac{37}{96}\epsilon^4\Big)\Big[\frac{3+2\omega_{\rm{BD}}}{4+2\omega_{\rm{BD}}}+\frac{(1-2s_1)(1-2s_2)}{4+2\omega_{\rm{BD}}}\Big]^3.
\end{equation}
From the rate of energy loss, we can calculate the rate of orbital period loss as it is done in Section. \ref{sec2}.
\begin{figure}[h]
\includegraphics[height=8cm]{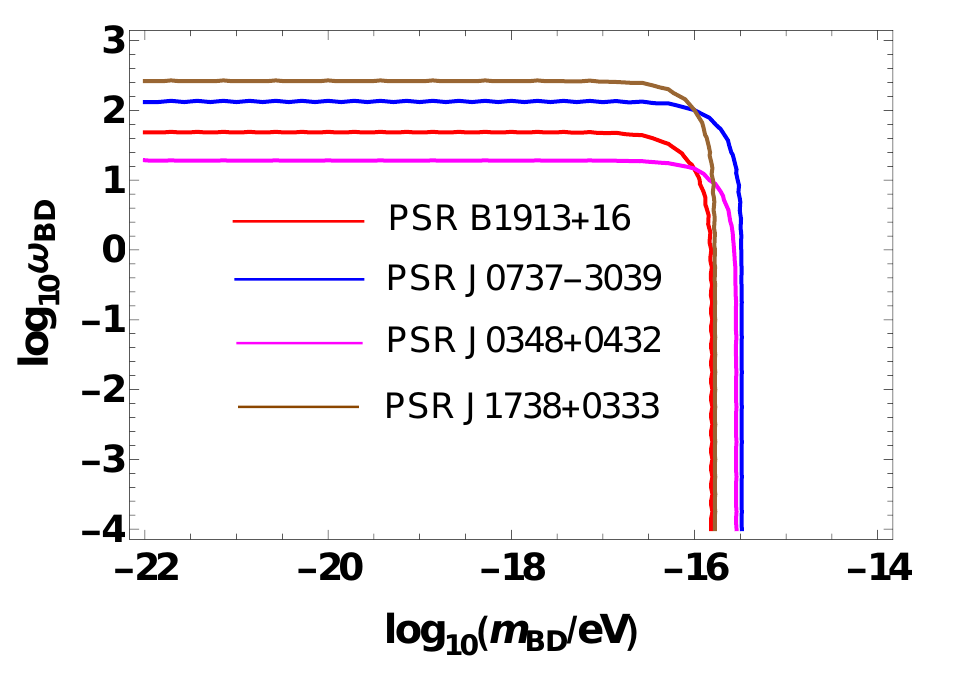}
\caption{\it Constraints on BD parameters $(\omega_{\rm{BD}}, m_{\rm{BD}})$ from orbital period loss of compact binary systems due to BD scalar mediated fifth force.}
\label{bdplot1}
\end{figure} 

In Fig. \ref{bdplot1} we obtain lower bounds on the BD coupling parameter $(\omega_{\rm{BD}})$ from the orbital period loss of compact binary systems. We consider the same four compact binary systems as mentioned in Section. \ref{sec2}. Here, we consider that the orbital period loss of the compact binary systems only decreases due to the BD scalar mediated fifth force together with the GW radiation. Here, we have chosen the values of the sensitivities as $0.2$ for pulsar/NS, and $10^{-4}$ for a WD. We do not need the condition $s_1\neq s_2$ to get the bounds on $\omega_{\rm{BD}}$ from BD scalar mediated fifth force, contributing to the orbital period loss of binary systems. The brown, blue, red, and magenta lines denote the variation of $\omega_{\rm{BD}}$ with $m_{\rm{BD}}$ for PSR J1738+0333, PSR J0737-3039, PSR B1913+16, and PSR J0348+0432. The regions below these lines are excluded. The lower bounds on the BD coupling are
\begin{eqnarray}
\omega_{\rm{BD}} &> & 47 \quad \text{for PSR B1913+16}, \\
\omega_{\rm{BD}} &> & 127 \quad \text{for PSR J0737-3039}, \\
\omega_{\rm{BD}} &> & 19 \quad \text{for PSR J0348+0432}, \\
\omega_{\rm{BD}} &> & 266 \quad \text{for PSR J1738+0333}. \label{bdf}
\end{eqnarray}
We obtain the stronger bound on $\omega_{\rm{BD}}$ as $\omega_{\rm{BD}}>266$ from PSR J1738+0333. The range of the Brans-Dicke scalar mediated fifth force is constrained by the distance between the two stars in the binary system. The above bounds on $\omega_{\rm{BD}}$ are only valid for mass of the BD scalar $m_{\rm{BD}}\lesssim 9.19\times 10^{-17}~\rm{eV}~(\lambda\gtrsim 2.15\times 10^6 ~\rm{km})$.
\subsection{Constraints on fifth force strength from Nordtvedt effect} 
Compact objects with a significant amount of self gravitational energy do not follow the geodesic described by the background metric. The non zero difference of accelerations of a pair of bodies of different self gravitational energies towards a third body is parametrized by $\eta_N$, called the Nordtvedt parameter. This effect is called the Nordtvedt effect which is a direct consequence of the violation of the Strong Equivalence Principle (SEP). The Nordtvedt effect or equivalently the violation of SEP is detectable in the Earth-Moon-Sun system from Lunar Laser Ranging (LLR) \cite{Hofmann2010}. If the accelerations of Earth and Moon towards the Sun are different then the SEP is violated which causes shifting in the lunar orbit towards the Sun. The Nordtvedt parameter can also be probed from the Mercury-Earth-Sun system \cite{Genova2018} and PSR J0337+1715 \cite{Archibald:2018oxs,Ransom:2014xla}. 

Suppose, we consider a three object system with masses $m_1$, $m_2$, and $m_3$ where object $1$ and object $2$ are close to each other and object $3$ is far away from object $1$ and object $2$. Hence, the accelerations of objects $1$ and $2$ towards object $3$ are
\begin{equation}
\begin{split}
\mathbf{a}_1=-\frac{G m_2}{r_{12}^2}\Big(\frac{3+2\omega_{\rm{BD}}}{4+2\omega_{\rm{BD}}}\Big)\hat{r}_{12}-\frac{Gm_3}{4\epsilon_{13}r^2}\frac{1}{2+\omega_{\rm{BD}}}(1-2s_1)(1-2s_3)\Big[(1+\epsilon_{13})e^{-m_{\rm{BD}}r(1+\epsilon_{13})}-\\
(1-\epsilon_{13})e^{-m_{\rm{BD}}r(1-\epsilon_{13})}+m_{BD}r\{(1+\epsilon_{13})^2e^{-m_{\rm{BD}}r(1+\epsilon_{13})}-(1-\epsilon_{13})^2e^{-m_{\rm{BD}}r(1-\epsilon_{13})}\}\Big]\hat{r},
\end{split}
\label{nv1}
\end{equation}
and 
\begin{equation}
\begin{split}
\mathbf{a}_2=+\frac{G m_1}{r_{12}^2}\Big(\frac{3+2\omega_{\rm{BD}}}{4+2\omega_{\rm{BD}}}\Big)\hat{r}_{12}-\frac{Gm_3}{4\epsilon_{23}r^2}\frac{1}{2+\omega_{\rm{BD}}}(1-2s_2)(1-2s_3)\Big[(1+\epsilon_{23})e^{-m_{\rm{BD}}r(1+\epsilon_{23})}-\\
(1-\epsilon_{23})e^{-m_{\rm{BD}}r(1-\epsilon_{23})}+m_{BD}r\{(1+\epsilon_{23})^2e^{-m_{\rm{BD}}r(1+\epsilon_{23})}-(1-\epsilon_{23})^2e^{-m_{\rm{BD}}r(1-\epsilon_{23})}\}\Big]\hat{r},
\end{split}
\label{nv2}
\end{equation}
where we consider $r_{12}\ll r_{13}\sim r_{23}\sim r$, $\epsilon_{12}\ll\epsilon_{13}\sim \epsilon_{23}\sim\epsilon$, and $m_{\rm{BD}}r_{12}\sim 0$. Now the difference in accelerations of object $1$ and object $2$ towards object $3$ is $\mathbf{a}_{12}=\mathbf{a}_1-\mathbf{a}_2$ and 
\begin{equation}
\mathbf{a}_{12}=-\frac{G_{\rm{eff}}(m_1+m_2)}{r^2_{12}}\hat{r}_{12}-\frac{G_{\rm{eff}}m_3}{r^2}(s_2-s_1)\eta_N\hat{r},
\label{nv3}
\end{equation}
where the Nordtvedt parameter is
\begin{equation}
\begin{split}
\eta_N=\frac{1}{(3+2\omega_{\rm{BD}})\epsilon}(1-2s_3)\Big[\{(1+\epsilon)e^{-m_{\rm{BD}}r(1+\epsilon)}-(1-\epsilon)e^{-m_{\rm{BD}}r(1-\epsilon)}\}+\\
m_{BD}r\{(1+\epsilon)^2e^{-m_{\rm{BD}}r(1+\epsilon)}-(1-\epsilon)^2e^{-m_{\rm{BD}}r(1-\epsilon)}\}\Big].
\end{split}
\label{nv4}
\end{equation}
The first term on the right hand side of Eq. \ref{nv3} is the rescaled Newtonian acceleration and the second term arises due to the effect of the fifth force. This term measures the violation of SEP. If the two objects in the inner binary are the same then they will have the same self gravitational energies which imply $s_1=s_2$ and there will be no SEP violation. If the object $3$ is a BH $(s_3=\frac{1}{2})$, then also $\eta_N=0$ and there will be no SEP violation. 
\begin{figure}[h]
\includegraphics[height=8cm]{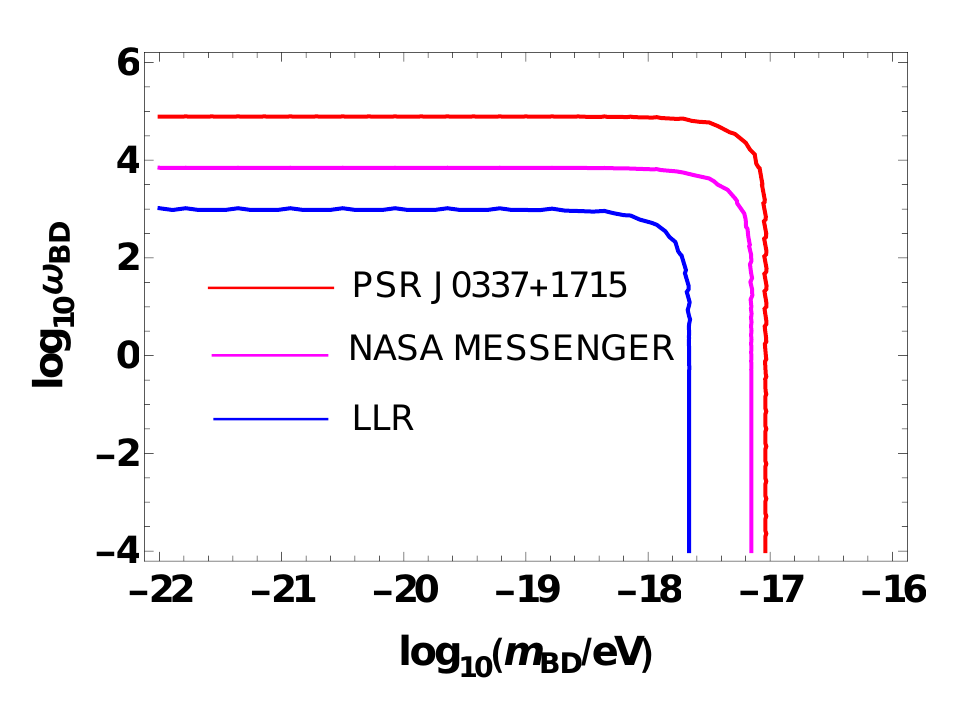}
\caption{\it Constraints on BD parameters $(\omega_{\rm{BD}}, m_{\rm{BD}})$ from Nordtvedt effect due to BD scalar mediated fifth force.}
\label{bdplot2}
\end{figure}  
In FIG. \ref{bdplot2} we obtain constraints on $\omega_{\rm{BD}}$ from Nordtvedt effect. We consider pulsar-WD-WD (PSR J0337+1715), Earth-Sun-Moon (LLR), and Mercury-Earth-Sun (NASA MESSENGER) systems to constrain the BD parameter. The Lunar Laser Ranging (LLR) experiment measures the Nordtvedt parameter as $\eta_N=(0.6\pm 5.2)\times 10^{-4}$ \cite{Hofmann2010} whereas the NASA MESSENGER mission measures the Nordtvedt parameter for Mercury-Earth-Sun system as $\eta_N=(-6.6\pm 7.2)\times 10^{-5}$ \cite{Genova2018}. For PSR J0337+1715, WSRT, GBT, and AO telescope measure the Nordtvedt parameter with an uncertainty $\sigma_{(s_2-s_1)\eta_N}=2.5\times 10^{-6}$ \cite{Archibald:2018oxs,Ransom:2014xla}. The lower bounds $(95\% ~\rm{CL})$ on BD coupling are
\begin{eqnarray}
\omega_{\rm{BD}} &> & 75858 \quad \text{for PSR J0337+1715},\label{bds} \\
\omega_{\rm{BD}} &> & 1069 \quad \text{for LLR}, \\
\omega_{\rm{BD}} &> & 7244 \quad \text{for NASA MESSENGER}.  
\end{eqnarray} 
We obtain the stronger lower bound on $\omega_{\rm{BD}}$ for PSR J0337+1715 from Nordtvedt effect as $\omega_{\rm{BD}}>75858$. The bounds are only valid for $m_{\rm{BD}}\lesssim 1.35\times 10^{-18}~\rm{eV}$. The bounds are stronger than that are discussed in \cite{Alsing:2011er,Seymour:2019tir}. This bound on $\omega_{\rm{BD}}$ obtained from the Nordtvedt effect for PSR J0337+1715  is stronger than the bound obtained from the orbital period loss of compact binary systems by two orders of magnitude. The bounds get stronger if one includes the effect of eccentricity.  
\section{Conclusion and Discussion}\label{con}
In this paper, we obtain constraints on fifth force strength and gauge coupling of ultralight scalar and vector particles from the orbital period loss of compact binary systems and coalescence of two NSs for the GW170817 event. The search for ultralight particles is important as they can be promising candidates for Fuzzy Dark Matter (FDM). The study of FDM is important as it can solve the small scale structure problems in the universe and evade the DM direct detection constraints. In deriving these constraints, we include the effect of eccentricity as the inclusion of eccentricity can enhance the GW energy loss by an order of magnitude for the Hulse-Taylor binary. If the mass of the ultralight particle is less than the inverse of the binary separation of two stars and if both the compact stars contain DM particles then long range Yukawa type fifth force can mediate between the two stars of the binary system. The fifth force can contribute to the orbital period loss of compact binary systems however, its contribution should be no larger than the uncertainty in the measurement of orbital period loss. Compared with the experimental results, we obtain constraints on the fifth force strength from two NS-NS and two NS-WD binary systems. We obtain a stronger constraint on the fifth force strength as $\alpha\lesssim 1.11\times 10^{-3}$ from PSR J1738+0333. This constraint is valid for the mass of the fifth force mediator $M_{Z^\prime}\lesssim9.19\times 10^{-17}~\rm{eV}$ (see Eq.\ref{e16}, FIG.\ref{plot1}). 

Besides the fifth force, ultralight particles can also radiate from the binary system if there is a dark charge to the mass asymmetry between the two stars of the binary system and the mass of the ultralight particle is less than the orbital frequency of the binary system. The radiation of ultralight particles can also contribute to the orbital period loss of compact binary systems. Compared with the experimental results, we obtain constraints on gauge couplings for the radiation of scalar and vector particles. The constraints on gauge couplings become stronger for a large number of dark charge particles and large values of fifth force strength. We obtain stronger constraint on vector gauge coupling as $g_V\lesssim 2.29\times 10^{-20}$ and scalar gauge coupling as $g_S\lesssim 3.06\times 10^{-20}$ for $\alpha=0.9$ with $10^{55}$ number of dark charge particles in the NS. These constraints are valid for the mass of the gauge boson $M_{Z^\prime,S}\lesssim 1.35\times 10^{-19}~\rm{eV}$ (see TABLE \ref{tableI}, FIG.\ref{plot2}).

We also obtain constraints on the axion decay constant from the orbital period loss of compact binary systems. If compact stars are immersed in a low mass axionic potential then axions can mediate a fifth force between two stars in the binary system. The pseudoscalar axions can also radiate from the binary system if the mass of the axion is less than the orbital frequency of the binary system. The orbital period loss due to the fifth force effect rules out axions with decay constant $1.69\times 10^{14}~\rm{GeV}\lesssim f_a\lesssim 3.16\times 10^{17}~\rm{GeV}$ for masses $m_a\lesssim 2.51\times 10^{-16}~\rm{eV}$ (see FIG.\ref{pcap3}). The combined effect of fifth force and radiation rules out axions with decay constant $7.94\times 10^{10}~\rm{GeV}\lesssim f_a\lesssim 3.16\times 10^{17}~\rm{GeV}$ for $\alpha=0.9$ and $m_a\lesssim 2.51\times 10^{-16}\rm{eV}$ (see FIG.\ref{radaxion}). The combined effect of the fifth force and radiation gives stronger constraints on $f_a$ compared to the case of only radiation.

We obtain constraints on the fifth force and radiation strength from the coalescence of two NS in the GW170817 event. In deriving these bounds we include the effect of eccentricity. Since, the eccentricity changes with time for an inspiral binary, we consider time variation of eccentricity with fixed initial values of eccentricity. We consider that there is no charge to the mass asymmetry of the two stars in the binary system to eliminate the radiation effect. The GW frequency increases with increasing eccentricity and fifth force coupling. The frequency also increases with decreasing the mediator mass. Hence, the frequency with smaller values of $\alpha, \epsilon_0$ and larger values of $M_{Z^\prime}$ may stay inside the LIGO frequency band and these parameters can be probed from direct detection of GW. Here, we also obtain that GW frequency with $M_{Z^\prime}=1.98\times 10^{-12}~\rm{eV}, \alpha=0.025$, and $\epsilon_0=10^{-6}, 0.1$ falls into the LIGO sensitivity band in the entire time domain. However, the shifts in coalescence time relative to the gravity only scenario are different for different eccentric orbits. The coalescence time is shifted by $\sim 1.8~\rm{s}$ and $~2.7\rm{s}$ relative to the gravity only scenario for $\epsilon_0=10^{-6}$ and $\epsilon_0=0.1$ respectively for $\alpha\lesssim 0.025$ and $M_{Z^\prime}=1.98\times 10^{-12}~\rm{eV}$. Hence, for $M_{Z^\prime}=1.98\times 10^{-12}~\rm{eV}$, and $\epsilon_0=10^{-6}, 0.1$, LIGO can probe the fifth force strength $\alpha\lesssim 0.025$ (see FIG.\ref{dplotfirst}). 

We also eliminate the effect of the fifth force by considering that one of the compact stars does not contain a dark charge. In that case, we can constrain the ultralight particles from the radiation effect. The GW frequency with the radiation strength $\beta\lesssim 10^{-3}$ falls into the LIGO sensitivity band through the entire time domain and can be probed from direct detection of GW. However, the shifts in coalescence time compared to the gravity only scenario is different for different eccentric orbits. The shift is $9.7~\rm{s}$ for $\epsilon_0\sim 10^{-6}$ and $10.6~\rm{s}$ for $\epsilon_0\sim 0.1$ with $\beta=10^{-3}$ and $M_{Z^\prime}=1.98\times 10^{-14}~\rm{eV}$. The frequency also increases with increasing $\beta$ and shifts towards the left of the LIGO sensitivity band. Hence, large values of $\beta$ cannot be probed from LIGO/Virgo data. For the radiation case, smaller values of $M_{Z^\prime}$ can be probed from LIGO/Virgo data compared to the fifth force case. Hence, the ultralight particle radiation is switched on before the waveform comes into the LIGO frequency band. The frequency of the GW for vector radiation is larger than the scalar radiation. Hence, a particular value of $\beta$ which cannot be probed for vector radiation can be probed for scalar radiation from the LIGO/Virgo data (see FIG.\ref{dplotthird}).

We also obtain the variation of GW amplitude with time for different values of strengths of a fifth force, radiation and initial values of eccentricity. The amplitude increases with increasing $\alpha$, $\beta$, and $\epsilon_0$. The contribution of $\beta$ in amplitude measurement is larger than $\alpha$. Hence, amplitude with very large values of $\beta$ stay outside of the LIGO frequency band and cannot be probed from GW (see FIG.\ref{amplitude}).  

In FIG.\ref{cap1} we qualitatively obtain the capture of DM mass fraction by a compact binary system. A binary system can capture more DM particles than an isolated NS. Also, we obtain a stronger bound on DM mass fraction if the DM particles are captured by a shorter distance. In deriving the bounds, we do not consider any capture mechanism. 
 
We obtain constraints on BD coupling from the orbital period loss of compact binary system due to BD scalar mediated fifth force in Brans-Dicke's theory of gravity. We obtain strongest bound on $\omega_{\rm{BD}}$ as $\omega_{\rm{BD}}>266$ for PSR J1738+0333 (see Eq.\ref{bdf}, FIG.\ref{bdplot1}). The nonzero value of the Nordtvedt parameter is a measure of the violation of SEP. We obtain $\omega_{\rm{BD}}>75858$ from the Nordtvedt effect for the system PSR J0337+1715 (see Eq.\ref{bds}, FIG.\ref{bdplot2}). This bound is two orders stronger than the bound obtained from orbital period loss by compact binary systems. Similarly as done in Einstein's gravity, we can derive constraints on $\omega_{\rm{BD}}$ from the Shapiro time delay, and periastron advance for an eccentric orbit. We can also obtain constraints on BD coupling strength from indirect and direct detection of gravitational waves by including the radiation of ultralight particles and eccentricity with effect as described in section \ref{bd}. We will address these cases in a separate publication.

Our method of calculating the GW frequency is general and can be used for arbitrary eccentric compact binary systems (NS-NS, NS-WD, NS-BH, etc). The eccentricity factor is important as it can enhance the GW frequency. For example, the value of orbital period loss increases by one order of magnitude for the Hulse-Taylor binary system if one includes the effect of eccentricity. Hence, particular values of fifth force parameters $(\alpha, \beta, M_{Z^\prime,S})$ which cannot be probed from LIGO/Virgo data with $\epsilon_0=0$ scenario, can be probed from $\epsilon_0\neq 0$ scenario and vice versa. Also, the shift in the coalescence time from gravity only scenario is larger for orbits with higher values of eccentricity.

Our procedure of obtaining the constraints on fifth force and radiation strength for arbitrary eccentricity can be easily generalized to second and third generation GW detectors such as Advanced LIGO, Einstein Telescope, KAGRA, Cosmic Explorer, LISA etc. Our bounds on new force parameters can become strengthened with these future observations. It will be interesting to compute the waveform and phase of GW in presence of fifth force parameters and eccentricity both in Einstein's and alternative theories of gravity.   

We also envisage that beyond the observables discussed in this paper, the dynamics of the dark sector can potentially leave its imprints on GW from binary neutron star inspirals in several other ways. For example, the dark sector particles could instead modify the properties of the NS like tidal deformability \cite{Nelson:2018xtr} or due to particles produced via BH superradiance \cite{Day:2019bbh} effects. However, it is needless to say that these effects will not alter the results presented in our analysis for probing the fifth force effects. These scenarios may only be present as additional new effects which we plan to study in the future.
\section*{Acknowledgements}
T.K.P would like to thank Ranjan Laha for useful discussions. 
\appendix
\section{Terms responsible for the energy loss of binary systems}
\label{appendix1}
\begin{equation}
\begin{split}
    \mathcal{E}_1=-\frac{GMm}{2r^2}\frac{dr}{dt}\Big[1+\frac{\alpha}{2\epsilon}\Big\{(1+\epsilon)e^{-M_{Z^\prime}r(1+\epsilon)}-(1-\epsilon)e^{-M_{Z^\prime}r(1-\epsilon)}+M_{Z^\prime}r\Big((1+\epsilon)^2 e^{-M_{Z^\prime}r(1+\epsilon)}-\\
    (1-\epsilon)^2 e^{-M_{Z^\prime}r(1-\epsilon)}\Big)+M^2_{Z^\prime}r^2\Big((1+\epsilon)^3 e^{-M_{Z^\prime}r(1+\epsilon)}-(1-\epsilon)^3 e^{-M_{Z^\prime}r(1-\epsilon)}\Big)\Big\}\Big]+\mathcal{O}(\alpha^2),
    \end{split}
\end{equation}
\begin{equation}
\begin{split}
    \mathcal{E}_2=-\frac{GMm\alpha}{4r\epsilon}\times \frac{304}{15}G\mu \Omega^4
r^2 \epsilon (1-\epsilon^2)^{-5/2}\Big(1+\frac{121}{304}\epsilon^2\Big)\Big[\frac{1}{\epsilon}(1+\epsilon)e^{-M_{Z^\prime}r(1+\epsilon)}-\\
\frac{1}{\epsilon}(1-\epsilon)e^{-M_{Z^\prime}r(1-\epsilon)}+M_{Z^\prime}r(1+\epsilon)e^{-M_{Z^\prime}r(1+\epsilon)}-e^{-M_{Z^\prime}r(1+\epsilon)}+M_{Z^\prime}r(1-\epsilon)e^{-M_{Z^\prime}r(1-\epsilon)}-\\
e^{-M_{Z^\prime}r(1-\epsilon)}\Big]+\mathcal{O}(\alpha^2),
\end{split}
\end{equation}
and
\begin{equation}
    \begin{split}
      \mathcal{E}_3=-\frac{GMm\alpha M_{Z^\prime}}{4\epsilon}\times \frac{304}{15}G\mu \Omega^4 r^2\epsilon (1-\epsilon^2)^{-5/2}\Big(1+\frac{121}{304}\epsilon^2\Big)\Big[\frac{1}{\epsilon}(1+\epsilon)^2e^{-M_{Z^\prime}r(1+\epsilon)}-\\
      \frac{1}{\epsilon}(1-\epsilon)^2e^{-M_{Z^\prime}r(1-\epsilon)}-2(1+\epsilon)e^{-M_{Z^\prime}r(1+\epsilon)}+M_{Z^\prime}r(1+\epsilon)^2e^{-M_{Z^\prime}r(1+\epsilon)}-2(1-\epsilon)e^{-M_{Z^\prime}r(1-\epsilon)}+\\
      M_{Z^\prime}r(1-\epsilon)^2e^{-M_{Z^\prime}r(1-\epsilon)}\Big]+\mathcal{O}(\alpha^2),
    \end{split}
\end{equation}
\begin{equation}
\begin{split}
    \mathcal{E}_4= -\frac{GMm}{2r^2}\Big[1+\frac{\alpha}{2\epsilon}\Big\{(1+\epsilon)e^{-M_{Z^\prime}r(1+\epsilon)}-(1-\epsilon)e^{-M_{Z^\prime}r(1-\epsilon)}+M_{Z^\prime}r\Big((1+\epsilon)^2 e^{-M_{Z^\prime}r(1+\epsilon)}-\\
    (1-\epsilon)^2 e^{-M_{Z^\prime}r(1-\epsilon)}\Big)+M^2_{Z^\prime}r^2\Big((1+\epsilon)^3 e^{-M_{Z^\prime}r(1+\epsilon)}-(1-\epsilon)^3 e^{-M_{Z^\prime}r(1-\epsilon)}\Big)\Big\}\Big]+\mathcal{O}(\alpha^2).
    \end{split}
\end{equation}
\begin{equation}
\begin{split}
   \mathcal{\xi}_1= -\frac{3m_1}{r^4}-\frac{m_1}{r^3}\frac{\alpha}{2\epsilon}\Big[\frac{3}{r}\{(1+\epsilon)e^{-M_{Z^\prime}r(1+\epsilon)}-(1-\epsilon)e^{-M_{Z^\prime}r(1-\epsilon)}\}+\\
   M_{Z^\prime}\{(1+\epsilon)^2e^{-M_{Z^\prime}r(1+\epsilon)}-(1-\epsilon)^2e^{-M_{Z^\prime}r(1-\epsilon)}\}\Big]-\frac{m_1\alpha M_{Z^\prime}}{\epsilon r^3}\Big[(1+\epsilon)^2e^{-M_{Z^\prime}r(1+\epsilon)}-\\
   (1-\epsilon)^2e^{-M_{Z^\prime}r(1-\epsilon)}+\frac{1}{2}r M_{Z^\prime}\{(1+\epsilon)^3e^{-M_{Z^\prime}r(1+\epsilon)}-(1-\epsilon)^3e^{-M_{Z^\prime}r(1-\epsilon)}\}\Big]+\mathcal{O}(\alpha^2)
   \end{split}
\end{equation}
and 
\begin{equation}
    \begin{split}
       \mathcal{\xi}_2= -\frac{m_1}{r^3}\times\frac{304}{15}G\mu\Omega^4 r^2\epsilon(1-\epsilon^2)^{-\frac{5}{2}}\Big(1+\frac{121}{304}\epsilon^2\Big)\frac{\alpha}{2\epsilon}\Big[\frac{1}{\epsilon}(1+\epsilon)e^{-M_{Z^\prime}r(1+\epsilon)}-\\
       \frac{1}{\epsilon}(1-\epsilon)e^{-M_{Z^\prime}r(1-\epsilon)}+M_{Z^\prime}r(1+\epsilon)e^{-M_{Z^\prime}r(1+\epsilon)}+M_{Z^\prime}r(1-\epsilon)e^{-M_{Z^\prime}r(1-\epsilon)}-e^{-M_{Z^\prime}r(1+\epsilon)}-\\
       e^{-M_{Z^\prime}r(1-\epsilon)}\Big]-\frac{m_1\alpha M_{Z^\prime}}{2\epsilon r^2}\frac{304}{15}G\mu\Omega^4 r^2\epsilon(1-\epsilon^2)^{-5/2}\Big(1+\frac{121}{304}\epsilon^2\Big)\Big[\frac{1}{\epsilon}(1+\epsilon)^2e^{-M_{Z^\prime}r(1+\epsilon)}-\\
       \frac{1}{\epsilon}(1-\epsilon)^2e^{-M_{Z^\prime}r(1-\epsilon)}+M_{Z^\prime}r(1+\epsilon)^2e^{-M_{Z^\prime}r(1+\epsilon)}+M_{Z^\prime}r(1-\epsilon)^2e^{-M_{Z^\prime}r(1-\epsilon)}-\\
       2(1+\epsilon)e^{-M_{Z^\prime}r(1+\epsilon)}-2(1-\epsilon)e^{-M_{Z^\prime}r(1-\epsilon)}\Big]+\mathcal{O}(\alpha^2),
    \end{split}
\end{equation}
\begin{equation}
\begin{split}
    K_1=\frac{96}{5}\frac{G^4(M+m)^2 Mm}{r^7}(1-\epsilon^2)^{-7/2}\Big(1+\frac{73}{24}\epsilon^2+\frac{37}{96}\epsilon^4\Big)\Big[1+a_1-b_1+\\
    \frac{3\alpha}{2\epsilon}\{(1+\epsilon)e^{-M_{Z^\prime}r(1+\epsilon)}-(1-\epsilon)e^{-M_{Z^\prime}r(1-\epsilon)}\}+\frac{3\alpha M_{Z^\prime}r}{2\epsilon}\{(1+\epsilon)^2e^{-M_{Z^\prime}r(1+\epsilon)}-\\
    (1-\epsilon)^2e^{-M_{Z^\prime}r(1-\epsilon)}\}\Big]+\mathcal{O}(\alpha^2).
    \end{split}
    \label{ki1}
\end{equation}
\begin{equation}
    \begin{split}
        a_1=\frac{r\alpha}{6\epsilon}\Big[\frac{3}{r}\{(1+\epsilon)e^{-M_{Z^\prime}r(1+\epsilon)}-(1-\epsilon)e^{-M_{Z^\prime}r(1-\epsilon)}\}+M_{Z^\prime}\{(1+\epsilon)^2e^{-M_{Z^\prime}r(1+\epsilon)}-\\
        (1-\epsilon)^2e^{-M_{Z^\prime}r(1-\epsilon)}\}\Big]+\frac{\alpha M_{Z^\prime}r}{3\epsilon}\Big[(1+\epsilon)^2e^{-M_{Z^\prime}r(1+\epsilon)}-(1-\epsilon)^2e^{-M_{Z^\prime}r(1-\epsilon)}+\\
        \frac{1}{2}r M_{Z^\prime}\{(1+\epsilon)^3e^{-M_{Z^\prime}r(1+\epsilon)}- (1-\epsilon)^3e^{-M_{Z^\prime}r(1-\epsilon)}\}\Big]+\mathcal{O}(\alpha^2)
    \end{split}
\end{equation}
\begin{equation}
    \begin{split}
        b_1=\frac{\alpha}{2\epsilon}\Big[(1+\epsilon)e^{-M_{Z^\prime}r(1+\epsilon)}-(1-\epsilon)e^{-M_{Z^\prime}r(1-\epsilon)}+M_{Z^\prime}r\Big((1+\epsilon)^2 e^{-M_{Z^\prime}r(1+\epsilon)}-\\
    (1-\epsilon)^2 e^{-M_{Z^\prime}r(1-\epsilon)}\Big)+M^2_{Z^\prime}r^2\Big((1+\epsilon)^3 e^{-M_{Z^\prime}r(1+\epsilon)}-(1-\epsilon)^3 e^{-M_{Z^\prime}r(1-\epsilon)}\Big)\Big]+\mathcal{O}(\alpha^2).
    \end{split}
\end{equation}
\begin{equation}
    \begin{split}
        K^S_2=\frac{g^2}{4\pi}\frac{G^2(M+m)Mm}{r^6}\Big(\frac{Q}{M}-\frac{q}{m}\Big)^2\sum_{n>n_{0\alpha}^S}2n^2\Big[{J^\prime_n}^2(n\epsilon)+\Big(\frac{1-\epsilon^2}{\epsilon^2}\Big)J_n^2(n\epsilon)\Big]\Big(1-\frac{{n_{0\alpha}^S}^2}{n^2}\Big)^\frac{3}{2},
    \end{split}
    \label{ki2}
\end{equation}
\begin{equation}
\begin{split}
     K^V_2=\frac{g^2}{2\pi}\frac{G^2(M+m)Mm}{r^6}\Big(\frac{Q}{M}-\frac{q}{m}\Big)^2\sum_{n>n_{0\alpha}^V}2n^2\Big[{J^\prime_n}^2(n\epsilon)+\Big(\frac{1-\epsilon^2}{\epsilon^2}\Big)J_n^2(n\epsilon)\Big]\sqrt{1-\frac{{n_{0\alpha}^V}^2}{n^2}}\times\\
     \Big(1+\frac{{n_{0\alpha}^V}^2}{2n^2}\Big).
     \end{split}
\end{equation}
\begin{equation}
    \begin{split}
        K_3=\frac{76\alpha}{5}\frac{G^4(M+m)^2mM}{r^7}(1-\epsilon^2)^{-5/2}\Big(1+\frac{121}{304}\epsilon^2\Big)\Big[\frac{1}{\epsilon}(1+\epsilon)e^{-M_{Z^\prime}r(1+\epsilon)}-\\
        \frac{1}{\epsilon}(1-\epsilon)e^{-M_{Z^\prime}r(1-\epsilon)}+M_{Z^\prime}r(1+\epsilon)e^{-M_{Z^\prime}r(1+\epsilon)}-e^{-M_{Z^\prime}r(1+\epsilon)}+M_{Z^\prime}r(1-\epsilon)e^{-M_{Z^\prime}r(1-\epsilon)}-\\
e^{-M_{Z^\prime}r(1-\epsilon)}\Big]+\mathcal{O}(\alpha^2).
\label{ki3}
    \end{split}
\end{equation}
\begin{equation}
    \begin{split}
        K_4=\frac{76}{5}\alpha M_{Z^\prime}\frac{G^4(M+m)^2Mm}{r^6}(1-\epsilon^2)^{-5/2}\Big(1+\frac{121}{304}\epsilon^2\Big)\Big[\frac{1}{\epsilon}(1+\epsilon)^2e^{-M_{Z^\prime}r(1+\epsilon)}-\\
      \frac{1}{\epsilon}(1-\epsilon)^2e^{-M_{Z^\prime}r(1-\epsilon)}-2(1+\epsilon)e^{-M_{Z^\prime}r(1+\epsilon)}+M_{Z^\prime}r(1+\epsilon)^2e^{-M_{Z^\prime}r(1+\epsilon)}-2(1-\epsilon)e^{-M_{Z^\prime}r(1-\epsilon)}+\\
      M_{Z^\prime}r(1-\epsilon)^2e^{-M_{Z^\prime}r(1-\epsilon)}\Big]+\mathcal{O}(\alpha^2),
    \end{split}
    \label{ki4}
\end{equation}
\begin{equation}
    \begin{split}
        L_1=\frac{96}{5}G^{5/3}\mathcal{M}_{\rm{ch}}^{5/3}\Omega_f^{11/3}(1-\epsilon^2)^{-7/2}\Big(1+\frac{73}{24}\epsilon^2+\frac{37}{96}\epsilon^4\Big)\Big[1+a_1-b_1+\\
        \frac{3\alpha}{2\epsilon}\{(1+\epsilon)e^{-M_{Z^\prime}r(1+\epsilon)}-(1-\epsilon)e^{-M_{Z^\prime}r(1-\epsilon)}\}+\frac{3\alpha M_{Z^\prime}r}{2\epsilon}\{(1+\epsilon)^2e^{-M_{Z^\prime}r(1+\epsilon)}-\\
    (1-\epsilon)^2e^{-M_{Z^\prime}r(1-\epsilon)}\}-\frac{7\alpha}{6\epsilon}\{(1+\epsilon)e^{-M_{Z^\prime}(1+\epsilon)m_1(m_1\Omega_f)^{-\frac{2}{3}}}-(1-\epsilon)e^{-M_{Z^\prime}(1-\epsilon)m_1(m_1\Omega_f)^{-\frac{2}{3}}}\}-\\
    \frac{7\alpha M_{Z^\prime}}{6\epsilon}\Big(\frac{m_1}{\Omega_f^2}\Big)^\frac{1}{3}\{(1+\epsilon)^2e^{-M_{Z^\prime}(1+\epsilon)m_1(m_1\Omega_f)^{-\frac{2}{3}}}-(1-\epsilon)^2e^{-M_{Z^\prime}(1-\epsilon)m_1(m_1\Omega_f)^{-\frac{2}{3}}}\}\Big]\Big]+\mathcal{O}(\alpha^2) ,
    \end{split}
\end{equation}

\begin{equation}
    \begin{split}
        L^S_2=\frac{g^2}{4\pi}\frac{\Omega_f^3\mathcal{M}_{\rm{ch}}^{5/3}}{(M+m)^{2/3}}\delta\sum_{n>n_{0\alpha}^S}2n^2\Big[{J^\prime_n}^2(n\epsilon)+\Big(\frac{1-\epsilon^2}{\epsilon^2}\Big)J_n^2(n\epsilon)\Big]\Big(1-\frac{{n_{0\alpha}^S}^2}{n^2}\Big)^\frac{3}{2}+\mathcal{O(\delta\alpha)},
    \end{split}
\end{equation}
\begin{equation}
    \begin{split}
        L^V_2=\frac{g^2}{2\pi}\frac{\Omega_f^3\mathcal{M}_{\rm{ch}}^{5/3}}{(M+m)^{2/3}}\delta\sum_{n>n_{0\alpha}^V}2n^2\Big[{J^\prime_n}^2(n\epsilon)+\Big(\frac{1-\epsilon^2}{\epsilon^2}\Big)J_n^2(n\epsilon)\Big]\sqrt{1-\frac{{n_{0\alpha}^V}^2}{n^2}}\Big(1+\frac{{n_{0\alpha}^V}^2}{2n^2}\Big)+\mathcal{O(\delta\alpha)},
    \end{split}
\end{equation}
\begin{equation}
    \begin{split}
        L_3=\frac{76\alpha}{5}G^{5/3}\mathcal{M}_{\rm{ch}}^{5/3}\Omega_f^{11/3}(1-\epsilon^2)^{-5/2}\Big(1+\frac{121}{304}\epsilon^2\Big)\Big[\frac{1}{\epsilon}(1+\epsilon)e^{-M_{Z^\prime}r(1+\epsilon)}-\\
        \frac{1}{\epsilon}(1-\epsilon)e^{-M_{Z^\prime}r(1-\epsilon)}+M_{Z^\prime}r(1+\epsilon)e^{-M_{Z^\prime}r(1+\epsilon)}-e^{-M_{Z^\prime}r(1+\epsilon)}+M_{Z^\prime}r(1-\epsilon)e^{-M_{Z^\prime}r(1-\epsilon)}-\\
e^{-M_{Z^\prime}r(1-\epsilon)}\Big]+\mathcal{O}(\alpha^2),
    \end{split}
\end{equation}
\begin{equation}
\begin{split}
    L_4=\frac{76}{5}\alpha M_{Z^\prime}G^2\mathcal{M}_{\rm{ch}}^{5/3}(M+m)^{1/3}\Omega_f^3(1-\epsilon^2)^{-5/2}\Big(1+\frac{121}{304}\epsilon^2\Big)\Big[\frac{1}{\epsilon}(1+\epsilon)^2e^{-M_{Z^\prime}r(1+\epsilon)}-\\
    \frac{1}{\epsilon}(1-\epsilon)^2e^{-M_{Z^\prime}r(1-\epsilon)}-2(1+\epsilon)e^{-M_{Z^\prime}r(1+\epsilon)}+M_{Z^\prime}r(1+\epsilon)^2e^{-M_{Z^\prime}r(1+\epsilon)}-\\
    2(1-\epsilon)e^{-M_{Z^\prime}r(1-\epsilon)}+M_{Z^\prime}r(1-\epsilon)^2e^{-M_{Z^\prime}r(1-\epsilon)}\Big]+\mathcal{O}(\alpha^2),
      \end{split}
\end{equation}
\begin{equation}
    \begin{split}
        L_5=-\frac{76}{15}G^{5/3}\alpha\mathcal{M}_{\rm{ch}}^{5/3}\Omega_f^{11/3}(1-\epsilon^2)^{-5/2}\Big(1+\frac{121}{304}\epsilon^2\Big)\Big[\frac{1}{\epsilon}(1+\epsilon)e^{-M_{Z^\prime}r(1+\epsilon)}-\\
       \frac{1}{\epsilon}(1-\epsilon)e^{-M_{Z^\prime}r(1-\epsilon)}+M_{Z^\prime}r(1+\epsilon)e^{-M_{Z^\prime}r(1+\epsilon)}+M_{Z^\prime}r(1-\epsilon)e^{-M_{Z^\prime}r(1-\epsilon)}-e^{-M_{Z^\prime}r(1+\epsilon)}-\\
       e^{-M_{Z^\prime}r(1-\epsilon)}\Big]-\frac{76}{15}G^2\mathcal{M}_{\rm{ch}}^{5/3}(M+m)^{1/3}\alpha M_{Z^\prime}\Omega_f^3(1-\epsilon^2)^{-5/2}\Big(1+\frac{121}{304}\epsilon^2\Big)\Big[\frac{1}{\epsilon}(1+\epsilon)^2e^{-M_{Z^\prime}r(1+\epsilon)}-\\
       \frac{1}{\epsilon}(1-\epsilon)^2e^{-M_{Z^\prime}r(1-\epsilon)}+M_{Z^\prime}r(1+\epsilon)^2e^{-M_{Z^\prime}r(1+\epsilon)}+M_{Z^\prime}r(1-\epsilon)^2e^{-M_{Z^\prime}r(1-\epsilon)}-\\
       2(1+\epsilon)e^{-M_{Z^\prime}r(1+\epsilon)}-2(1-\epsilon)e^{-M_{Z^\prime}r(1-\epsilon)}\Big]+\mathcal{O}(\alpha^2).
    \end{split}
\end{equation}
\bibliographystyle{utphys}
\bibliography{zultralight}

\providecommand{\href}[2]{#2}\begingroup\raggedright\begin{thebibliography}{10}

\bibitem{CMS:2012qbp}
{\bfseries CMS} Collaboration, S.~Chatrchyan {\em et~al.}, ``{Observation of a
  New Boson at a Mass of 125 GeV with the CMS Experiment at the LHC},''
  \href{http://dx.doi.org/10.1016/j.physletb.2012.08.021}{{\em Phys. Lett. B}
  {\bfseries 716} (2012) 30--61},
  \href{http://arxiv.org/abs/1207.7235}{{\ttfamily arXiv:1207.7235 [hep-ex]}}.

\bibitem{ATLAS:2012yve}
{\bfseries ATLAS} Collaboration, G.~Aad {\em et~al.}, ``{Observation of a new
  particle in the search for the Standard Model Higgs boson with the ATLAS
  detector at the LHC},''
  \href{http://dx.doi.org/10.1016/j.physletb.2012.08.020}{{\em Phys. Lett. B}
  {\bfseries 716} (2012) 1--29},
  \href{http://arxiv.org/abs/1207.7214}{{\ttfamily arXiv:1207.7214 [hep-ex]}}.

\bibitem{LIGOScientific:2016aoc}
{\bfseries LIGO Scientific, Virgo} Collaboration, B.~P. Abbott {\em et~al.},
  ``{Observation of Gravitational Waves from a Binary Black Hole Merger},''
  \href{http://dx.doi.org/10.1103/PhysRevLett.116.061102}{{\em Phys. Rev.
  Lett.} {\bfseries 116} no.~6, (2016) 061102},
  \href{http://arxiv.org/abs/1602.03837}{{\ttfamily arXiv:1602.03837 [gr-qc]}}.

\bibitem{ParticleDataGroup:2020ssz}
{\bfseries Particle Data Group} Collaboration, P.~A. Zyla {\em et~al.},
  ``{Review of Particle Physics},''
  \href{http://dx.doi.org/10.1093/ptep/ptaa104}{{\em PTEP} {\bfseries 2020}
  no.~8, (2020) 083C01}.

\bibitem{Sakharov:1967dj}
A.~D. Sakharov, ``{Violation of CP Invariance, C asymmetry, and baryon
  asymmetry of the universe},''
  \href{http://dx.doi.org/10.1070/PU1991v034n05ABEH002497}{{\em Pisma Zh. Eksp.
  Teor. Fiz.} {\bfseries 5} (1967) 32--35}.

\bibitem{Peccei:1977hh}
R.~D. Peccei and H.~R. Quinn, ``{CP Conservation in the Presence of
  Instantons},'' \href{http://dx.doi.org/10.1103/PhysRevLett.38.1440}{{\em
  Phys. Rev. Lett.} {\bfseries 38} (1977) 1440--1443}.

\bibitem{Planck:2015fie}
{\bfseries Planck} Collaboration, P.~A.~R. Ade {\em et~al.}, ``{Planck 2015
  results. XIII. Cosmological parameters},''
  \href{http://dx.doi.org/10.1051/0004-6361/201525830}{{\em Astron. Astrophys.}
  {\bfseries 594} (2016) A13},
  \href{http://arxiv.org/abs/1502.01589}{{\ttfamily arXiv:1502.01589
  [astro-ph.CO]}}.

\bibitem{Planck:2018vyg}
{\bfseries Planck} Collaboration, N.~Aghanim {\em et~al.}, ``{Planck 2018
  results. VI. Cosmological parameters},''
  \href{http://dx.doi.org/10.1051/0004-6361/201833910}{{\em Astron. Astrophys.}
  {\bfseries 641} (2020) A6}, \href{http://arxiv.org/abs/1807.06209}{{\ttfamily
  arXiv:1807.06209 [astro-ph.CO]}}. [Erratum: Astron.Astrophys. 652, C4
  (2021)].

\bibitem{Fischbach:1985tk}
E.~Fischbach, D.~Sudarsky, A.~Szafer, C.~Talmadge, and S.~H. Aronson,
  ``{Reanalysis of the Eotvos Experiment},''
  \href{http://dx.doi.org/10.1103/PhysRevLett.56.3}{{\em Phys. Rev. Lett.}
  {\bfseries 56} (1986) 3}. [Erratum: Phys.Rev.Lett. 56, 1427 (1986)].

\bibitem{Joyce:2014kja}
A.~Joyce, B.~Jain, J.~Khoury, and M.~Trodden, ``{Beyond the Cosmological
  Standard Model},''
  \href{http://dx.doi.org/10.1016/j.physrep.2014.12.002}{{\em Phys. Rept.}
  {\bfseries 568} (2015) 1--98},
  \href{http://arxiv.org/abs/1407.0059}{{\ttfamily arXiv:1407.0059
  [astro-ph.CO]}}.

\bibitem{Wald:1984rg}
R.~M. Wald,
  \href{http://dx.doi.org/10.7208/chicago/9780226870373.001.0001}{{\em {General
  Relativity}}}.
\newblock Chicago Univ. Pr., Chicago, USA, 1984.

\bibitem{Hulse:1974eb}
R.~A. Hulse and J.~H. Taylor, ``{Discovery of a pulsar in a binary system},''
  \href{http://dx.doi.org/10.1086/181708}{{\em Astrophys. J. Lett.} {\bfseries
  195} (1975) L51--L53}.

\bibitem{Weisberg:2016jye}
J.~M. Weisberg and Y.~Huang, ``{Relativistic Measurements from Timing the
  Binary Pulsar PSR B1913+16},''
  \href{http://dx.doi.org/10.3847/0004-637X/829/1/55}{{\em Astrophys. J.}
  {\bfseries 829} no.~1, (2016) 55},
  \href{http://arxiv.org/abs/1606.02744}{{\ttfamily arXiv:1606.02744
  [astro-ph.HE]}}.

\bibitem{Hook:2017psm}
A.~Hook and J.~Huang, ``{Probing axions with neutron star inspirals and other
  stellar processes},'' \href{http://dx.doi.org/10.1007/JHEP06(2018)036}{{\em
  JHEP} {\bfseries 06} (2018) 036},
  \href{http://arxiv.org/abs/1708.08464}{{\ttfamily arXiv:1708.08464
  [hep-ph]}}.

\bibitem{Huang:2018pbu}
J.~Huang, M.~C. Johnson, L.~Sagunski, M.~Sakellariadou, and J.~Zhang,
  ``{Prospects for axion searches with Advanced LIGO through binary mergers},''
  \href{http://dx.doi.org/10.1103/PhysRevD.99.063013}{{\em Phys. Rev. D}
  {\bfseries 99} no.~6, (2019) 063013},
  \href{http://arxiv.org/abs/1807.02133}{{\ttfamily arXiv:1807.02133
  [hep-ph]}}.

\bibitem{Kopp:2018jom}
J.~Kopp, R.~Laha, T.~Opferkuch, and W.~Shepherd, ``{Cuckoo\textquoteright{}s
  eggs in neutron stars: can LIGO hear chirps from the dark sector?},''
  \href{http://dx.doi.org/10.1007/JHEP11(2018)096}{{\em JHEP} {\bfseries 11}
  (2018) 096}, \href{http://arxiv.org/abs/1807.02527}{{\ttfamily
  arXiv:1807.02527 [hep-ph]}}.

\bibitem{Poddar:2021ose}
T.~K. Poddar, ``{Constraints on ultralight axions, vector gauge bosons, and
  unparticles from geodetic and frame-dragging measurements},''
  \href{http://dx.doi.org/10.1140/epjc/s10052-022-10956-z}{{\em Eur. Phys. J.
  C} {\bfseries 82} no.~11, (2022) 982},
  \href{http://arxiv.org/abs/2111.05632}{{\ttfamily arXiv:2111.05632
  [hep-ph]}}.

\bibitem{Poddar:2020qft}
T.~K. Poddar and S.~Mohanty, ``{Probing the angle of birefringence due to long
  range axion hair from pulsars},''
  \href{http://dx.doi.org/10.1103/PhysRevD.102.083029}{{\em Phys. Rev. D}
  {\bfseries 102} no.~8, (2020) 083029},
  \href{http://arxiv.org/abs/2003.11015}{{\ttfamily arXiv:2003.11015
  [hep-ph]}}.

\bibitem{Poddar:2021sbc}
T.~K. Poddar, ``{Constraints on axionic fuzzy dark matter from light bending
  and Shapiro time delay},''
  \href{http://dx.doi.org/10.1088/1475-7516/2021/09/041}{{\em JCAP} {\bfseries
  09} (2021) 041}, \href{http://arxiv.org/abs/2104.09772}{{\ttfamily
  arXiv:2104.09772 [hep-ph]}}.

\bibitem{KumarPoddar:2020kdz}
T.~Kumar~Poddar, S.~Mohanty, and S.~Jana, ``{Constraints on long range force
  from perihelion precession of planets in a gauged $L_e-L_{\mu,\tau}$
  scenario},'' \href{http://dx.doi.org/10.1140/epjc/s10052-021-09078-9}{{\em
  Eur. Phys. J. C} {\bfseries 81} no.~4, (2021) 286},
  \href{http://arxiv.org/abs/2002.02935}{{\ttfamily arXiv:2002.02935
  [hep-ph]}}.

\bibitem{KumarPoddar:2019ceq}
T.~Kumar~Poddar, S.~Mohanty, and S.~Jana, ``{Vector gauge boson radiation from
  compact binary systems in a gauged $L_\mu-L_\tau$ scenario},''
  \href{http://dx.doi.org/10.1103/PhysRevD.100.123023}{{\em Phys. Rev. D}
  {\bfseries 100} no.~12, (2019) 123023},
  \href{http://arxiv.org/abs/1908.09732}{{\ttfamily arXiv:1908.09732
  [hep-ph]}}.

\bibitem{KumarPoddar:2019jxe}
T.~Kumar~Poddar, S.~Mohanty, and S.~Jana, ``{Constraints on ultralight axions
  from compact binary systems},''
  \href{http://dx.doi.org/10.1103/PhysRevD.101.083007}{{\em Phys. Rev. D}
  {\bfseries 101} no.~8, (2020) 083007},
  \href{http://arxiv.org/abs/1906.00666}{{\ttfamily arXiv:1906.00666
  [hep-ph]}}.

\bibitem{LIGOScientific:2017vwq}
{\bfseries LIGO Scientific, Virgo} Collaboration, B.~P. Abbott {\em et~al.},
  ``{GW170817: Observation of Gravitational Waves from a Binary Neutron Star
  Inspiral},'' \href{http://dx.doi.org/10.1103/PhysRevLett.119.161101}{{\em
  Phys. Rev. Lett.} {\bfseries 119} no.~16, (2017) 161101},
  \href{http://arxiv.org/abs/1710.05832}{{\ttfamily arXiv:1710.05832 [gr-qc]}}.

\bibitem{Taylor:1982zz}
J.~H. Taylor and J.~M. Weisberg, ``{A new test of general relativity:
  Gravitational radiation and the binary pulsar PS R 1913+16},''
  \href{http://dx.doi.org/10.1086/159690}{{\em Astrophys. J.} {\bfseries 253}
  (1982) 908--920}.

\bibitem{Weisberg:1984zz}
J.~M. Weisberg and J.~H. Taylor, ``{Observations of Post-Newtonian Timing
  Effects in the Binary Pulsar PSR 1913+16},''
  \href{http://dx.doi.org/10.1103/PhysRevLett.52.1348}{{\em Phys. Rev. Lett.}
  {\bfseries 52} (1984) 1348--1350}.

\bibitem{Kramer:2006nb}
M.~Kramer {\em et~al.}, ``{Tests of general relativity from timing the double
  pulsar},'' \href{http://dx.doi.org/10.1126/science.1132305}{{\em Science}
  {\bfseries 314} (2006) 97--102},
  \href{http://arxiv.org/abs/astro-ph/0609417}{{\ttfamily
  arXiv:astro-ph/0609417}}.

\bibitem{Antoniadis:2013pzd}
J.~Antoniadis {\em et~al.}, ``{A Massive Pulsar in a Compact Relativistic
  Binary},'' \href{http://dx.doi.org/10.1126/science.1233232}{{\em Science}
  {\bfseries 340} (2013) 6131},
  \href{http://arxiv.org/abs/1304.6875}{{\ttfamily arXiv:1304.6875
  [astro-ph.HE]}}.

\bibitem{Freire:2012mg}
P.~C.~C. Freire, N.~Wex, G.~Esposito-Farese, J.~P.~W. Verbiest, M.~Bailes,
  B.~A. Jacoby, M.~Kramer, I.~H. Stairs, J.~Antoniadis, and G.~H. Janssen,
  ``{The relativistic pulsar-white dwarf binary PSR J1738+0333 II. The most
  stringent test of scalar-tensor gravity},''
  \href{http://dx.doi.org/10.1111/j.1365-2966.2012.21253.x}{{\em Mon. Not. Roy.
  Astron. Soc.} {\bfseries 423} (2012) 3328},
  \href{http://arxiv.org/abs/1205.1450}{{\ttfamily arXiv:1205.1450
  [astro-ph.GA]}}.

\bibitem{Preskill:1982cy}
J.~Preskill, M.~B. Wise, and F.~Wilczek, ``{Cosmology of the Invisible
  Axion},'' \href{http://dx.doi.org/10.1016/0370-2693(83)90637-8}{{\em Phys.
  Lett. B} {\bfseries 120} (1983) 127--132}.

\bibitem{Abbott:1982af}
L.~F. Abbott and P.~Sikivie, ``{A Cosmological Bound on the Invisible Axion},''
  \href{http://dx.doi.org/10.1016/0370-2693(83)90638-X}{{\em Phys. Lett. B}
  {\bfseries 120} (1983) 133--136}.

\bibitem{Dine:1982ah}
M.~Dine and W.~Fischler, ``{The Not So Harmless Axion},''
  \href{http://dx.doi.org/10.1016/0370-2693(83)90639-1}{{\em Phys. Lett. B}
  {\bfseries 120} (1983) 137--141}.

\bibitem{Weinberg:1977ma}
S.~Weinberg, ``{A New Light Boson?},''
  \href{http://dx.doi.org/10.1103/PhysRevLett.40.223}{{\em Phys. Rev. Lett.}
  {\bfseries 40} (1978) 223--226}.

\bibitem{Wilczek:1977pj}
F.~Wilczek, ``{Problem of Strong $P$ and $T$ Invariance in the Presence of
  Instantons},'' \href{http://dx.doi.org/10.1103/PhysRevLett.40.279}{{\em Phys.
  Rev. Lett.} {\bfseries 40} (1978) 279--282}.

\bibitem{Peccei:1977ur}
R.~D. Peccei and H.~R. Quinn, ``{Constraints Imposed by CP Conservation in the
  Presence of Instantons},''
  \href{http://dx.doi.org/10.1103/PhysRevD.16.1791}{{\em Phys. Rev. D}
  {\bfseries 16} (1977) 1791--1797}.

\bibitem{Svrcek:2006yi}
P.~Svrcek and E.~Witten, ``{Axions In String Theory},''
  \href{http://dx.doi.org/10.1088/1126-6708/2006/06/051}{{\em JHEP} {\bfseries
  06} (2006) 051}, \href{http://arxiv.org/abs/hep-th/0605206}{{\ttfamily
  arXiv:hep-th/0605206}}.

\bibitem{Inoue:2008zp}
Y.~Inoue, Y.~Akimoto, R.~Ohta, T.~Mizumoto, A.~Yamamoto, and M.~Minowa,
  ``{Search for solar axions with mass around 1 eV using coherent conversion of
  axions into photons},''
  \href{http://dx.doi.org/10.1016/j.physletb.2008.08.020}{{\em Phys. Lett. B}
  {\bfseries 668} (2008) 93--97},
  \href{http://arxiv.org/abs/0806.2230}{{\ttfamily arXiv:0806.2230
  [astro-ph]}}.

\bibitem{CAST:2008ixs}
{\bfseries CAST} Collaboration, E.~Arik {\em et~al.}, ``{Probing eV-scale
  axions with CAST},''
  \href{http://dx.doi.org/10.1088/1475-7516/2009/02/008}{{\em JCAP} {\bfseries
  02} (2009) 008}, \href{http://arxiv.org/abs/0810.4482}{{\ttfamily
  arXiv:0810.4482 [hep-ex]}}.

\bibitem{Hannestad:2005df}
S.~Hannestad, A.~Mirizzi, and G.~Raffelt, ``{New cosmological mass limit on
  thermal relic axions},''
  \href{http://dx.doi.org/10.1088/1475-7516/2005/07/002}{{\em JCAP} {\bfseries
  07} (2005) 002}, \href{http://arxiv.org/abs/hep-ph/0504059}{{\ttfamily
  arXiv:hep-ph/0504059}}.

\bibitem{Melchiorri:2007cd}
A.~Melchiorri, O.~Mena, and A.~Slosar, ``{An improved cosmological bound on the
  thermal axion mass},''
  \href{http://dx.doi.org/10.1103/PhysRevD.76.041303}{{\em Phys. Rev. D}
  {\bfseries 76} (2007) 041303},
  \href{http://arxiv.org/abs/0705.2695}{{\ttfamily arXiv:0705.2695
  [astro-ph]}}.

\bibitem{Hannestad:2008js}
S.~Hannestad, A.~Mirizzi, G.~G. Raffelt, and Y.~Y.~Y. Wong, ``{Cosmological
  constraints on neutrino plus axion hot dark matter: Update after WMAP-5},''
  \href{http://dx.doi.org/10.1088/1475-7516/2008/04/019}{{\em JCAP} {\bfseries
  04} (2008) 019}, \href{http://arxiv.org/abs/0803.1585}{{\ttfamily
  arXiv:0803.1585 [astro-ph]}}.

\bibitem{Hamann:2009yf}
J.~Hamann, S.~Hannestad, G.~G. Raffelt, and Y.~Y.~Y. Wong, ``{Isocurvature
  forecast in the anthropic axion window},''
  \href{http://dx.doi.org/10.1088/1475-7516/2009/06/022}{{\em JCAP} {\bfseries
  06} (2009) 022}, \href{http://arxiv.org/abs/0904.0647}{{\ttfamily
  arXiv:0904.0647 [hep-ph]}}.

\bibitem{Semertzidis:1990qc}
Y.~Semertzidis, R.~Cameron, G.~Cantatore, A.~C. Melissinos, J.~Rogers,
  H.~Halama, A.~Prodell, F.~Nezrick, C.~Rizzo, and E.~Zavattini, ``{Limits on
  the Production of Light Scalar and Pseudoscalar Particles},''
  \href{http://dx.doi.org/10.1103/PhysRevLett.64.2988}{{\em Phys. Rev. Lett.}
  {\bfseries 64} (1990) 2988--2991}.

\bibitem{Cameron:1993mr}
R.~Cameron {\em et~al.}, ``{Search for nearly massless, weakly coupled
  particles by optical techniques},''
  \href{http://dx.doi.org/10.1103/PhysRevD.47.3707}{{\em Phys. Rev. D}
  {\bfseries 47} (1993) 3707--3725}.

\bibitem{Robilliard:2007bq}
C.~Robilliard, R.~Battesti, M.~Fouche, J.~Mauchain, A.-M. Sautivet,
  F.~Amiranoff, and C.~Rizzo, ``{No light shining through a wall},''
  \href{http://dx.doi.org/10.1103/PhysRevLett.99.190403}{{\em Phys. Rev. Lett.}
  {\bfseries 99} (2007) 190403},
  \href{http://arxiv.org/abs/0707.1296}{{\ttfamily arXiv:0707.1296 [hep-ex]}}.

\bibitem{GammeVT-969:2007pci}
{\bfseries GammeV (T-969)} Collaboration, A.~S. Chou, W.~C. Wester, III,
  A.~Baumbaugh, H.~R. Gustafson, Y.~Irizarry-Valle, P.~O. Mazur, J.~H. Steffen,
  R.~Tomlin, X.~Yang, and J.~Yoo, ``{Search for axion-like particles using a
  variable baseline photon regeneration technique},''
  \href{http://dx.doi.org/10.1103/PhysRevLett.100.080402}{{\em Phys. Rev.
  Lett.} {\bfseries 100} (2008) 080402},
  \href{http://arxiv.org/abs/0710.3783}{{\ttfamily arXiv:0710.3783 [hep-ex]}}.

\bibitem{Sikivie:2007qm}
P.~Sikivie, D.~B. Tanner, and K.~van Bibber, ``{Resonantly enhanced
  axion-photon regeneration},''
  \href{http://dx.doi.org/10.1103/PhysRevLett.98.172002}{{\em Phys. Rev. Lett.}
  {\bfseries 98} (2007) 172002},
  \href{http://arxiv.org/abs/hep-ph/0701198}{{\ttfamily arXiv:hep-ph/0701198}}.

\bibitem{Kim:1986ax}
J.~E. Kim, ``{Light Pseudoscalars, Particle Physics and Cosmology},''
  \href{http://dx.doi.org/10.1016/0370-1573(87)90017-2}{{\em Phys. Rept.}
  {\bfseries 150} (1987) 1--177}.

\bibitem{Cheng:1987gp}
H.-Y. Cheng, ``{The Strong CP Problem Revisited},''
  \href{http://dx.doi.org/10.1016/0370-1573(88)90135-4}{{\em Phys. Rept.}
  {\bfseries 158} (1988) 1}.

\bibitem{Rosenberg:2000wb}
L.~J. Rosenberg and K.~A. van Bibber, ``{Searches for invisible axions},''
  \href{http://dx.doi.org/10.1016/S0370-1573(99)00045-9}{{\em Phys. Rept.}
  {\bfseries 325} (2000) 1--39}.

\bibitem{Hertzberg:2008wr}
M.~P. Hertzberg, M.~Tegmark, and F.~Wilczek, ``{Axion Cosmology and the Energy
  Scale of Inflation},''
  \href{http://dx.doi.org/10.1103/PhysRevD.78.083507}{{\em Phys. Rev. D}
  {\bfseries 78} (2008) 083507},
  \href{http://arxiv.org/abs/0807.1726}{{\ttfamily arXiv:0807.1726
  [astro-ph]}}.

\bibitem{Visinelli:2009zm}
L.~Visinelli and P.~Gondolo, ``{Dark Matter Axions Revisited},''
  \href{http://dx.doi.org/10.1103/PhysRevD.80.035024}{{\em Phys. Rev. D}
  {\bfseries 80} (2009) 035024},
  \href{http://arxiv.org/abs/0903.4377}{{\ttfamily arXiv:0903.4377
  [astro-ph.CO]}}.

\bibitem{Battye:1994au}
R.~A. Battye and E.~P.~S. Shellard, ``{Axion string constraints},''
  \href{http://dx.doi.org/10.1103/PhysRevLett.73.2954}{{\em Phys. Rev. Lett.}
  {\bfseries 73} (1994) 2954--2957},
  \href{http://arxiv.org/abs/astro-ph/9403018}{{\ttfamily
  arXiv:astro-ph/9403018}}. [Erratum: Phys.Rev.Lett. 76, 2203--2204 (1996)].

\bibitem{Yamaguchi:1998gx}
M.~Yamaguchi, M.~Kawasaki, and J.~Yokoyama, ``{Evolution of axionic strings and
  spectrum of axions radiated from them},''
  \href{http://dx.doi.org/10.1103/PhysRevLett.82.4578}{{\em Phys. Rev. Lett.}
  {\bfseries 82} (1999) 4578--4581},
  \href{http://arxiv.org/abs/hep-ph/9811311}{{\ttfamily arXiv:hep-ph/9811311}}.

\bibitem{Hagmann:2000ja}
C.~Hagmann, S.~Chang, and P.~Sikivie, ``{Axion radiation from strings},''
  \href{http://dx.doi.org/10.1103/PhysRevD.63.125018}{{\em Phys. Rev. D}
  {\bfseries 63} (2001) 125018},
  \href{http://arxiv.org/abs/hep-ph/0012361}{{\ttfamily arXiv:hep-ph/0012361}}.

\bibitem{Karkevandi:2021ygv}
D.~R. Karkevandi, S.~Shakeri, V.~Sagun, and O.~Ivanytskyi, ``{Bosonic dark
  matter in neutron stars and its effect on gravitational wave signal},''
  \href{http://dx.doi.org/10.1103/PhysRevD.105.023001}{{\em Phys. Rev. D}
  {\bfseries 105} no.~2, (2022) 023001},
  \href{http://arxiv.org/abs/2109.03801}{{\ttfamily arXiv:2109.03801
  [astro-ph.HE]}}.

\bibitem{Nelson:2018xtr}
A.~Nelson, S.~Reddy, and D.~Zhou, ``{Dark halos around neutron stars and
  gravitational waves},''
  \href{http://dx.doi.org/10.1088/1475-7516/2019/07/012}{{\em JCAP} {\bfseries
  07} (2019) 012}, \href{http://arxiv.org/abs/1803.03266}{{\ttfamily
  arXiv:1803.03266 [hep-ph]}}.

\bibitem{Chatziioannou:2020pqz}
K.~Chatziioannou, ``{Neutron star tidal deformability and equation of state
  constraints},'' \href{http://dx.doi.org/10.1007/s10714-020-02754-3}{{\em Gen.
  Rel. Grav.} {\bfseries 52} no.~11, (2020) 109},
  \href{http://arxiv.org/abs/2006.03168}{{\ttfamily arXiv:2006.03168 [gr-qc]}}.

\bibitem{Nishizawa:2016jji}
A.~Nishizawa, E.~Berti, A.~Klein, and A.~Sesana, ``{eLISA eccentricity
  measurements as tracers of binary black hole formation},''
  \href{http://dx.doi.org/10.1103/PhysRevD.94.064020}{{\em Phys. Rev. D}
  {\bfseries 94} no.~6, (2016) 064020},
  \href{http://arxiv.org/abs/1605.01341}{{\ttfamily arXiv:1605.01341 [gr-qc]}}.

\bibitem{Nishizawa:2016eza}
A.~Nishizawa, A.~Sesana, E.~Berti, and A.~Klein, ``{Constraining stellar binary
  black hole formation scenarios with eLISA eccentricity measurements},''
  \href{http://dx.doi.org/10.1093/mnras/stw2993}{{\em Mon. Not. Roy. Astron.
  Soc.} {\bfseries 465} no.~4, (2017) 4375--4380},
  \href{http://arxiv.org/abs/1606.09295}{{\ttfamily arXiv:1606.09295
  [astro-ph.HE]}}.

\bibitem{Breivik:2016ddj}
K.~Breivik, C.~L. Rodriguez, S.~L. Larson, V.~Kalogera, and F.~A. Rasio,
  ``{Distinguishing Between Formation Channels for Binary Black Holes with
  LISA},'' \href{http://dx.doi.org/10.3847/2041-8205/830/1/L18}{{\em Astrophys.
  J. Lett.} {\bfseries 830} no.~1, (2016) L18},
  \href{http://arxiv.org/abs/1606.09558}{{\ttfamily arXiv:1606.09558
  [astro-ph.GA]}}.

\bibitem{Randall:2017jop}
L.~Randall and Z.-Z. Xianyu, ``{Induced Ellipticity for Inspiraling Binary
  Systems},'' \href{http://dx.doi.org/10.3847/1538-4357/aaa1a2}{{\em Astrophys.
  J.} {\bfseries 853} no.~1, (2018) 93},
  \href{http://arxiv.org/abs/1708.08569}{{\ttfamily arXiv:1708.08569 [gr-qc]}}.

\bibitem{Randall:2018qna}
L.~Randall and Z.-Z. Xianyu, ``{An Analytical Portrait of Binary Mergers in
  Hierarchical Triple Systems},''
  \href{http://dx.doi.org/10.3847/1538-4357/aad7fe}{{\em Astrophys. J.}
  {\bfseries 864} no.~2, (2018) 134},
  \href{http://arxiv.org/abs/1802.05718}{{\ttfamily arXiv:1802.05718 [gr-qc]}}.

\bibitem{Randall:2018lnh}
L.~Randall and Z.-Z. Xianyu, ``{A Direct Probe of Mass Density Near Inspiraling
  Binary Black Holes},'' \href{http://dx.doi.org/10.3847/1538-4357/ab20c6}{{\em
  Astrophys. J.} {\bfseries 878} no.~2, (2019) 75},
  \href{http://arxiv.org/abs/1805.05335}{{\ttfamily arXiv:1805.05335 [gr-qc]}}.

\bibitem{KAGRA:2013rdx}
{\bfseries KAGRA, LIGO Scientific, Virgo, VIRGO} Collaboration, B.~P. Abbott
  {\em et~al.}, ``{Prospects for observing and localizing gravitational-wave
  transients with Advanced LIGO, Advanced Virgo and KAGRA},''
  \href{http://dx.doi.org/10.1007/s41114-020-00026-9}{{\em Living Rev. Rel.}
  {\bfseries 21} no.~1, (2018) 3},
  \href{http://arxiv.org/abs/1304.0670}{{\ttfamily arXiv:1304.0670 [gr-qc]}}.

\bibitem{Hild:2009ns}
S.~Hild, S.~Chelkowski, A.~Freise, J.~Franc, N.~Morgado, R.~Flaminio, and
  R.~DeSalvo, ``{A Xylophone Configuration for a third Generation Gravitational
  Wave Detector},'' \href{http://dx.doi.org/10.1088/0264-9381/27/1/015003}{{\em
  Class. Quant. Grav.} {\bfseries 27} (2010) 015003},
  \href{http://arxiv.org/abs/0906.2655}{{\ttfamily arXiv:0906.2655 [gr-qc]}}.

\bibitem{Will:2014kxa}
C.~M. Will, ``{The Confrontation between General Relativity and Experiment},''
  \href{http://dx.doi.org/10.12942/lrr-2014-4}{{\em Living Rev. Rel.}
  {\bfseries 17} (2014) 4}, \href{http://arxiv.org/abs/1403.7377}{{\ttfamily
  arXiv:1403.7377 [gr-qc]}}.

\bibitem{Will:1993hxu}
C.~M. Will, \href{http://dx.doi.org/10.1017/CBO9780511564246}{{\em {Theory and
  Experiment in Gravitational Physics}}}.
\newblock 1993.

\bibitem{Clifton:2011jh}
T.~Clifton, P.~G. Ferreira, A.~Padilla, and C.~Skordis, ``{Modified Gravity and
  Cosmology},'' \href{http://dx.doi.org/10.1016/j.physrep.2012.01.001}{{\em
  Phys. Rept.} {\bfseries 513} (2012) 1--189},
  \href{http://arxiv.org/abs/1106.2476}{{\ttfamily arXiv:1106.2476
  [astro-ph.CO]}}.

\bibitem{Esposito-Farese:2009ouh}
G.~Esposito-Farese, ``{Motion in alternative theories of gravity},''
  \href{http://dx.doi.org/10.1007/978-90-481-3015-3_17}{{\em Fundam. Theor.
  Phys.} {\bfseries 162} (2011) 461--489},
  \href{http://arxiv.org/abs/0905.2575}{{\ttfamily arXiv:0905.2575 [gr-qc]}}.

\bibitem{Fujii:2003pa}
Y.~Fujii and K.~Maeda, \href{http://dx.doi.org/10.1017/CBO9780511535093}{{\em
  {The scalar-tensor theory of gravitation}}}.
\newblock Cambridge Monographs on Mathematical Physics. Cambridge University
  Press, 7, 2007.

\bibitem{Brans:1961sx}
C.~Brans and R.~H. Dicke, ``{Mach's principle and a relativistic theory of
  gravitation},'' \href{http://dx.doi.org/10.1103/PhysRev.124.925}{{\em Phys.
  Rev.} {\bfseries 124} (1961) 925--935}.

\bibitem{Alsing:2011er}
J.~Alsing, E.~Berti, C.~M. Will, and H.~Zaglauer, ``{Gravitational radiation
  from compact binary systems in the massive Brans-Dicke theory of gravity},''
  \href{http://dx.doi.org/10.1103/PhysRevD.85.064041}{{\em Phys. Rev. D}
  {\bfseries 85} (2012) 064041},
  \href{http://arxiv.org/abs/1112.4903}{{\ttfamily arXiv:1112.4903 [gr-qc]}}.

\bibitem{Perivolaropoulos:2009ak}
L.~Perivolaropoulos, ``{PPN Parameter gamma and Solar System Constraints of
  Massive Brans-Dicke Theories},''
  \href{http://dx.doi.org/10.1103/PhysRevD.81.047501}{{\em Phys. Rev. D}
  {\bfseries 81} (2010) 047501},
  \href{http://arxiv.org/abs/0911.3401}{{\ttfamily arXiv:0911.3401 [gr-qc]}}.

\bibitem{Will:1989sk}
C.~M. Will and H.~W. Zaglauer, ``{Gravitational Radiation, Close Binary
  Systems, and the Brans-dicke Theory of Gravity},''
  \href{http://dx.doi.org/10.1086/168016}{{\em Astrophys. J.} {\bfseries 346}
  (1989) 366}.

\bibitem{Berti:2012bp}
E.~Berti, L.~Gualtieri, M.~Horbatsch, and J.~Alsing, ``{Light scalar field
  constraints from gravitational-wave observations of compact binaries},''
  \href{http://dx.doi.org/10.1103/PhysRevD.85.122005}{{\em Phys. Rev. D}
  {\bfseries 85} (2012) 122005},
  \href{http://arxiv.org/abs/1204.4340}{{\ttfamily arXiv:1204.4340 [gr-qc]}}.

\bibitem{Seymour:2019tir}
B.~C. Seymour and K.~Yagi, ``{Probing Massive Scalar Fields from a Pulsar in a
  Stellar Triple System},''
  \href{http://dx.doi.org/10.1088/1361-6382/ab9933}{{\em Class. Quant. Grav.}
  {\bfseries 37} no.~14, (2020) 145008},
  \href{http://arxiv.org/abs/1908.03353}{{\ttfamily arXiv:1908.03353 [gr-qc]}}.

\bibitem{Alexander:2018qzg}
S.~Alexander, E.~McDonough, R.~Sims, and N.~Yunes, ``{Hidden-Sector
  Modifications to Gravitational Waves From Binary Inspirals},''
  \href{http://dx.doi.org/10.1088/1361-6382/aaeb5c}{{\em Class. Quant. Grav.}
  {\bfseries 35} no.~23, (2018) 235012},
  \href{http://arxiv.org/abs/1808.05286}{{\ttfamily arXiv:1808.05286 [gr-qc]}}.

\bibitem{Poddar:2021yjd}
T.~K. Poddar, S.~Mohanty, and S.~Jana, ``{Gravitational radiation from binary
  systems in massive graviton theories},''
  \href{http://dx.doi.org/10.1088/1475-7516/2022/03/019}{{\em JCAP} {\bfseries
  03} (2022) 019}, \href{http://arxiv.org/abs/2105.13335}{{\ttfamily
  arXiv:2105.13335 [gr-qc]}}.

\bibitem{Taylor:1993an}
J.~H. Taylor, ``{Pulsar timing and relativistic gravity},''
  \href{http://dx.doi.org/10.1088/0264-9381/10/S/017}{{\em Class. Quant. Grav.}
  {\bfseries 10} (1993) S167--S174}.

\bibitem{Blum:2014vsa}
K.~Blum, R.~T. D'Agnolo, M.~Lisanti, and B.~R. Safdi, ``{Constraining Axion
  Dark Matter with Big Bang Nucleosynthesis},''
  \href{http://dx.doi.org/10.1016/j.physletb.2014.07.059}{{\em Phys. Lett. B}
  {\bfseries 737} (2014) 30--33},
  \href{http://arxiv.org/abs/1401.6460}{{\ttfamily arXiv:1401.6460 [hep-ph]}}.

\bibitem{Arvanitaki:2010sy}
A.~Arvanitaki and S.~Dubovsky, ``{Exploring the String Axiverse with Precision
  Black Hole Physics},''
  \href{http://dx.doi.org/10.1103/PhysRevD.83.044026}{{\em Phys. Rev. D}
  {\bfseries 83} (2011) 044026},
  \href{http://arxiv.org/abs/1004.3558}{{\ttfamily arXiv:1004.3558 [hep-th]}}.

\bibitem{Arvanitaki:2014wva}
A.~Arvanitaki, M.~Baryakhtar, and X.~Huang, ``{Discovering the QCD Axion with
  Black Holes and Gravitational Waves},''
  \href{http://dx.doi.org/10.1103/PhysRevD.91.084011}{{\em Phys. Rev. D}
  {\bfseries 91} no.~8, (2015) 084011},
  \href{http://arxiv.org/abs/1411.2263}{{\ttfamily arXiv:1411.2263 [hep-ph]}}.

\bibitem{Croon:2017zcu}
D.~Croon, A.~E. Nelson, C.~Sun, D.~G.~E. Walker, and Z.-Z. Xianyu,
  ``{Hidden-Sector Spectroscopy with Gravitational Waves from Binary Neutron
  Stars},'' \href{http://dx.doi.org/10.3847/2041-8213/aabe76}{{\em Astrophys.
  J. Lett.} {\bfseries 858} no.~1, (2018) L2},
  \href{http://arxiv.org/abs/1711.02096}{{\ttfamily arXiv:1711.02096
  [hep-ph]}}.

\bibitem{Zaglauer:1992bp}
H.~W. Zaglauer, ``{Neutron stars and gravitational scalars},''
  \href{http://dx.doi.org/10.1086/171537}{{\em Astrophys. J.} {\bfseries 393}
  (1992) 685--696}.

\bibitem{Hofmann2010}
F.~Hofmann, J.~M{\"u}ller, and L.~Biskupek, ``Lunar laser ranging test of the
  nordtvedt parameter and a possible variation in the gravitational constant,''
  \href{http://dx.doi.org/10.1051/0004-6361/201015659}{{\em Astronomy and
  Astrophysics} {\bfseries 522} (Nov, 2010) L5}.
  \url{https://ui.adsabs.harvard.edu/abs/2010A&A...522L...5H}.

\bibitem{Genova2018}
A.~Genova, E.~Mazarico, S.~Goossens, F.~G. Lemoine, G.~A. Neumann, D.~E. Smith,
  and M.~T. Zuber, ``Solar system expansion and strong equivalence principle as
  seen by the nasa messenger mission,''
  \href{http://dx.doi.org/10.1038/s41467-017-02558-1}{{\em Nature
  Communications} {\bfseries 9} no.~1, (Jan, 2018) 289}.
  \url{https://doi.org/10.1038/s41467-017-02558-1}.

\bibitem{Archibald:2018oxs}
A.~M. Archibald, N.~V. Gusinskaia, J.~W.~T. Hessels, A.~T. Deller, D.~L.
  Kaplan, D.~R. Lorimer, R.~S. Lynch, S.~M. Ransom, and I.~H. Stairs,
  ``{Universality of free fall from the orbital motion of a pulsar in a stellar
  triple system},'' \href{http://dx.doi.org/10.1038/s41586-018-0265-1}{{\em
  Nature} {\bfseries 559} no.~7712, (2018) 73--76},
  \href{http://arxiv.org/abs/1807.02059}{{\ttfamily arXiv:1807.02059
  [astro-ph.HE]}}.

\bibitem{Ransom:2014xla}
S.~M. Ransom {\em et~al.}, ``{A millisecond pulsar in a stellar triple
  system},'' \href{http://dx.doi.org/10.1038/nature12917}{{\em Nature}
  {\bfseries 505} (2014) 520}, \href{http://arxiv.org/abs/1401.0535}{{\ttfamily
  arXiv:1401.0535 [astro-ph.SR]}}.

\bibitem{Day:2019bbh}
F.~V. Day and J.~I. McDonald, ``{Axion superradiance in rotating neutron
  stars},'' \href{http://dx.doi.org/10.1088/1475-7516/2019/10/051}{{\em JCAP}
  {\bfseries 10} (2019) 051}, \href{http://arxiv.org/abs/1904.08341}{{\ttfamily
  arXiv:1904.08341 [hep-ph]}}.

\end{thebibliography}\endgroup
\end{document}